\documentclass[twocolumn,twocolappendix,astrosymb]{aastex631}

\usepackage{amsmath}
\usepackage[angle-symbol-over-decimal,unit-mode=text,table-alignment-mode=format,table-number-alignment=center,uncertainty-mode=separate,print-unity-mantissa=false,list-final-separator={, and },list-units=single]{siunitx}
\usepackage{xspace}
\usepackage{natbib}
\usepackage{multirow}
\usepackage{CJK}
\usepackage{color}
\usepackage{xparse}
\hypersetup{bookmarksnumbered=true} 

\ProvideDocumentCommand\unit{om}{\si[#1]{#2}}
\ProvideDocumentCommand\qty{omm}{\SI[#1]{#2}{#3}}
\ProvideDocumentCommand\qtylist{omm}{\SIlist[#1]{#2}{#3}}
\ProvideDocumentCommand\qtyrange{ommm}{\SIrange[#1]{#2}{#3}{#4}}

\IfFileExists{latexml.sty}
    {\usepackage{latexml}}
    {\newif\iflatexml\latexmlfalse}

\makeatletter
\renewcommand\ion[2]{\texorpdfstring{\text{#1\,\textsc{\@roman{#2}}}}{#1 \@roman{#2}}}
\makeatother
\newcommand*{\HI}{\ion{H}{1}\xspace}
\newcommand*{\sbHI}[1][]{\tsb{\HI\if\relax\detokenize{#1}\relax\else,#1\fi}\xspace}
\newcommand*{\tsb}[1]{\ensuremath{_{\text{#1}}}}

\newcommand*{\RHI}[1][]{\ensuremath{R\sbHI[#1]}\xspace}
\newcommand*{\MHI}{\ensuremath{M\sbHI}\xspace}
\newcommand*{\NHI}{\ensuremath{N\sbHI}\xspace}
\newcommand*{\fHI}{\ensuremath{f\sbHI}\xspace}
\newcommand*{\MHIint}{\ensuremath{M\sbHI[int]}\xspace}
\newcommand*{\NHIlim}{\ensuremath{N\sbHI[lim]}\xspace}

\newcommand*{\uMsun}{\unit{\Msun}\xspace}

\newlength{\myhalfimgsize}
\newcommand*\myplotone[1]{%
  \centering
  \leavevmode
  \includegraphics[width={%
    \ifdim\textwidth=\linewidth%
      2\myhalfimgsize%
    \else\ifdim\linewidth>\myhalfimgsize%
      \myhalfimgsize%
    \else%
      \linewidth
    \fi\fi}]{#1}%
}%

\IfFileExists{newtx.sty}{%
    \usepackage{newtx}
  }{%
    \usepackage{newtxtext,newtxmath}
  }
  \usepackage{CJK}
  \AtBeginDocument{\DeclareSIUnit[quantity-product={}]\degree{\text{\textdegree}}}

\DeclareSIUnit\yr{yr}
\DeclareSIUnit\Gyr{\giga\yr}
\DeclareSIUnit\Myr{\mega\yr}
\DeclareSIUnit\pc{pc}
\DeclareSIUnit\Msun{M\ensuremath{_\odot}}
\DeclareSIUnit\kpc{\kilo\pc}
\DeclareSIUnit\Mpc{\mega\pc}
\DeclareSIUnit\dex{dex}
\DeclareSIUnit\deg{deg}
\DeclareSIUnit\mag{mag}
\DeclareSIUnit\jansky{Jy}
\DeclareSIUnit\beam{beam}
\DeclareSIUnit\pixel{pixel}

\def\kms{\unit{\km\per\s}\xspace}
\def\cmsq{\unit{\per\square\centi\metre}\xspace}

\def\pc{\si{\pc}\xspace}
\def\jyb{\unit{\jansky\per\beam}\xspace}
\def\jypix{\unit{\jansky\per\pixel}\xspace}
\def\kpc{\si{\kilo\pc}\xspace}
\def\mpc{\si{\mega\pc}\xspace}
\def\cm{\si{\centi\metre}\xspace}

\def\MHz{\si{\mega\hertz}\xspace}
\def\mjy{\si{\milli\jansky}\xspace}

\def\hub{\unit{\km.\s^{-1}.\mpc^{-1}}\xspace}

\begin{document}

\begin{CJK*}{UTF8}{gbsn}

\defcitealias{wang2024}{W24}
\defcitealias{wangFEASTSIGMCooling2023b}{W23}

\title{FEASTS Combined with Interferometry (IV): Mapping \HI Emission to a limit of $\NHI=10^{17.7}$ \cmsq in Seven Edge-on Galaxies}
\correspondingauthor{Jing Wang}
\email{jwang\_astro@pku.edu.cn}

\author{Dong Yang (杨冬)}
\affiliation{Kavli Institute for Astronomy and Astrophysics, Peking University, Beijing 100871, People's Republic of China}

\author{Jing Wang (王菁)}
\affiliation{Kavli Institute for Astronomy and Astrophysics, Peking University, Beijing 100871, People's Republic of China}

\author{Zhijie Qu (屈稚杰)}
\affiliation{Department of Astronomy and Astrophysics, The University of Chicago, 5640 S. Ellis Avenue, Chicago, IL 60637, USA}

\author{Zezhong Liang (梁泽众)}
\affiliation{Kavli Institute for Astronomy and Astrophysics, Peking University, Beijing 100871, People's Republic of China}

\author{Xuchen Lin (林旭辰)}
\affiliation{Kavli Institute for Astronomy and Astrophysics, Peking University, Beijing 100871, People's Republic of China}

\author{Simon Weng}
\affiliation{Sydney Institute for Astronomy, School of Physics A28, University of Sydney, NSW 2006, Australia}
\affiliation{ARC Centre of Excellence for All-Sky Astrophysics in 3 Dimensions (ASTRO 3D), Australia}
\affiliation{ATNF, CSIRO Space and Astronomy,  PO Box 76, Epping, NSW 1710, Australia}

\author{Xinkai Chen (陈新凯)}
\affiliation{Kavli Institute for Astronomy and Astrophysics, Peking University, Beijing 100871, People's Republic of China}

\author{Barbara Catinella}
\affiliation{International Centre for Radio Astronomy Research, University of Western Australia, 35 Stirling Highway, Crawley, WA 6009, Australia}
\affiliation{ARC Centre of Excellence for All-Sky Astrophysics in 3 Dimensions (ASTRO 3D), Australia}

\author{Luca Cortese}
\affiliation{International Centre for Radio Astronomy Research, University of Western Australia, 35 Stirling Highway, Crawley, WA 6009, Australia}
\affiliation{ARC Centre of Excellence for All-Sky Astrophysics in 3 Dimensions (ASTRO 3D), Australia}

\author{D. B. Fisher}
\affiliation{Centre for Astrophysics and Supercomputing, Swinburne University of Technology, P.O. Box 218, Hawthorn, VIC 3122, Australia}
\affiliation{ARC Centre of Excellence for All-Sky Astrophysics in 3 Dimensions (ASTRO 3D), Australia}

\author{Luis C. Ho (何子山)}
\affiliation{Kavli Institute for Astronomy and Astrophysics, Peking University, Beijing 100871, People's Republic of China}

\author{Yingjie Jing (景英杰)}
\affiliation{National Astronomical Observatories, Chinese Academy of Sciences, 20A Datun Road, Chaoyang District, Beijing, China}

\author{Fangzhou Jiang (姜方周)}
\affiliation{Kavli Institute for Astronomy and Astrophysics, Peking University, Beijing 100871, People's Republic of China}

\author{Peng Jiang (姜鹏)}
\affiliation{National Astronomical Observatories, Chinese Academy of Sciences, 20A Datun Road, Chaoyang District, Beijing, China}

\author{Ziming Liu (刘孜铭)}
\affiliation{National Astronomical Observatories, Chinese Academy of Sciences, 20A Datun Road, Chaoyang District, Beijing, China}

\author{C\'eline P\'eroux}
\affiliation{European Southern Observatory, Karl-Schwarzschildstrasse 2, D-85748 Garching bei M{\"u}nchen, Germany}
\affiliation{Aix Marseille Universit\'e, CNRS, LAM (Laboratoire d'Astrophysique de Marseille) UMR 7326, 13388, Marseille, France}

\author{Li Shao (邵立)}
\affiliation{National Astronomical Observatories, Chinese Academy of Sciences, 20A Datun Road, Chaoyang District, Beijing, China}

\author{Lister Staveley-Smith}
\affiliation{International Centre for Radio Astronomy Research, University of Western Australia, 35 Stirling Highway, Crawley, WA 6009, Australia}
\affiliation{ARC Centre of Excellence for All-Sky Astrophysics in 3 Dimensions (ASTRO 3D), Australia}

\author{Q. Daniel Wang}
\affiliation{Department of Astronomy, University of Massachusetts, Amherst, MA 01003, USA}

\author{Jie Wang (王杰)}
\affiliation{National Astronomical Observatories, Chinese Academy of Sciences, 20A Datun Road, Chaoyang District, Beijing, China}

\received{2024 November 14}
\revised{2025 February 21}
\accepted{2025 February 22}

\submitjournal{The Astrophysical Journal}
\begin{abstract}
We present a statistical study of the neutral atomic hydrogen (\HI) gas extending into the circumgalactic medium perpendicular to the disk for 7 edge-on galaxies with inclinations above \ang{85} from the FEASTS program with a $3\sigma$ (20 \kms) column density (\NHI) depth of $5\times10^{17}$ \cmsq. 
We develop two photometric methods to separate the extraplanar \HI from the disk component, based on existing interferometric data and parametric modeling of the disk flux distribution respectively.
With both methods, the FEASTS data exhibit clear extended wings beyond the disk along the minor axis.
The extraplanar \HI accounts for 5\% to 20\% of the total \HI mass and extends to $20\text{-}50$ \kpc at $\NHI=10^{18}$ \cmsq. We find a tight positive correlation between vertical extensions of the extraplanar \HI and total \HI mass \MHI. 
The iso-density shape of \HI at $\NHI=10^{18}$ \cmsq has an average axis ratio of $0.56\pm0.11$. The off-disk \NHI profiles of these edge-on galaxies well represent the lower envelop of previous Lyman-$\alpha$ absorption measurements at low-redshift. Our results suggest that at $\NHI=5\times10^{17}$ \cmsq, the \HI extends considerably further than the known thin and thick disks in the vertical direction, but still remains much flattener than a spherical distribution, consistent with theoretical expectations that outflow, circulation, and accretion should have different impacts in these two directions. 
We show the tension of our results with Illustris and TNG predictions, highlighting the constraining power of our results for future simulations.
\end{abstract}


\section{Introduction}\label{sec:intro}
To a large extent, galaxies evolve by accreting, ejecting, and recycling their gas. The circumgalactic medium (CGM) serves as a venue for these processes, mediating the inflow and outflow of gas between the galaxy's interstellar medium (ISM) and its environment \citep{tumlinson2017}. Galaxies need to continuously accrete gas to sustain their star formation \citep{tacconi2010,peroux2020a}. Cosmological simulations predict the gas accretion works in both hot mode and cold mode \citep{keres2005b,keres2009,nelson2013}. In the cold mode accretion, gas at the temperature of $\sim 10^4$ \unit{\kelvin} often inflows in filamentary structures penetrating to the disk. In the hot mode accretion, the hot ($T \geq 10^6$ \unit{\kelvin}) virialized CGM gas cools down and fragments, which may later be accreted. Meanwhile, the feedback energy from stellar feedback and/or active galactic nuclei (AGNs) could drive galactic winds outward \citep{fabian2012}. Part of the outflow gas has velocities below the escaping velocity of galactic halos, indicating much of the gas will be falling back, or ``recycled'' in a galactic fountain \citep{fraternali2017a}. These multiphase gas flows interact with the hot virialized gas in the CGM, exchanging mass, energy and momentum \citep{faucher-giguere2023}. 

The neutral atomic hydrogen gas (\HI) is an important component of the galaxy.
While the \HI gas within the stellar disk can be transformed into molecular gas and form stars \citep{wangXGASSFuelingStar2020}, the \HI gas beyond the optical disk, and even extending far away into the CGM, can serve as the gas reservoir for future star formation (\citealp{lan2018,yu2022}), which indicates a quasi-equilibrium of \HI gas flows throughout the galaxy \citep{wangFEASTSIGMCooling2023b}. In addition, \HI could trace the cool dense gas phase in the CGM and/or ISM-CGM transition regime, which is more closely linked to the galaxy evolution \citep{tumlinson2017}. 
Throughout this paper, we generally refer to the extraplanar \HI gas as any features that are spatially deviating from the thin \HI disk, including both inflow/outflow gas and gas in equilibrium with the virialized CGM. \citet{marascoHALOGASPropertiesExtraplanar2019a} have shown the ubiquitous existence of thick \HI disks in nearby disk galaxies, whose distribution and kinematics are well explained by gas accretion through galactic fountains, as also inferred by the simulation \citep{grand2019}. \citet{martini2018} find a substantial component of \HI gas in the outflow of M82, which extends up to 10 \kpc along the minor axis. Additionally, extraplanar \HI gas can also trace the tidal interaction or mergers between galaxies \citep{wangFEASTSIGMCooling2023b}, which could trigger gas cooling and accretion. 

The \HI gas for the Milky Way has been studied for decades and serves as the cornerstone for our understandings of the extraplanar \HI in other galaxies. In the inner region of the Galaxy, the \HI density distribution along the vertical direction can be described by a thin and a thick Gaussian component, with an additional low-density exponential tail \citep{dickey1990}. In the outer region, the disk becomes more \HI dominated with an increase in the scale height and a strong warp \citep{kalberla2007}. Additionally, the compact ($\leq1$--$10\,\text{kpc}$) high-velocity \HI clouds (HVCs) with anomalous velocities are found several \kpc above the disk in the Milky Way, covering 15\% of the sky above a limiting column density of $2\times 10^{18}$ \cmsq \citep{westmeier2018}. These HVCs, usually forming filaments or clumps, are the coldest and densest component in the CGM and the volume filling factor is small \citep{putman2012}. Counterparts of HVCs have been identified in simulations, which suggest that they originate from the cooling of the hot CGM gas and/or satellite accretion \citep{ramesh2023,2024arXiv240604434L}.   

Interferometric observations are known to be not suitable for detecting the diffuse and extended extraplanar \HI due to the missing short-baselines problem and low sensitivity \citep{hogbom1974, thompson2017}. Indeed \cite{wangFEASTSIGMCooling2023b} and \cite{wang2024} (\citetalias{wangFEASTSIGMCooling2023b} and \citetalias{wang2024} hereafter) have statistically demonstrated that interferometric data tend to miss \HI flux, with the amount of missing \HI depending on the interferometric observational settings and decreasing as the angular size of \HI disk minor axis becomes smaller, transitioning from face-on to edge-on geometries. Edge-on galaxies are suitable candidates for mapping the extraplanar gas since the \HI extending into the CGM can be spatially disentangled from the \HI on the disk. Additionally, both simulations and observations have shown the modulation of gas flow rate direction with azimuthal angle, with the cold gas outflow mostly associated with the minor axis \citep{peroux2020, guo2023}. \citet{dasDetectionDiffuseEmission2020} and \citet{das2024} use Green Bank Telescope (GBT) observations to demonstrate the presence of \HI even 3 GBT beams away (\ang{;27.3}, corresponding to roughly 100 \kpc) from the disk for two edge-on galaxies NGC 891 and NGC 4565. However, the observations cover only several pointings. The diffuse low-density \HI gas in the CGM can also be detected by absorption in QSO spectra \citep{tumlinson2013,liang2014, borthakur2015, johnson2015, prochaskaCOSHalosSurveyMetallicities2017}, but such observations suffer from poor spatial coverage (mostly one pencil beam per galaxy, but see \citealt{bowen2019}). The comparison between emission and absorption results are influenced by the \HI morphology in the CGM \citep{mccourt2018}. Up to now, the extraplanar \HI gas distribution for nearby edge-on galaxies still remains unclear.

In this paper, we focus on the \HI distribution revealed by the FAST observation for seven edge-on galaxies: NGC 4244, NGC 4517, NGC 891, NGC 4565, NGC 1055, NGC 5907, NGC 5775. The \HI data come from the FAST Extended Atlas of Selected Targets Survey (FEASTS; PIs: Jing Wang \& Jie Wang). The FEASTS survey is an ongoing program aimed at mapping \HI in the disk as well as roughly 100 \kpc surroundings for roughly 100 nearby galaxies down to a few times $10^{17}$ \cmsq (see \citetalias{wang2024} for details). Because of the relatively small beam size and high sensitivity of the FAST telescope, the \HI gas distributions for our sample are continuously mapped and studied, especially along the minor axis. We will demonstrate the extensions of the \HI gas into the CGM.

\begin{deluxetable*}{cccccccccc}
    \small
    \tablecaption{The physical properties of our sample}\label{tab:basic}
    \tablehead{
        \colhead{Galaxy} & 
        \colhead{R.A.} & 
        \colhead{Decl.} & 
        \colhead{$v$} & 
        \colhead{Dist} & 
        \colhead{$i$} & 
        \colhead{PA} & 
        \colhead{$\log(\rm M_*/M_\odot)$} & 
        \colhead{$\log$ SFR} &
        \colhead{$R_{25}$}\\
        \colhead{} &  
        \colhead{} & 
        \colhead{} & 
        \colhead{(\kms)} &
        \colhead{(\mpc)} & 
        \colhead{(degree)} & 
        \colhead{(degree)} & 
        \colhead{} & 
        \colhead{(\unit{\Msun\per\yr})} &
        \colhead{(arcsec)}\\
        (1) & (2) & (3) & (4) & (5) & (6) & (7) & (8) & (9) & (10)
    }
    \startdata
    NGC 4244 & 12:17:29.7 & +37:48:25 & 246.0  & $4.24\pm 0.13$  & 88\tablenotemark{a} & 228 & 9.19  & -0.96& 973\\
    NGC 4517 & 12:32:45.6 & +00:06:54 & 1124.6 & $8.30\pm 0.44$  & 86\tablenotemark{b} & 263 & 9.97  & -0.4 & 547\\
    NGC 891  & 02:22:32.9 & +42:20:54 & 535.2  & $9.12\pm 0.34$  & 90\tablenotemark{c} & 202 & 10.65 & 0.25 & 791\\
    NGC 4565 & 12:36:20.8 & +25:59:15 & 1239.4 & $11.87\pm 0.11$ & 88\tablenotemark{d} & 316 & 10.87 & 0.03 & 996\\
    NGC 1055 & 02:41:45.7 & +00:26:35 & 988.0  & $12.62\pm 2.44$ & 86\tablenotemark{e} & 105 & 10.46 & 0.09 & 415\\
    NGC 5907 & 15:15:53.7 & +56:19:44 & 676.7  & $16.9\pm 0.39$  & 88\tablenotemark{f} & 335 & 10.82 & 0.32 & 673\\
    NGC 5775 & 14:53:57.7 & +03:32:40 & 1630.3 & $19.01\pm 3.5$  & 86\tablenotemark{g} & 146 & 10.43 & 0.42 & 223\\
    \enddata
    \tablecomments{
        Column~(1): galaxy name.
        Column~(2): Right Ascension.
        Column~(3): Declination.
        Column~(4): the heliocentric central radio velocity of \HI, from this work.
        Column~(5): luminosity distance, from \citet{tully2023}.
        Column~(6): inclination of the disk from kinematic modeling results: 
                    \tablenotemark{a}{from \cite{zschaechnerHALOGASOBSERVATIONSMODELING2011}},
                    \tablenotemark{b}{from \cite{2022aems.conf..408U}},
                    \tablenotemark{c}{from \cite{oosterlooColdGaseousHalo2007}},
                    \tablenotemark{d}{from \cite{zschaechnerHALOGASOBSERVATIONSMODELING2012}},
                    \tablenotemark{e}{from \cite{schechtman-rook2014}},
                    \tablenotemark{f}{from \cite{yimInterstellarMediumStar2014}},
                    \tablenotemark{g}{from \cite{irwin1994}}.
        Column~(7): the position angle of the galaxy disk, measured from north counterclockwise towards the receding side, obtained from NED.
        Column~(8): the stellar mass from \citet{leroy2019}, rescaled by the distance.
        Column~(9): the star formation rate from \citet{leroy2019}, rescaled by the distance.
        Column~(10): the optical radius $R_{25}$, from HyperLEDA \citep{paturel2003}.
    }
\end{deluxetable*}

\begin{deluxetable*}{c|cccc|cccccc}
    \small
    \tablecaption{The basic information for \HI\ observations\label{tab:info}}
    \tablehead{
        ~ & \multicolumn{4}{c|}{Single dish data (FEASTS)} & \multicolumn{6}{c}{Interferometric data} \\
        Galaxy & \colhead{$\sigma$} & \colhead{$\rm \NHIlim$} & \colhead{$\log \MHI$} & $\sigma_{\rm pointing}$ & \colhead{$\sigma$} & \colhead{$\rm W_{\rm ch}$} & \colhead{$\rm b_{\rm maj}$} & \colhead{$\rm b_{\rm min}$} & \colhead{$\rm \NHIlim$} & \colhead{$\log \MHI$} \\
        ~ & (\mjy/$B_{\rm F}$) & ($10^{18} \cmsq$) & ($M_\odot$) & (arcsec) & (\mjy/$B_{\rm int}$) & (\kms) & (arcsec) & (arcsec) & ($10^{18} \cmsq$) & ($M_\odot$) \\
        (1) & (2) & (3) & (4) & (5) & (6) & (7) & (8) & (9) & (10) & (11)
    }
    \startdata
    NGC 4244 & 0.90 & 0.44 & $9.27\pm0.04$  & 7.0 & 0.24 & 4.12 & 37.8 & 34.8 & 5.42 & 9.25 \\
    NGC 4517 & 1.26 & 0.62 & $9.35\pm0.04$  & 10.0 & / & / & / & / & / & / \\
    NGC 891  & 0.93 & 0.46 & $9.61\pm0.04$  & 3.0 & 0.16 & 8.24 & 35.1 & 32.2 & 6.12 & 9.57 \\
    NGC 4565 & 0.92 & 0.46 & $9.96\pm0.04$  & 6.7 & 0.22 & 4.12 & 43.7 & 33.7 & 4.54 & 9.95 \\
    NGC 1055 & 1.13 & 0.56 & $9.68\pm0.04$  & 6.2 & 2.12 & 5.15 & 16.7 & 14.8 & 288.9 & 9.63 \\
    NGC 5907 & 0.94 & 0.47 & $10.32\pm0.04$ & 9.9 & 0.28 & 20.61 & 15.4 & 13.7 & 88.6 & 10.28 \\
    NGC 5775 & 1.09 & 0.54 & $10.00\pm0.04$ & 7.9 & 0.34 & 52.77 & 21.3 & 14.8 & 118.3 & 9.91
    \enddata
    \tablecomments{
        Column~(1): galaxy name.
        Column~(2): rms level of the FEASTS cube in units of \mjy per FEASTS beam.
        Column~(3): \HI column density limit of the FEASTS cube, assuming $3\sigma$ detection and 20 \kms line widths.
        Column~(4): \HI mass detected by the FEASTS data. The uncertainties are dominated by the assumed 10\% calibration errors. 
        Column~(5): the pointing uncertainties of FEASTS observations according to recalibration of the WCS system against the interferometric data. The typical pointing uncertainty of \ang{;;10} \citep{2019SCPMA..6259502J} is used for NGC 4517 without interferometric data.
        Column~(6): rms level of the interferometric cube in units of \mjy per interferometric beam.
        Column~(7): channel widths of the interferometric cube. 
        Column~(8): major axis of the interferometric beam.
        Column~(9): minor axis of the interferometric beam.
        Column~(10): \HI column density limit of the interferometric cube, assuming $3\sigma$ detection and 20 \kms line widths.
        Column~(11): \HI mass detected by the interferometric data.
    }
\end{deluxetable*}

This paper is organized as follows. In Section \ref{sec:data}, we introduce the basic information for the sample and compare the \HI data from FEASTS with interferometric observations. In Section \ref{sec:ana}, we describe the analysis methods used for our sample. We extract profiles to characterize the \HI distribution and present models used to fit these profiles. The results are presented in Section \ref{sec:result}. We highlight the significant extraplanar \HI in the outer regions along the minor axis detected in FEASTS, as revealed by both the model fitting results and comparisons with interferometric data. We also explore the dependencies of the \HI gas extensions on galaxy properties and study the axis ratio of the extraplanar \HI gas. In Section \ref{sec:discuss} we compare the FEASTS data with QSO absorption results and theoretical predictions of hydrodynamical simulations. 
Summaries are provided in Section \ref{sec:summary}. Throughout this paper, we adopt the cosmology from \cite{planckcollaboration2016}, notably $H_0 = 67.7$ \hub, $\Omega_m (z=0) = 0.309$, and $\Omega_\Lambda(z=0) = 0.691$.

\section{Sample}\label{sec:data}
\subsection{The FEASTS data}
In this work we focus on a subset of 7 nearly edge-on galaxies obtained in the FEASTS program to study the extraplanar \HI. The main physical properties for these galaxies, including their inclinations $>\ang{85}$, are shown in Table \ref{tab:basic}. We highlight that the inclinations are all derived from three dimensional kinematic modeling results, and are thus highly accurate. The FAST \HI imaging observations were conducted in 2021 and 2022 with the observing project ID as PT2021\_0071. During the observations, roughly 1 deg by 1 deg field around each target is scanned 3 times along both Dec and RA directions using on-the-fly mapping mode. The 19-beam receiver is used as backend with the feeds rotated \ang{23.4}/\ang{53.4} during Dec/RA scans to satisfy Nyquist sampling. A 10-\unit{\kelvin} noise diode is periodically injected for flux calibration. We refer readers to \citetalias{wangFEASTSIGMCooling2023b} for more details of the observational settings.

The raw FAST data are first cut to a frequency slice of 10 \MHz centered at the target velocity and then reduced by a pipeline developed by the FEASTS team \citepalias{wangFEASTSIGMCooling2023b}. The reduction procedures are standard, including radio frequency interference (RFI) flagging, calibration, gridding and continuum subtraction. The parameters during reduction are optimized specifically for the FEASTS data. The Full-Width-Half-Maximum (FWHM) of the beam for the final data cube is \ang{;3.24}, slightly larger than the raw FAST beam due to the gridding process. The pixel width is \ang{;;30} and the channel width is 1.61 \kms. 

We use the Smooth \& Clip Finder in the SoFiA software \citep{2015MNRAS.448.1922S,westmeierSoFiAAutomatedParallel2021} to generate the source mask. Compared to \citetalias{wang2024}, our parameters for the source finder are slightly looser, with a smoothing kernel of 0, 3, 5, 7 pixels along the x/y direction and 0, 3, 7, 15, 31 along the z direction. The detection threshold is set to be $3.8\sigma$ and the reliability threshold to be 0.99. We use a larger smoothing kernel along the z axis and a lower detection threshold to avoid missing any weak flux in the outer regions. The rms level and the $3\sigma$ column density detection limit of \HI assuming 20 \kms line width for the final cube are listed in Table \ref{tab:info}. We calculate the moment 0 and moment 1 maps using the source mask. When converting the unit from \jyb to \jypix, we apply a correction factor of 1.064, as in \citetalias{wang2024}, to account for the discrepancy in beam area between the actual beam and the Gaussian beam. We confirm that there is not much flux beyond SoFiA masks, by stacking the spectra beyond the masks but within the virial radius of galaxies. We also tried stacking spetra within apertures of $\sim \ang{;9}$ perpendicular to the disks beyond the masks (see Section \ref{sec:discuss_extension}), and reached a similar conclusion.

\begin{figure*}
        \centering
        \includegraphics[width=0.32\linewidth]{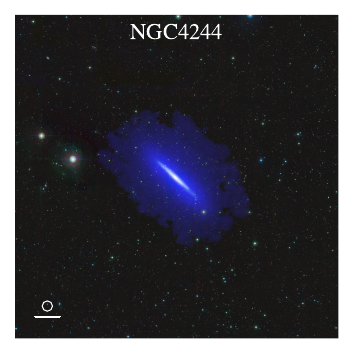}%
        \includegraphics[width=0.32\linewidth]{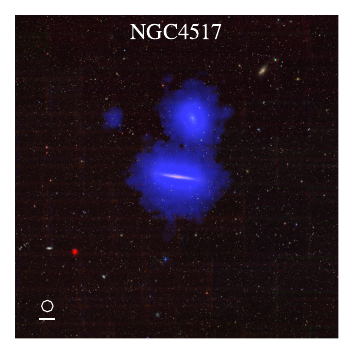}\\
        \includegraphics[width=0.32\linewidth]{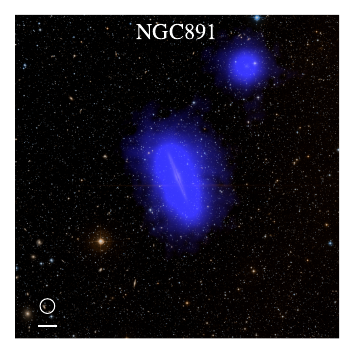}%
        \includegraphics[width=0.32\linewidth]{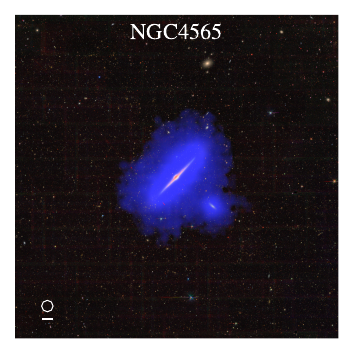}%
        \includegraphics[width=0.32\linewidth]{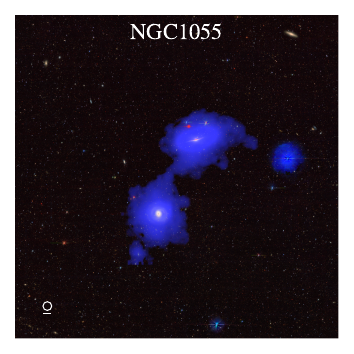}\\
        \includegraphics[width=0.32\linewidth]{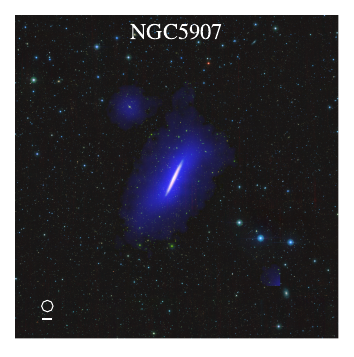}%
        \includegraphics[width=0.32\linewidth]{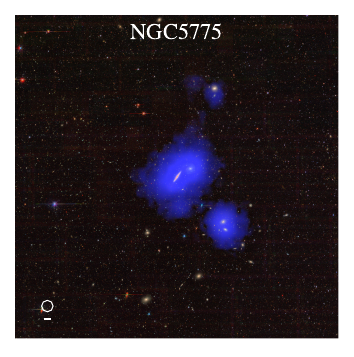}
       \caption{The false-color atlas of our sample showing the FEASTS \HI clouds in blue overlaid on the optical images. The small circle and horizontal bar at the lower left corner correspond to the FEASTS beam and a length of 10 \kpc, respectively. The optical background images are from the Legacy Survey \citep{dey2019}.}\label{fig:atlas}
\end{figure*}

\begin{figure}[!ht]
    \plotone{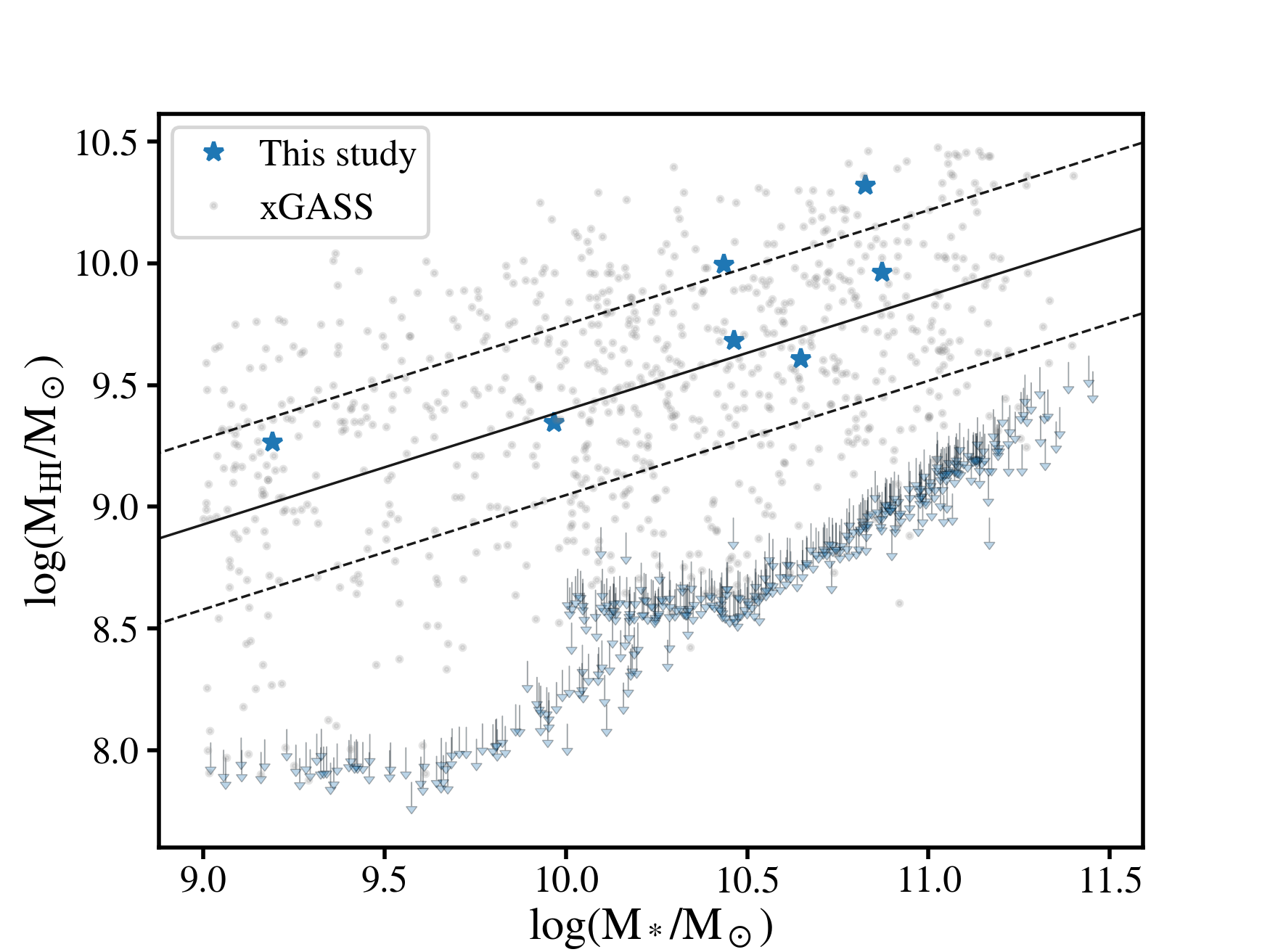}
    \caption{The stellar mass and \HI mass distribution for our edge-on sample. The xGASS \citep{catinella2018} sample is shown in the background as grey dots for comparison, with the mean relation and scatter displayed as solid and dashed lines respectively.}\label{fig:sample}
\end{figure}

Figure \ref{fig:atlas} shows an atlas of false-color images of our sample, with \HI emission overlaid on optical images. Except for NGC 4244, all galaxies have one or more companions. It is worth pointing out that, these targeted galaxies are not selected to have companions, yet \HI-bearing companions seem to be prevalent around these \HI-rich galaxies, as also noticed in \citetalias{wang2024}, being also consistent with the finding that strong \HI absorption targets have often more than one satellite \citep{weng2023a}. Figure \ref{fig:sample} shows the stellar mass and \HI mass distribution for our sample, which lies mostly within the $M_*$\text{-}\MHI main sequence \citep{catinella2018}.

\subsection{The interferometric data}
We assemble the interferometric data for our sample. As shown in \citet{thompson2017}, the interferometric data could miss \HI flux due to the lack of short baselines and limited sensitivity. Comparing the interferometric data with the FEASTS data can help separate the distribution of the dense \HI and reveal the nature of the diffuse \HI. 

The interferometric data for our sample come from various surveys and instruments. We use data from HALOGAS \citep{healdWesterborkHydrogenAccretion2011a} survey for NGC 4244, NGC 891 and NGC 4565; data from CHANGES survey \citep{zhengCHANGESXXVHI2022} for NGC 5775; and VLA archival data for NGC 1055 and NGC 5907, which we reprocess using CASA \citep{mcmullin2007}. The reduction procedure includes standard steps: RFI flagging, calibration, continuum subtraction, imaging and cleaning. We use the auto-multithresh method \citep{kepley2020} in CASA, setting the clean threshold to 3 times the rms derived from the dirty cube. No interferometric data are available for NGC 4517. 

We also use SoFiA to generate the source mask for the interferometric data. The smoothing kernel parameters vary according to the data quality and the detection threshold is set at $4\sigma$. The basic information and rms levels for the interferometric data are also listed in Table~\ref{tab:info}.

\section{Quantifying the extraplanar \HI}\label{sec:ana}
In this paper, we refer to the extraplanar \HI gas extending into the CGM as features that extend beyond the \HI disk. We characterize the extraplanar \HI by first identifying an \HI component closely associated with the disk and then subtracting this component from the total \HI. This approach is necessary to reduce the contamination from PSF scattered light from the disk, given the relatively large FWHM of the FAST beam.

The characterization of \HI related to the disk is conducted in two ways. In the first way, we use the interferometric data detected flux to approximate the disk (Section \ref{sec:simu}). Compared to the \HI gas on the disk, the possible extraplanar \HI gas extending into the CGM is believed to have a lower column density and a much larger vertical extension. Both attributes make it difficult to detect with interferometric data, which as a result tends to capture the flux of \HI gas on the disk. However, the interferometric data for our sample come from different surveys with varying detection limits, making it challenging to uniformly define the extraplanar \HI gas. Moreover, the \HI detected in interferometric data may contain some extraplanar \HI, especially when the data are from HALOGAS \citep{marascoHALOGASPropertiesExtraplanar2019a}. Thus, in the second way we estimate the \HI disk characteristic thickness from the interferometric data and then model the disk flux distribution along the minor axis directly from the FEASTS data (Section \ref{sec:model}). 

Both the interferometry missed \HI and the residual \HI of the disk model indicate the relatively diffuse, extraplanar \HI. The former depends more on the details of interferometric observations, while the latter depends more on the assumed disk model. However, we will show that both measures consistently reveal a far-extending extraplanar \HI into the CGM.

\subsection{Simulating the FEASTS observation of the interferometry detected \HI}\label{sec:simu}
For each galaxy with interferometric data, we calibrate the flux levels and WCS system of the FEASTS data using the interferometric data as the reference dataset, following the approach in \citetalias{wang2024}. The pointing uncertainties of our sample are listed in Table \ref{tab:info} with a median value of \ang{;;6.9}. For NGC 4517, the galaxy without interferometric data, we use the typical pointing uncertainty of \ang{;;10} during FAST observation \citep{2019SCPMA..6259502J}. We derive the missed \HI by subtracting the interferometric data from the FEASTS data. Before subtraction, the datasets are convolved with each other's beam to achieve the same resolution. The FEASTS data retain a similar resolution as roughly $\ang{;3.24}$ after convolution, as interferometric beams are much smaller, so we continue to refer to them as ``FEASTS''. We developed a FEASTS observing simulator to convolve the interferometric data with the FEASTS beam. The convolved data cubes are labeled as ``INTCONV'' in the following analysis. Details of the FEASTS observing simulator are further explained in Section \ref{sec:FAST_simu}.

\subsubsection{A FAST 19-beam observing simulator}\label{sec:FAST_simu}
\begin{figure}[!ht]
    \plotone{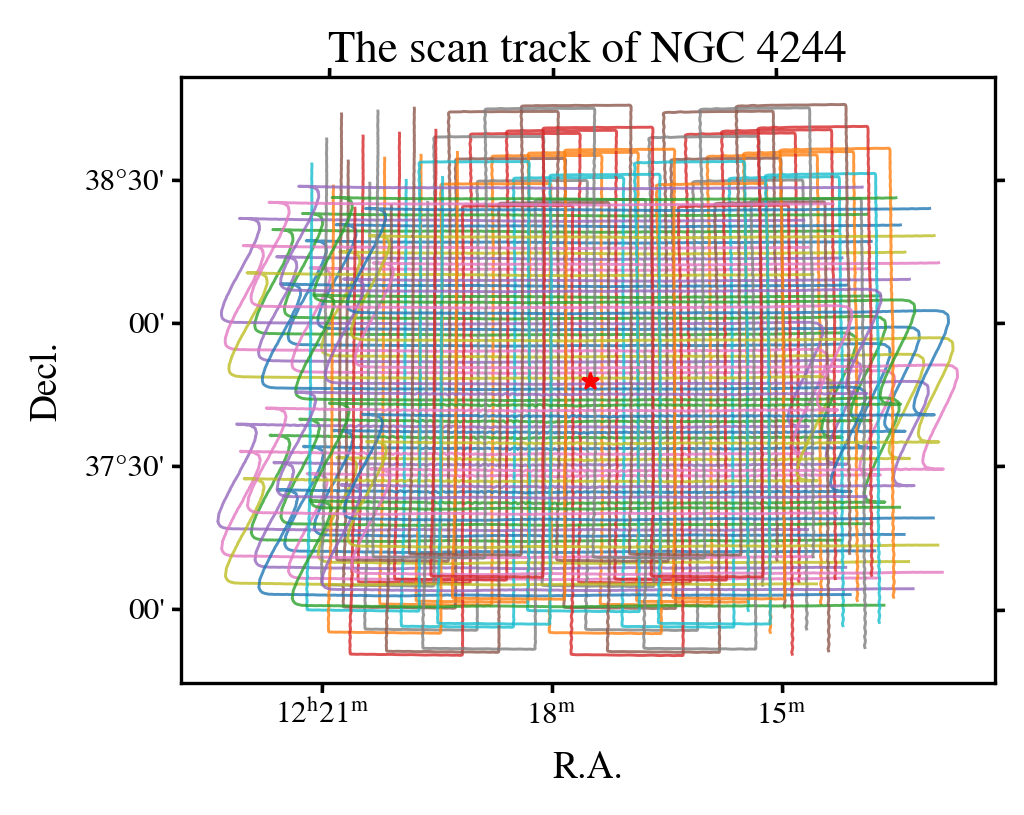}
    \caption{The vertical and horizontal scan track of the 19 beams for NGC 4244 with different colors corresponding to different beams. The red star marks the galaxy center of NGC 4244.}\label{fig:track}
\end{figure}

The FEASTS data are observed using the FAST 19-beam receiver, arranged in a hexagonal pattern and rotated by {\ang{23.4}} (\ang{53.4}) during scans along Dec (RA) directions. In this observing mode, each point in the sky is scanned using different beams, each with slightly different patterns, making it difficult to define a uniform beam for the final data cube (see Figure~\ref{fig:track} for the scan track of the galaxy NGC 4244). \citetalias{wang2024} used an average beam as the effective beam, which does not account for the variation in beam patterns for each sampling. As we focus on the weak diffuse \HI gas extending into the CGM, the different shapes, areas and sidelobe levels of the various beam patterns may introduce uncertainties when we try to separate it from the \HI in the bright disk. It becomes challenging to distinguish between the extraplanar \HI flux and the sidelobe contamination from the galaxy disk.

To address this issue, we have developed a FAST 19-beam observing simulator. We use the interferometric data as the sky model and convolve it with the 38 beam patterns (19 beams and two rotation angles) individually to generate 38 data cubes. Every beam pattern has its sampling stripes on the sky in a complete observation. We calculate the contributing fraction from every beam pattern for each pixel based on its sampling positions. The simulated flux is then calculated by summing the flux from the 38 data cubes, weighted by the contributing fraction. The resulting summed-up data cube is re-projected onto the FEASTS data to match the coordinates. The simulator allows us to mimic the observation of the interferometric data in a manner similar to FEASTS, accounting for the varying beam patterns and samplings. 

We apply the observing simulator to all the interferometric data for our sample. We assess the difference made by the FAST observing simulator by comparing its results with those obtained from directly convolving interferometric data with the average beam, as shown in the Appendix \ref{append:simulator}. 
In particular, we measure $\NHI$ profiles perpendicular to the disks for both treatments and examine the difference (see Figure \ref{fig:INTCONV_diff}). There are considerable offsets between these two profiles within 4 arcmin. The maximum offset reaches $10^{19.5}$ \cmsq, partly because the column densities (thus normalizations) are high in these edge-on systems. The offset profiles are closest to the profiles of FAST-simulator convolved interferometric data (the INTCONV profile defined in Section \ref{sec:simu}) at intermediate radius of 4\text{-}6 arcmin, where the relative difference is -0.6 dex, or the deviation is 25\% of the true value. This value serves as a systematic uncertainty for future analysis that convolves images with the average beam of FAST instead of through the FAST observing simulator. On the other hand, most of the $\NHI$ profiles of excess \HI detected in the FEASTS data with respect to the interferometric data (derived in Section \ref{sec:excess_mom0}) are relatively far from the offset profiles, thus are robust against the systematic uncertainty even if it were produced through convolving with an imperfect average beam.

Because the edge-on galaxies have the highest column densities, the experiment here demonstrates the worst situation of uncertainties in beam (side-lobe) scattering. Our conclusion of the beam uncertainty affecting relatively little the outlying excess \HI with respect to interferometric observations also applies to the outlying extraplanar \HI with respect to a parameterized disk model (Section \ref{sec:model}), because in both case the brightest part of images are removed.

In this study, we mainly use the simulator to obtain differences between FAST and interferometry images, and leave the more powerful forward modeling of \HI disk structures with the simulator to a future study.

\subsection{Modeling the \HI distributions of FEASTS}\label{sec:model}
The edge-on geometry of our sample allows for spatial separation of extraplanar \HI gas from the \HI gas on the disk along the minor axis. 
To increase the signal-to-noise ratio and minimize the uncertainty introduced by radial flux variation in modeling, 
we derive the average column density profile of the \HI along each galaxy's minor axis for FEASTS data, INTCONV data and interferometric data. This profile is referred to hereafter as the z-profile, and the modeling processes are conducted directly on this profile. In this manner, we focus on the average \HI vertical distribution, assuming a uniform characteristic height along the major axis under the FEASTS resolution, which is not strongly disfavored by the moment 0 morphologies in Figure \ref{fig:atlas}.

The z-profiles of FEASTS data contain flux from \HI on the disk and potential extraplanar \HI gas. The typical characteristic thickness for the \HI disk is $\leq 1$ \kpc (see, for example, \citealt{bacchini2019}; \citealt{zheng2022}), corresponding to tens of arcseconds at the distance of our sample. It is expected that the \HI gas on the disk will be smoothed out and mixed with any potential extraplanar \HI under the FEASTS resolution. Therefore, it is challenging to constrain the characteristic thickness of the disk component and set a clear boundary between the disk and the CGM solely using the FEASTS z-profiles. Instead, we derive the characteristic thickness of the disk from z-profiles of interferometric data. Then we fix the characteristic thickness of the disk but let the flux peak vary when modeling the FEASTS z-profiles. This approach allows for simultaneous modeling of the flux of the disk component and the profiles of the extraplanar gas. Details of the fitting process are provided in next Section \ref{sec:fitting_multi}. 

\subsubsection{Fitting Gaussian models for \HI disks and extraplanar \HI} \label{sec:fitting_multi}
To derive the z-profiles for the sample, we first rotate the moment 0 map so that the major axis aligns with the x-axis, positioning the receding part of the velocity field on the positive side. The rotated moment 0 map is then averaged along the x-axis dividing by the mean pixel number along x-axis, assuming the extraplanar \HI shares a similar extent along the x-axis as the disk.

We derive the z-profile of the interferometric data and model it with a single Gaussian function. The Gaussian function is convolved with the 1D beam shape, which is derived by integrating the 2D interferometric beam along the major axis, similar to the operation on the moment 0 map. The $\sigma$ value of the Gaussian function is defined as the disk thickness $z_0$ which is used when modeling the FEASTS z-profiles. We test the influence of the interferometric data depth in estimating the $z_0$ and find that $z_0$ decreases by $\leq 30\%$ when the data detection limit is manually increased from $3\times 10^{19}$ \cmsq to $5 \times 10^{20}$ \cmsq for the HALOGAS data subset (see Appendix \ref{append:h_disk} for details). So $z_0$ can be relatively uniformly estimated for our sample. For NGC 4517 without interferometric data, the $z_0$ is derived by the best-fit relation between $z_0$ and \MHIint in the logarithm space (see Appendix \ref{append:h_disk} for details).

We use a two sided dual Gaussian model (Eq.\ref{eq:triplegauss}) to fit the FEASTS z-profiles. The inner Gaussian component (hereafter labeled as InnerG) represents the disk \HI flux, while the outer Gaussian component captures the potential extraplanar \HI gas. The $\sigma$ value of the inner Gaussian component is kept fixed to $z_0$ during fitting. The outer Gaussian component on the two sides of the disk is modeled separately to account for the possible asymmetric distribution of the extended extraplanar \HI. Similarly the model is convolved with 1D FEASTS beam, integrated from the 2D average beam from \citetalias{wang2024}, during fitting the FEASTS z-profiles. Unlike interferometric z-profiles, FEASTS z-profiles are fitted in the logarithm space to better capture the profile shape in the outer regions. 
The dual Gaussian model fitting results for the FEASTS z-profiles include the PSF-free model and the best-fit profiles (i.e., the former convolved with the FEASTS average beam), with the fitting residuals usually within 0.1 dex. There are some systematical bias for the residuals to be negative at small radius, suggesting that the inner disks are over-estimated and outer components slightly under-estimated (see Section \ref{sec:z-pro}). The errors of PSF-free model profiles are estimated using bootstrapping method, considering both the fitting uncertainties and the errors of $z_0$. For the galaxies NGC 1055, NGC 5907, and NGC 5775, we also include the error introduced by underestimation of $z_0$ due to shallower interferometric data (see Appendix \ref{append:h_disk} for details).

\begin{equation}\label{eq:triplegauss}
    \NHI(z)=\left\{
    \begin{array}{ll}
    A_0e^{-z^2/2z_0^2} + A_1e^{-z^2/2z_{1,\rm lower}^2}, & z<0,\\
    A_0e^{-z^2/2z_0^2} + A_2e^{-z^2/2z_{1,\rm upper}^2}, & z\geq 0,
    \end{array}
    \right.
\end{equation}

It is worth mentioning that historically, the vertical distribution of \HI has typically been modeled with either an exponential function \citep[e.g.,][]{yimInterstellarMediumStar2014} or a Gaussian function \citep[e.g.,][]{koyama2009}, but these models are mostly applied to the disk. The distribution for the extraplanar \HI component does not necessarily follow the disk pattern. \citet{oosterlooColdGaseousHalo2007} use an empirical function to measure the characteristic height of \HI halo, which is zero at the disk plane and declines exponentially outward, minimizing the spatial coexistence of the \HI disk and halo. However, it is difficult to apply this method to our sample due to the resolution limitation of FAST\@. 
Actually, the disk flux is blended with the extraplanar \HI, the distribution shape of which varies from galaxy to galaxy (see Section \ref{sec:z-pro}). Therefore, it is difficult to uniformly fit the disk for extraplanar \HI with a single function for our sample. In this context, we use the Gaussian functions primarily to provide a general characterization of the overall distribution of \HI with the aim of constraining the high-column density inner part with a best effort, so that its scattered light to the outer extraplanar \HI dominated part can be removed.
We have also tested the exponential function and the S\'ersic function to fit z-profiles, but the results are less satisfactory with little or no improvement in fitting the outer regions. Moreover, the characteristic height derived from Gaussian function can be directly compared to the FEASTS beam. For consistency, we also measure the $z_0$ from the interferometric z-profiles using a Gaussian function.

\subsection{Depicting the extraplanar \HI}\label{sec:extrp_HI}
We calculate the $\text{FEASTS}-\text{INTCONV}$ and $\text{FEASTS}-\text{InnerG}*\rm B_F$ z-profiles to depict the extraplanar \HI gas. The INTCONV (interferometric data convolved with FEASTS beam through the FEASTS observing simulator) and InnerG$*\rm B_F$ (the inner-disk Gaussian model convolved with the average FEASTS beam) profiles are complementary to each other in representing the \HI disk. The INTCONV profile is capable of capturing dense structures detectable by interferometric data but may miss flux due to limitations in sensitivity or short spacing. The InnerG$*\rm B_F$ profile can trace the high-density regime of the FEASTS z-profile but is based on the model assumption of the disk z-profile. 
It is possible that INTCONV profiles contain some truly extraplanar flux, particularly when the HALOGAS data are involved, so our estimation may be a lower limit. We refer to the first type of extraplanar \HI as the excess \HI, and the second type as the outer \HI. We quantify the extraplanar \HI distribution from the outer \HI and use the excess \HI distribution as an independent sanity check for the outer \HI results: if the excess \HI extends as far as the outer \HI, then the outer \HI z-profile is likely reasonable.

We calculate the amount of the excess \HI by summing the flux of the three dimensional $\text{FEASTS}-\text{INTCONV}$ data cube using the same source mask as the FEASTS data. The relative fraction of excess \HI is then derived by dividing by the total FEASTS \HI mass. We calculate the moment 0 map of the residual cube as well as the z-profiles to show the distribution of the excess \HI. Since the InnerG$*\rm B_F$ profiles are directly fitted to the FEASTS z-profiles, we can obtain the relative fraction of the outer \HI by dividing the integral of the $\text{FEASTS}-\text{InnerG}*\rm B_F$ by the integral of FEASTS z-profiles. The \HI flux of the target galaxy should be separated from companions when necessary. We use a three-dimensional watershed algorithm to separate the \HI flux of NGC 1055 from its companion M 77 and the flux of NGC 4565 from NGC 4562, which automatically segments the voxels for each galaxy after manually setting markers \citep{huang2024}. We do not apply this method to NGC 5775 as the \HI flux is significantly mixed with NGC 5774 under FEASTS resolution. We do not exclude companions which have low \HI mass ($\MHI<3\times 10^{8}M_\odot$) or faint optical brightness. These companion galaxies are typically difficult to identify at higher redshifts in \HI-21\cm studies, and their \HI will soon be stripped or merged into the target galaxies.

Unlike the \HI disk, the shape of the excess and outer \HI profiles is weakly influenced by the FEASTS beam. We convolve the outer (excess) \HI profiles with the FEASTS beam and find the characteristic thickness only increases by a median value of 6.3\% (5.9\%) (see Appendix \ref{appendix:beam_smoothing} for details). Their shapes are also close to the outer parts of the PSF-free model, except for the left side of NGC 1055. Thus, we can use these excess and outer \HI profiles to reasonably characterize the distribution of extraplanar \HI into the CGM. We quantify the extension of the extraplanar \HI by deriving the characteristic height $z_{18}$, where the column density of the extraplanar \HI z-profile reaches $10^{18}$ \cmsq. We eliminate the smoothing effects of FEASTS beam using the function $z_{18}=\sqrt{z_{18,\rm obs}^2-\rm FWHM^2}$, which provides a reasonable estimate of the intrinsic height although it is imperfect technically. This is because the excess and outer \HI profiles typically peak at $10^{19.5}$ \cmsq (see Section \ref{sec:z-pro}), suggesting that $10^{18}$ \cmsq corresponds to 3\% level of the peak, or $2.6\sigma$ (roughly 1.1 FWHM) away from the center, for a Gaussian distribution. The photometric uncertainties of $z_{18}$ should be close to the pointing uncertainties listed in Table \ref{tab:info}.

\section{Results}\label{sec:result} 
We first present the fraction of excess \HI and outer \HI, and inspect the image of excess \HI in Section \ref{sec:excess_mom0}. 
Then we show the direct measurements of z-profiles from various images and modeling results in Section \ref{sec:z-pro}. Finally we investigate the vertical distributions of the excess \HI and outer \HI in Section \ref{sec:z18_relation} and study its morphology by measuring the axis ratio in Section \ref{sec:axis_ratio}. 

\subsection{The amount and distribution of the extraplanar \HI}\label{sec:excess_mom0}
\begin{figure*}
    \includegraphics{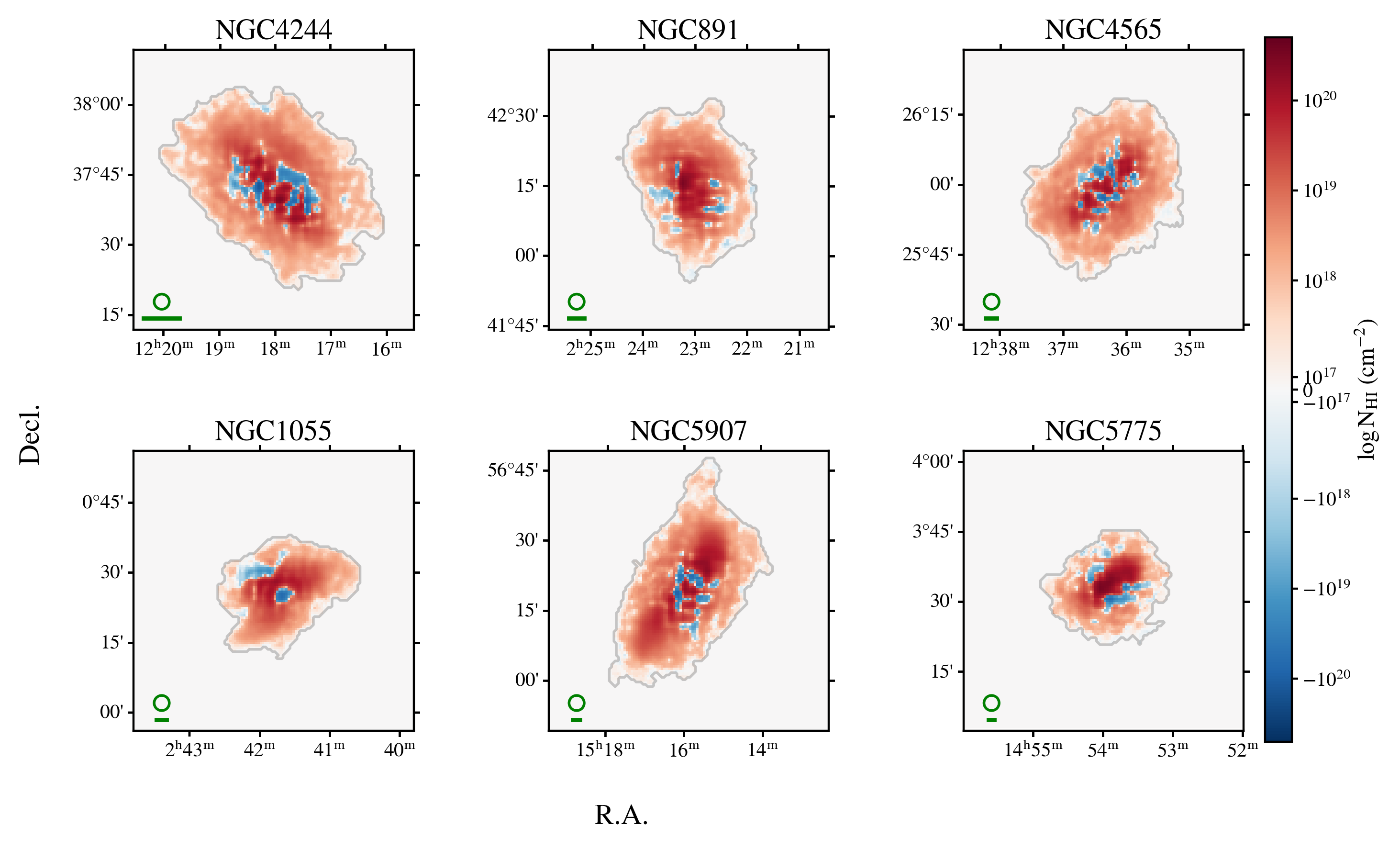}
    \caption{The moment 0 map of the excess \HI for roughly \ang{1} $\times$ \ang{1} regions centered at each galaxy. The gray contours show the mask for the main source from SoFiA and 3D watershed method. The green circle and horizontal bar at the lower left corner correspond to the FEASTS beam and a length of 10 \kpc, respectively. The colorbar is symmetric: red regions show where the FEASTS value is larger, and blue regions show where the INTCONV data (interferometric data convolved with FEASTS beam through the FEASTS observing simulator) value is larger.}\label{fig:res_map}
\end{figure*}

\begin{deluxetable*}{c|cccccccc|ccccc}
    \tabletypesize{\scriptsize}
    \tablecaption{The amount, fraction and $z_{18}$ values for excess and outer \HI for our sample}\label{tab:model_data}
    \tablehead{
        ~ & \multicolumn{8}{c|}{Excess \HI} & \multicolumn{5}{c}{Outer \HI} \\
        Galaxy & 
        \colhead{$\log \MHI$ ($\fHI$)} &
        $f_\text{short-spacing}$&
        \multicolumn{2}{c}{$r_{\rm 18}$} &
        \multicolumn{2}{c}{$z_{\rm 18,lower}$} & 
        \multicolumn{2}{c|}{$z_{\rm 18,upper}$} &
        \colhead{$\log \MHI$ ($\fHI$)} &
        \multicolumn{2}{c}{$z_{\rm 18,lower}$} & 
        \multicolumn{2}{c}{$z_{\rm 18,upper}$} \\
        ~ & 
        $\rm M_\odot$ (/)  & 
        /&
        (\kpc) &
        (arcmin) &
        (\kpc) &
        (arcmin) & 
        (\kpc) &
        (arcmin) & 
        $\rm M_\odot$ (/) &
        (\kpc) &
        (arcmin) &
        (\kpc) &
        (arcmin) \\
        (1) & (2) & (3) & (4) & (5) & (6) & (7) & (8) & (9) & (10) & (11) & (12) & (13) & (14)
    }
    \startdata
    NGC 4244 & 8.0 (5.5\%)  & 86.1\% & 26.7 & 21.7 & 13.4 & 10.9 & 14.3 & 11.6 & 8.3 (11.4\%)  & $13.4\pm 0.6$  & $10.9\pm 0.4$  & $14.3\pm 0.4$  & $11.6\pm 0.3$ \\
    NGC 4517 & /            &   /    & /    & /    & /    & /    & /    & /    & 8.2 (7.3\%)   & $17.8\pm 1.1$  & $7.4\pm 0.5$   & $8.9\pm 4.1$   & $3.7\pm 1.7$ \\
    NGC 891  & 8.6 (10.1\%) & 94.2\% & 40.8 & 15.4 & 16.7 & 6.3  & 25.9 & 9.8  & 8.8 (14.8\%)  & $20.1\pm 1.7$  & $7.6\pm 0.6$   & $26.0\pm 0.9$  & $9.8\pm 0.3$ \\
    NGC 4565 & 8.7 (5.4\%)  & 95.0\% & 58.1 & 16.8 & 33.9 & 9.8  & 34.8 & 10.1 & 9.0 (11.7\%)  & $34.2\pm 1.5$  & $9.9\pm 0.4$   & $34.8\pm 1.4$  & $10.1\pm 0.4$ \\
    NGC 1055 & 8.7 (10.1\%) & 77.7\% & 46.2 & 12.6 & 9.4  & 2.6  & 32.4 & 8.8  & 8.8 (12.4\%)  & $16.6\pm 6.0$  & $4.5\pm 1.6$   & $32.4\pm 1.7$  & $8.8\pm 0.5$ \\
    NGC 5907 & 9.2 (8.2\%)  & 92.5\% & 95.7 & 19.5 & 41.7 & 8.5  & 43.6 & 8.9  & 9.6 (17.0\%)  & $42.9\pm 5.2$  & $8.7\pm 1.1$   & $44.5\pm 1.3$  & $9.0\pm 0.3$ \\
    NGC 5775 & 9.2 (17.5\%) & 51.6\% & 59.7 & 10.8 & 30.4 & 5.5  & /    & /    & 9.0 (10.9\%)  & $29.7\pm 16.2$ & $5.4\pm 2.9$   & /    & / \\
    \enddata
    \tablecomments{
        Column~(1): galaxy name.
        \iflatexml%
            See the PDF version for other columns.
        \else%
            Column~(2): the excess \HI mass and fraction over the total \HI.
            Column~(3): the excess \HI fraction missed due to short-spacing, see text for details.
            Column~(4\text{-}5): the semi-major axis size $r_{18}$ at $10^{18}$ \cmsq level. We note that $r_{18}$ differs from the more commonly used radius measured at a given surface density level (e.g., $R_{\rm 1}$ ($R_{\rm 001}$) at the 1 (0.01) \unit{\Msun\per\square\pc} level), because the column densities suffer from projection effects and the edge-on geometry of our sample makes it hard to deproject.
            Column~(6\text{-}9): the $z_{18}$ values for lower and upper side of the excess \HI z-profiles, where the profiles decline to $10^{18}$ \cmsq, corrected for the FEASTS beam smoothing effects.
            Column~(10): the outer \HI mass and fraction over the total \HI.
            Column~(11\text{-}14): the $z_{18}$ values for lower and upper side of the outer \HI z-profiles.
        \fi
    }
\end{deluxetable*}

We have calculated the excess \HI mass and fraction from the excess \HI data cube and outer \HI mass and fraction from z-profiles. These values are presented in Table \ref{tab:model_data}. 
As expected, $f_{\rm excess}$ correlates with the depth of interferometric data, ranging approximately between 5\% and 10\% for the deep HALOGAS subsample and between 8\% and 17\% for the other three galaxies. The $f_{\rm excess}$ values for our edge-on sample are generally lower than those for the less inclined galaxies in \citetalias{wang2024}. The $f_{\rm outer}$ values range from 7\% to 17\% and show no dependence on the depth of interferometric data. Instead they tend to trace interacting features, with larger values for NGC 891, NGC 1055 and NGC 5907.

We measure the fraction of excess \HI that is missed due to either spatial filtering or sensitivity limitation of the interferometric data in the same way as \citetalias{wangFEASTSIGMCooling2023b}. The major procedures are summarized here. The FEASTS data are firstly reprojected into the coordinate system of interferometric data and attenuated by the interferometric primary beam. Then we select voxels whose flux are higher than $3\sigma$ rms level of the INTCONV data. By summing up the flux of excess \HI within these voxels, we obtain the excess \HI flux that is missed due to large angular-scale spatial filtering and the ratio over the total excess \HI flux which is denoted as $f_\text{short-spacing}$. The results are also listed in Table \ref{tab:model_data}. We find that $f_\text{short-spacing}$ ranges from 51.6\% to 95.0\% with a median value of 89.3\%, thus most of the excess \HI is missed by interferometric data due to spatial filtering, even for the galaxy NGC 5775 with the shallowest interferometric data. This result is consistent with that of \citetalias{wangFEASTSIGMCooling2023b}, emphasizing the necessity of single-dish data in mapping the large angular-scale gas.

We present the excess \HI moment 0 maps of roughly $\ang{1}\times \ang{1}$ square regions centered on each galaxy for our sample in Figure~\ref{fig:res_map}. Only the target galaxy is shown, with the upper row for the HALOGAS subsample (NGC 4244, NGC 891 and NGC 4565) and the lower row for NGC 1055, NGC 5907 and NGC 5775. The excess \HI maps for the HALOGAS subsample exhibit noise patterns fluctuating around zero within the disk region, indicating that the HALOGAS data capture the \HI in the disk quite well. Therefore, the FEASTS and INTCONV data are consistent with each other and the excess \HI is dominated by noise. However, in the outer regions, there is clear excess \HI in FEASTS data. For the other three galaxies, the excess \HI exists in both disk regions and outside parts. The excess \HI follows the disk center and position angle and shares the same in-plane coverage as the interferometric data. On the other hand, it extends considerably further in the vertical direction. The column density of the excess \HI mostly ranges from $10^{18}$ \cmsq to $10^{19.5}$ \cmsq, well above the FEASTS detection limit. 

There is a tentative trend for the excess \HI to be prevalent in the tidal interacting regions. For NGC 891, the excess \HI assembles at the northwest corner where the filament structure is located \citep{oosterlooColdGaseousHalo2007,chastenet2024}. For NGC 1055, the excess \HI is partly located at the tidal bridge toward the nearby galaxy M 77. The bridge is prominently visible in FEASTS data (see Figure~\ref{fig:atlas}), but hardly seen in interferometric data. For NGC 5907, the excess \HI peaks on the two sides of the galaxy disk, where strong warps are identified \citep{sancisi1976}. In summary, features of tidal interaction in excess \HI are common within the sample, which is consistent with previous results of \citetalias{wang2024}.

We do not present the outer \HI images. They display similar features in the outskirts as excess \HI images. Their possible differences in the inner regions are inherited from the systematic difference in methods deriving them, which is not important as the focus of this paper is the extraplanar \HI extending into the CGM.

We also present the difference of moment 1 images between FEASTS and INTCONV data in Figure \ref{fig:mom1} in Appendix \ref{appendix:mom1}. They provide hints on kinematics of the excess \HI. It is clear that, the flux weighted mean velocity of total \HI detected by FEASTS is slower than that of the dense \HI detected by interferometers, by an extent of  $\sim40$ \kms, at the resolution of $\sim10$ \kpc. The lagging in velocity is consistent with the behavior of diffuse \HI in relatively face-on galaxies \citep{wang2024a}, and of extraplanar gas previously characterized with deep interferometry data \citep{marascoHALOGASPropertiesExtraplanar2019a}. However due to beam smearing, we leave more careful velocity-model based quantifications to future studies, using jointly deconvolved single-dish and interferometry data cubes \citep{2025arXiv250210672L}.

\subsection{The z-profiles of the total and extraplanar \HI}\label{sec:z-pro}

\begin{figure*}
    \centering
    \includegraphics[width=0.7\linewidth]{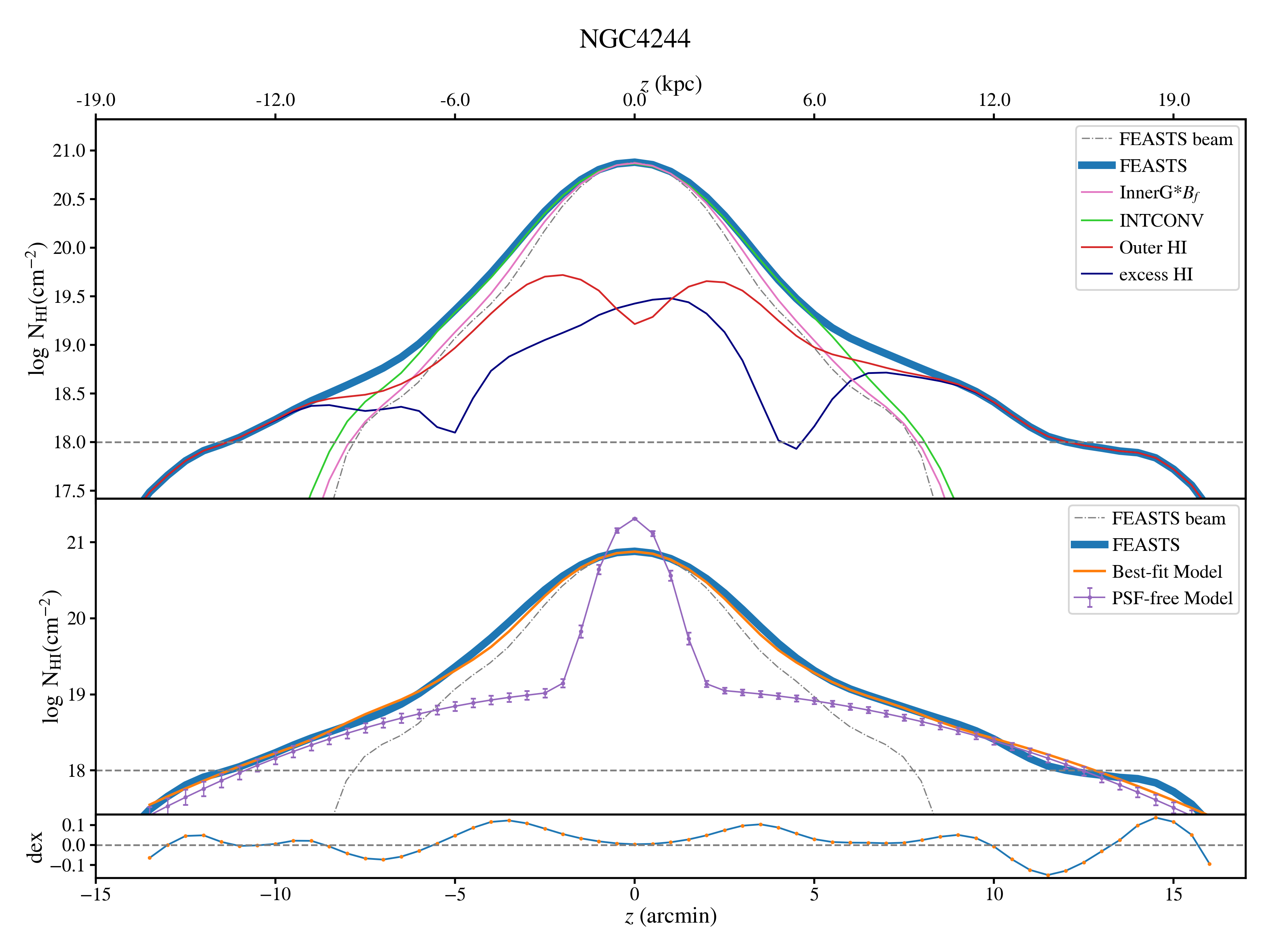}
    \caption{Upper panel: the z-profiles of FEASTS (blue), outer \HI (red) and excess \HI (dark blue) for NGC 4244. The InnerG$*\rm B_F$ (the inner-disk Gaussian model convolved with the average FEASTS beam) and INTCONV (interferometric data convolved with FEASTS beam through the FEASTS observing simulator) profiles are plotted as the pink line and the green line respectively for representing the disk. Middle panel: the best-fit model and PSF-free model profiles, shown as solid orange and purple lines respectively. The errors for the PSF-free model are calculated using bootstrapping method. Lower panel: the fitting residuals. Similar plots for the rest of the sample can be found in Figure \ref{fig:append_profile} in Appendix \ref{appendix:individual}.}\label{fig:all_profile}
\end{figure*}

We show the FEASTS z-profiles and the dual Gaussian model fitting results in the upper and middle panels of Figure \ref{fig:all_profile} for NGC 4244 and for the rest of the sample in Figure \ref{fig:append_profile} in Appendix \ref{appendix:individual}. The InnerG$*\rm B_F$ and INTCONV profiles are plotted as the pink line and the green line respectively in the upper panel to represent the disk flux. The excess \HI ($\text{FEASTS}-\text{INTCONV}$) and the outer \HI ($\text{FEASTS}-\text{InnerG}*\rm B_F$) are also plotted to represent the extraplanar gas.  

We list the best fitting parameters of the dual Gaussian model for our sample in Table \ref{tab:model_par}. The $z_1$ for the extraplanar \HI (typically \ang{;;250}) are generally larger than the FEASTS beam ($\sigma=\ang{;;84.5}$), while the $z_0$ for the disk are much smaller. Consequently, the InnerG$*\rm B_F$ profiles are usually similar to the FEASTS beam profiles, indicating that the disk is unresolved and thus largely unconstrained in the FEASTS data. In this meaning, using the $z_0$ derived from the interferometric data in the modeling represents a best-effort treatment of removing the PSF scattered light from the inner disk to the outlying \HI. In other words, the purpose of the modeling is not to recover the structure of the whole galaxy, but to remove the contamination in the outer \HI and make the outer \HI as resolved as possible. On the other hand, the INTCONV profiles are usually much broader than the FEASTS beam, especially for the HALOGAS subsample, indicating that the interferometric data may capture some extraplanar \HI flux near the disk.

\begin{deluxetable}{ccccccc}
    \tabletypesize{\scriptsize}
    \tablecaption{The two-side dual Gaussian fitting parameters}\label{tab:model_par}
    \tablehead{
        Galaxy & 
        \multicolumn{2}{c}{$z_{\rm 0,disk}$} & 
        \multicolumn{2}{c}{$z_{\rm 1,lower}$} & 
        \multicolumn{2}{c}{$z_{\rm 1,upper}$}\\
        ~ & 
        (\kpc) & 
        (arcsec) & 
        (\kpc) &
        (arcsec) & 
        (\kpc) &
        (arcsec) \\
        (1) & (2) & (3) & (4) & (5) & (6) & (7)
    }
    \startdata
    NGC 4244 & 0.68 & (32.9) & $6.0\pm 0.2$  & $292\pm9$   & $6.9\pm 0.1$  & $334\pm7$ \\
    NGC 4517 & 0.68 & (16.8) & $11.4\pm 0.4$ & $282\pm10$  & $11.6\pm 0.9$ & $289\pm23$ \\
    NGC 891  & 0.68 & (15.4) & $8.2\pm 0.5$  & $186\pm11$  & $9.5\pm 0.3$  & $216\pm6$ \\ 
    NGC 4565 & 1.08 & (18.8) & $14.5\pm 0.5$ & $252\pm8$   & $14.1\pm 0.4$ & $245\pm7$ \\ 
    NGC 1055 & 0.98 & (16.1) & $4.4\pm 1.7$  & $72\pm27$   & $13.2\pm 0.5$ & $215\pm8$ \\ 
    NGC 5907 & 1.91 & (23.3) & $20.7\pm 1.7$ & $253\pm21$  & $13.6\pm 0.3$ & $165\pm4$ \\ 
    NGC 5775 & 1.46 & (15.8) & $23.1\pm 4.8$ & $250\pm52$  & $20.3\pm 3$   & $220\pm32$ 
    \enddata
    \tablecomments{
        Column~(1): galaxy name.
        Column~(2\text{-}3): the \HI disk characteristic thickness derived from interferometric z-profiles.
        Column~(4\text{-}7): the $z_1$ values for lower and upper side of FEASTS z-profiles, representing the extraplanar \HI.
    }
\end{deluxetable}

Almost all the FEASTS z-profiles exhibit extended wings compared to the FEASTS beam, the InnerG$*\rm B_F$ profiles and the INTCONV profiles. We can see that the outer \HI profiles differ from the excess \HI profiles in the inner $10$ \kpc, depending on the depth of the interferometric data and the systematic uncertainties of the Gaussian model. However, in the outer region when $\NHI < 10^{18.4}$ \cmsq, or the vertical distance $z> 10\text{-}30$ \kpc, they share similar flattening trends and extend to similar radii, and are significantly flattener than the FEASTS beam and the inner part of the z-profiles, indicating their extended nature. The overall similarity of excess and outer \HI profiles in outer regions suggests that they may represent the similar extraplanar \HI distribution. It is worth noting that the FEASTS z-profiles are often asymmetric in the outer regions and extend to different distances, particularly for the interacting systems NGC 1055 and NGC 891.

\begin{figure*}
    \begin{tabular*}{\linewidth}{@{}c@{}c@{}}
        \includegraphics[width=0.5\linewidth]{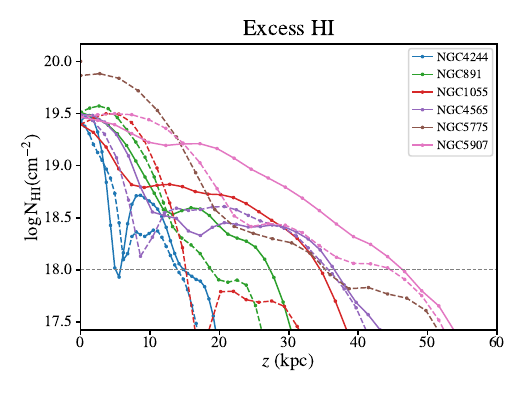} &
        \includegraphics[width=0.5\linewidth]{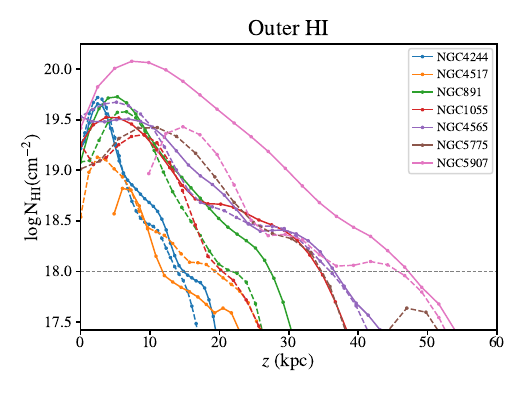} 
    \end{tabular*}
    \caption{The z-profiles for the excess \HI (left) and outer \HI (right). The galaxy labels are ordered according to increasing \HI mass. For each galaxy, the distributions for lower and upper side are plotted as dashed and solid lines respectively.}\label{fig:FEASTS_INT_pro}
\end{figure*}

We present the z-profiles of excess and outer \HI for our sample together in Figure~\ref{fig:FEASTS_INT_pro}, with the galaxies ordered by \HI mass. The distributions above and below the disk are plotted as solid and dashed lines respectively to account for the asymmetric distribution. 
The $z_{18}$ values for both the excess and the outer \HI are shown in Table \ref{tab:model_data}. The excess (outer) \HI extends to a maximum of 44 (45) \kpc for NGC 5907 and a median value of 30.4 (26.0) \kpc at the column density of $10^{18}$ \cmsq. While NGC 5907 may be perturbed by strong tidal interaction, we see a rough trend where more massive \HI disks extend further in the z direction. The extension will be investigated further in next Section \ref{sec:z18_relation}.

\subsection{The correlation of the characteristic height $z_{18}$ for the extraplanar \HI}\label{sec:z18_relation}

\begin{figure*}
    \begin{tabular*}{\linewidth}{@{}c@{}c@{}}
        \includegraphics[width=0.5\linewidth]{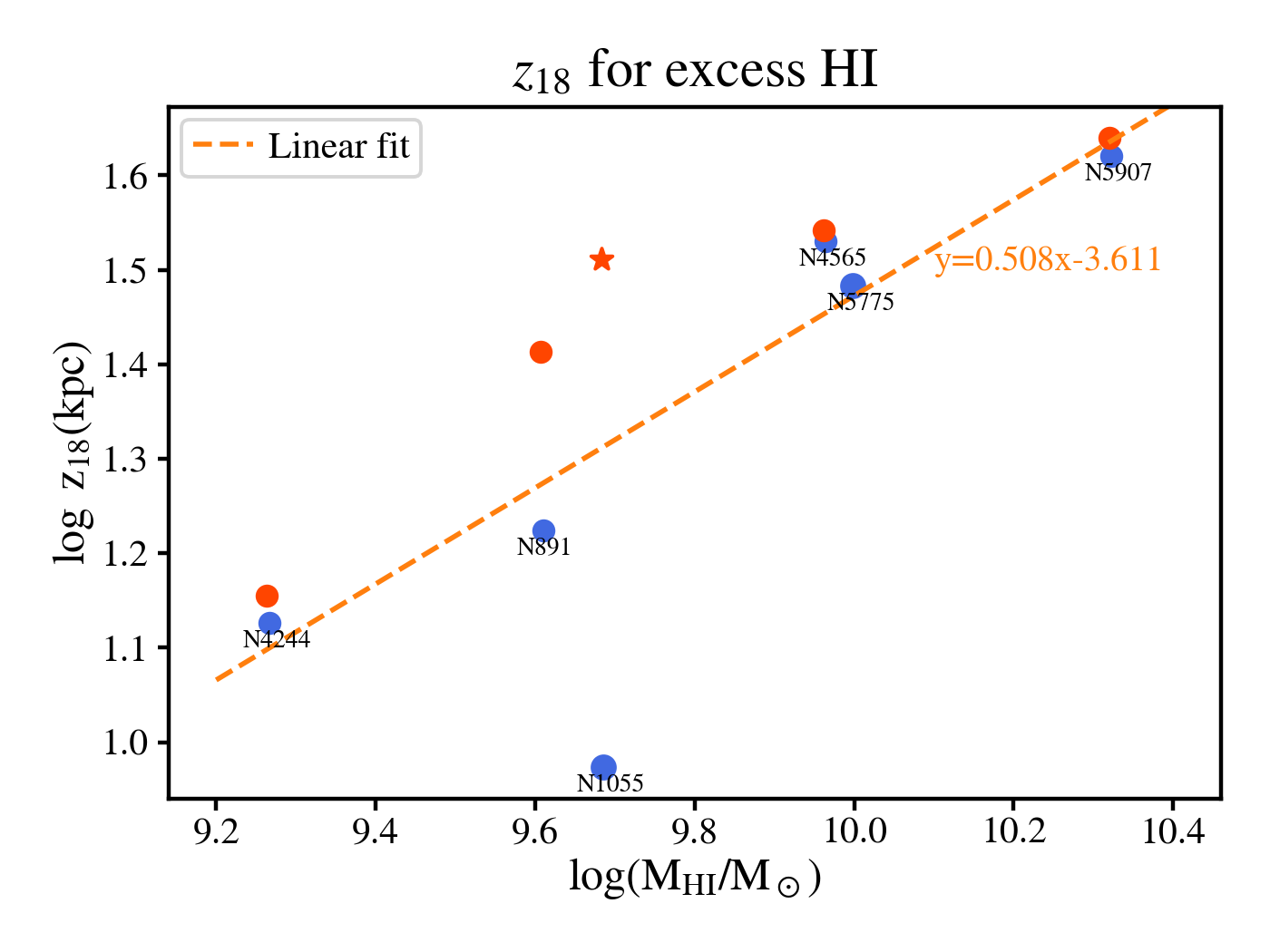} &
        \includegraphics[width=0.5\linewidth]{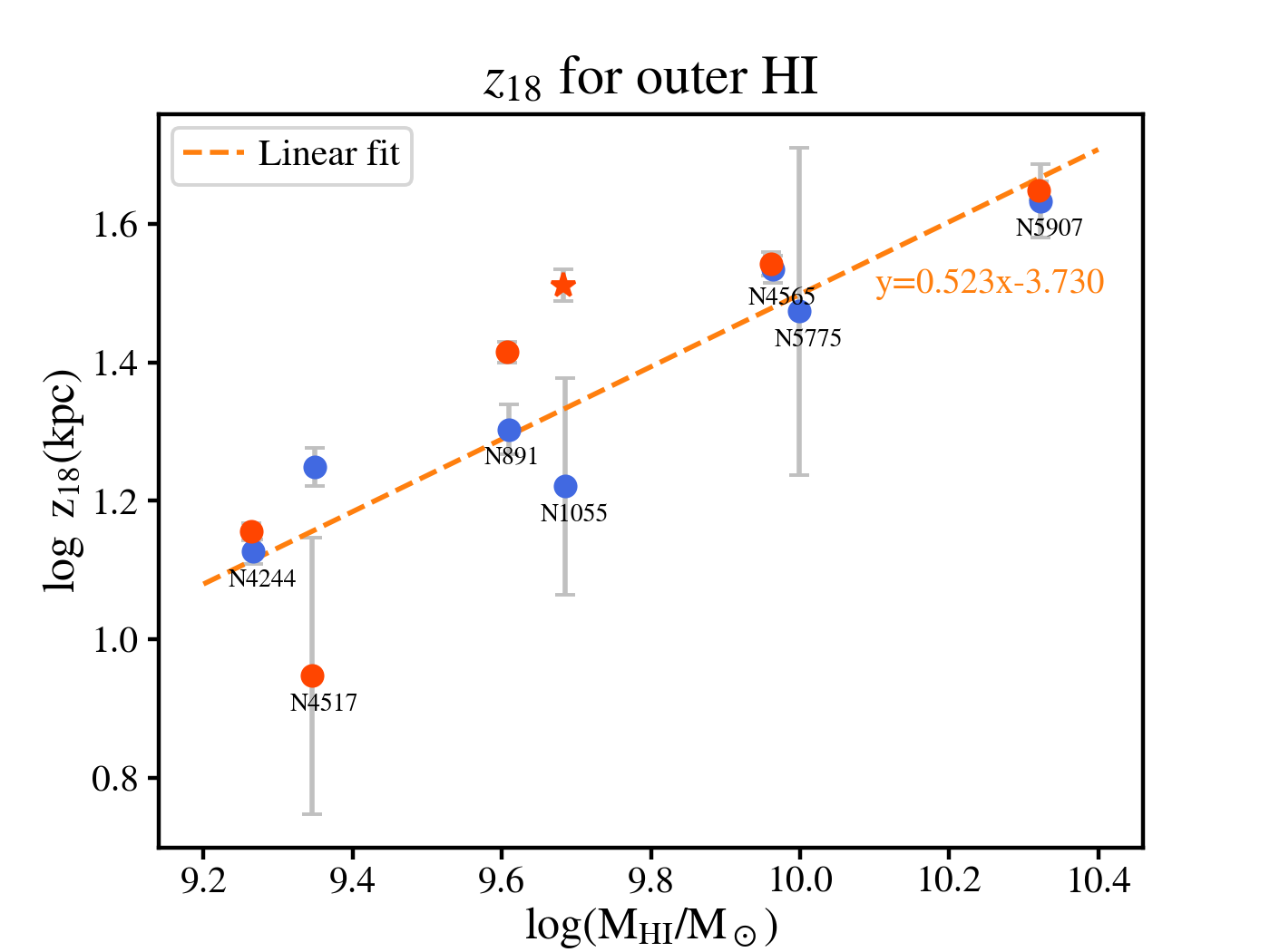}
    \end{tabular*}
    \caption{The relation between the characteristic height $z_{18}$ and the total \HI mass. The $z_{18}$ values are derived from excess \HI profiles (left) and from outer \HI profiles (right). The $z_{18}$ values below and above the disk are shown in blue and red dots, respectively. The star marks the tidal interacting side for NGC 1055, which is not included in calculating correlation coefficients. The $z_{18}$ errors for outer \HI are derived using the bootstrapping method, while the errors for excess \HI are not shown because they are significantly underestimated from the noise in data cubes. The r and p values for Spearman's, Pearson's and partial Pearson's correlation controlling the distance of the correlations are listed in the Table \ref{tab:correlation}. }\label{fig:scale_z18}
\end{figure*}

We study the correlation between the $z_{18}$ of excess and outer \HI profiles and other galaxy properties, including the \HI mass, stellar mass, and star formation rate (SFR). We also test correlations with the surface densities of these quantities, and stellar mass normalized specific values of them. In addition, we derive the partial correlation coefficients with the effect of luminosity distance controlled, in case there are still some resolution-dependent systematical uncertainties in our $z_{18}$ measurements. We list these correlation coefficients as well as their p-values in Table \ref{tab:correlation}. Only total $\MHI$ shows a significant correlation with $z_{18}$ derived from outer \HI profiles (excess \HI profiles), with Pearson's $r=0.91,p=4\times10^{-5}$ ($r=0.92,p=10^{-4}$), and partial correlation Pearson's $r=0.8,p=0.003$ ($r=0.8,p=0.01$). Other parameters all have Pearson's p-values or partial Pearson's p-values above 0.1. We show the trend between $z_{18}$ and $\MHI$ in Figure \ref{fig:scale_z18}.

\begin{deluxetable*}{cccccccccc}
    \tablecaption{The correlation coefficients}\label{tab:correlation}
    \tablehead{
        ~ & ~ & $\log \MHI$ & $\log \rm SFR$ & $\log M_*$ & $\log \Sigma_{\tsb{\scriptsize\HI}}$ & $\log \Sigma_{\rm SFR}$ & $\log \Sigma_{*}$ & $\log \rm sSFR$ & $\log (\MHI/M_*)$ \\
        ~ & ~ & (\unit{\Msun}) & (\unit{\Msun\per\yr}) & (\unit{\Msun}) & (\unit{\Msun\per\square\pc}) & (\unit{\Msun\per\square\pc\per\yr}) & (\unit{\Msun\per\square\pc}) & (\unit{\per\yr}) \\
        ~ & ~ & (1) & (2) & (3) & (4) & (5) & (6) & (7) & (8)
    }
    \startdata
    \multirow{2}{*}{Spearman} & $r$ & 0.90(0.83)      & 0.85(0.75)     & 0.72(0.52) & -0.03(-0.12) & 0.38(-0.12) & 0.01(-0.02) & -0.2(-0.15) & -0.7(-0.67) \\
    ~                         & $p$ & $<$0.001(0.003) & $<$0.001(0.01) & 0.008(0.1) & 0.9(0.7)     & 0.2(0.7)    & 1.0(1.0) & 0.5(0.7) & 0.01(0.03) \\
    \hline
    \multirow{2}{*}{Pearson}  & $r$ & 0.91(0.83)      & 0.77(0.64)   & 0.72(0.56)  & 0.12(0.05) & 0.41(0.30) & 0.07(-0.02) & -0.32(-0.26) & -0.49(-0.49) \\
    ~                         & $p$ & $<$0.001(0.003) & 0.003(0.04)  & 0.008(0.09) & 0.7(0.9)   & 0.2(0.4)   & 0.8(0.9) & 0.3(0.5) & 0.1(0.2) \\
    \hline
    \multirow{2}{*}{Partial Pearson} & $r$ & 0.83(0.78)  & 0.41(0.25) & 0.19(0.07) & -0.36(-0.36) & -0.48(-0.53) & -0.58(-0.61) & 0.07(0.1) & -0.39(-0.39) \\
    ~                                & $p$ & 0.002(0.01) & 0.2(0.5)   & 0.6(0.8)   & 0.3(0.3)     & 0.1(0.1)     & 0.6(0.8) & 0.8(0.8) & 0.2(0.3)
    \enddata
    \tablecomments{ The correlation coefficients between $\log \rm z_{18}$ and other galaxy properties, and the corresponding p-values. The values in and outside the parentheses are coefficients derived from the excess \HI and outer \HI profiles respectively. 
        Column~(1\text{-}3): \HI mass, star formation rate, and stellar mass. 
        Column~(4\text{-}6): The averaged surface density over the optical radius $R_{25}$ for \HI mass, star formation rate, and stellar mass.
        Column~(7): specific star formation rate.
        Column~(8): \HI mass to stellar mass ratio.}

\end{deluxetable*}

The origin of extraplanar gas has been broadly discussed. The galactic fountain driven by supernova feedback is believed to be one plausible mechanism to push the \HI above the disk \citep{bregman1980}. In this scenario, the SFR is expected to be more related to the total mass of the extraplanar gas rather than its characteristic height, according to \cite{marascoHALOGASPropertiesExtraplanar2019a}. Therefore SFR may not be the primary driver of the $z_{18}$ values.

The relation between $z_{18}$ and the total \HI mass may not be too surprising. \cite{randriamampandry2021} found the characteristic height of the \HI disk correlates with the \HI mass, and our results extend the relation to the extraplanar \HI. This resembles the \HI size-mass relation in the disk plane \citep{wangNewLessonsSizemass2016} but along the vertical direction, that is, larger \HI mass corresponds to both larger scale length in the disk and larger characteristic height perpendicular to the disk, indicating self-similar \HI profiles in these two directions.

\subsection{The axis ratio of the extraplanar \HI}\label{sec:axis_ratio}

We quantify the morphology of the extraplanar \HI gas at the first-order by measuring its axis ratio. We use two-dimensional elliptical shape to fit the contour at $10^{18}$ \cmsq level for excess \HI moment 0 map. The semi-major axis sizes $r_{\rm 18}$ of the elliptical models are listed in Table \ref{tab:model_data}. The resulting average minor-to-major axis ratio is $0.69\pm0.09$. We test measuring the axis ratios at different \NHI levels, and find the values to continuously decrease with \NHI, and reaches 0.35 when $\NHI \sim 10^{19.7}$\cmsq. This trend is qualitatively consistent with the \HI disk transiting from the thick to the thin one. Considering possible beam smoothing effects, the actual transition may be steeper than indicated by these measurements.

\begin{figure*}
    \begin{tabular*}{\linewidth}{@{}c@{}c@{}}
        \includegraphics[width=0.5\linewidth]{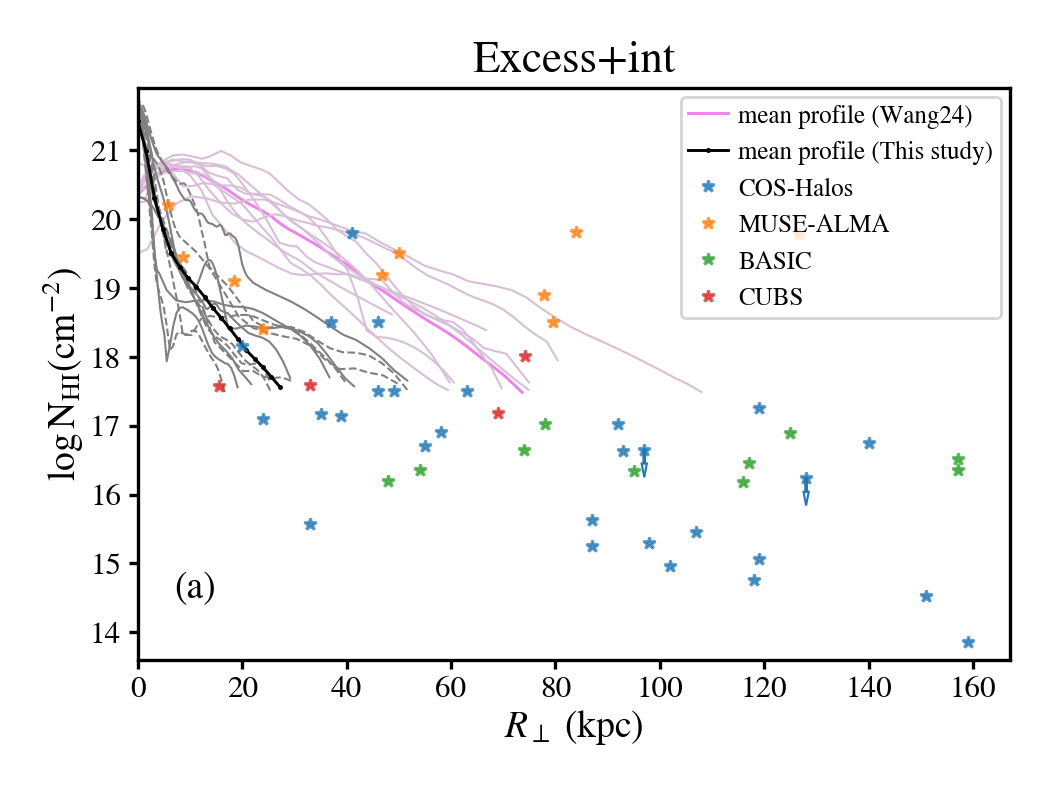} &
        \includegraphics[width=0.5\linewidth]{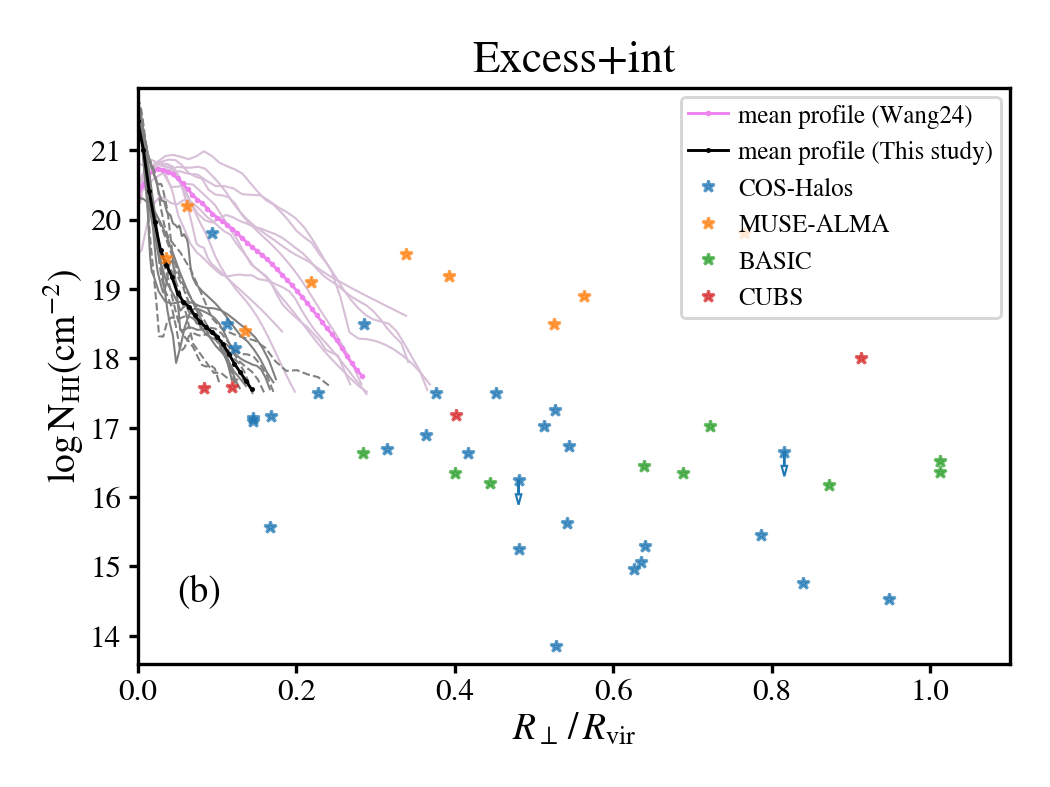} \\
        \includegraphics[width=0.5\linewidth]{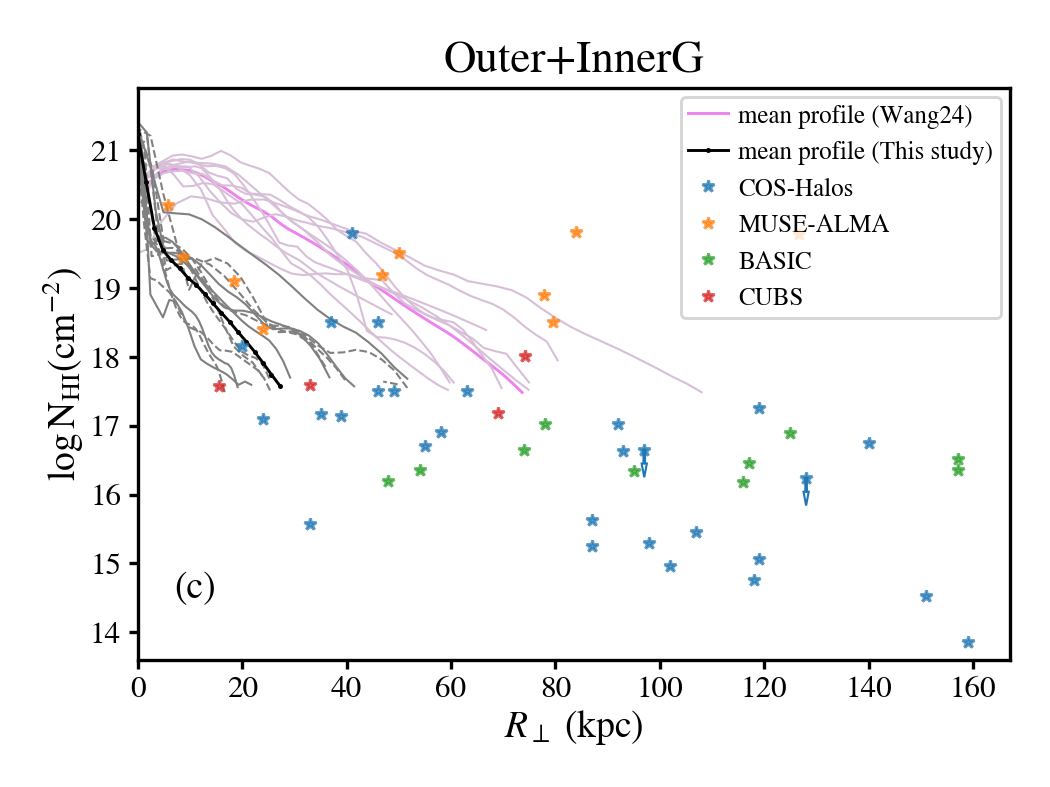} &
        \includegraphics[width=0.5\linewidth]{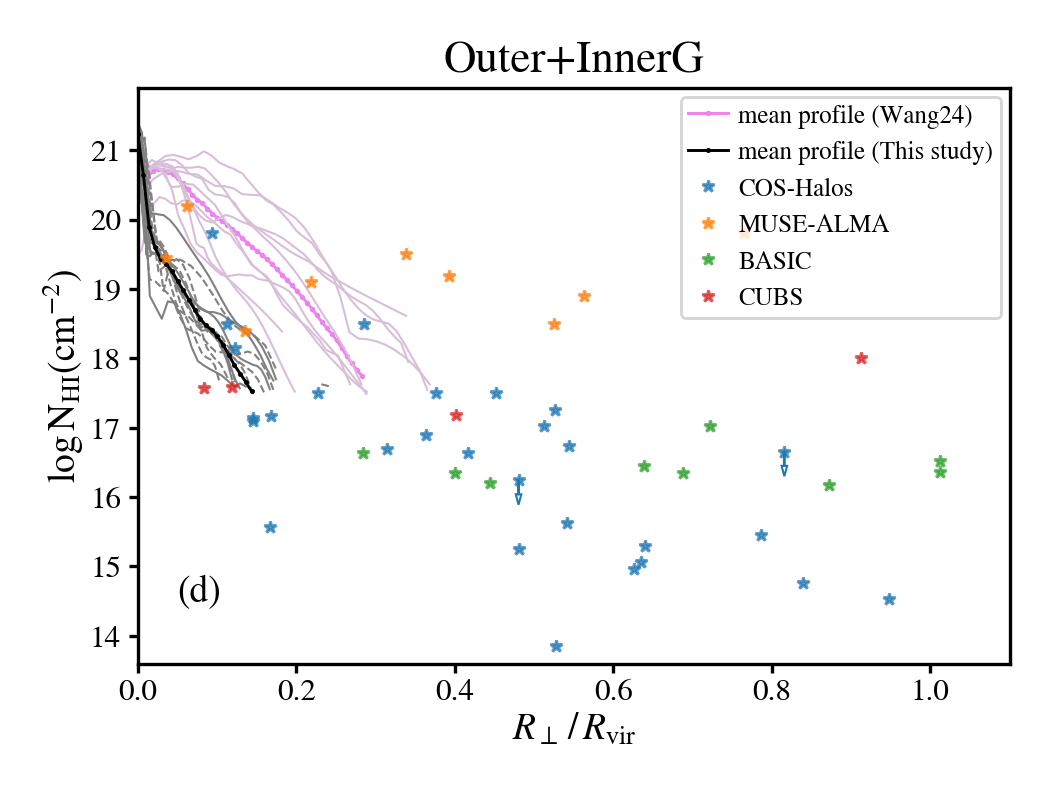}
    \end{tabular*}
    \caption{The $\NHI$ distributions for excess+int (top) and outer \HI+InnerG (bottom) profiles. The profiles along the disk plane for relatively face-on galaxies are also plotted for comparison \citep{wang2024a}. The radius is in the unit of \kpc on the left panel and normalized by $\rm R_{\rm vir}$ on the right panel. The thin pale lines are for each galaxy and the thick bold lines are the mean value in the logarithm space. The Ly $\alpha$ absorption results from COS-Halos \citep{werk2014}, MUSE-ALMA Halos \citep{weng2023}, BASIC \citep{berg2023}, and CUBS \citep{chen2020a} for star-forming galaxies with $\rm sSFR>10^{-11} \unit{\per\yr}$ are also plotted for comparison, see text for details.}\label{fig:modelled_pro}
\end{figure*}

Directly measuring the axis ratio from the excess \HI image as above may still suffer from some beam smoothing effect, which is not fully quantified yet. Therefore, we also estimate the axis ratio by comparing the scale height of the extraplanar \HI from our edge-on sample with the scale length of the planar \HI from relatively face-on FEASTS galaxies in \citet{wang2024a}. To do this, we reconstruct the total \HI z-profile by summing up the excess \HI profile and the interferometric profile (excess+int profile for short hereafter) for our sample. The result is shown in the top row of Figure \ref{fig:modelled_pro}. Mathematically, this reconstruction is similar to the linear combination method of \citet{wang2024a}, or of the Miriad/immerge \citep{sault1995} and CASA/feather \citep{mcmullin2007}. We also reconstruct a second version of the total \HI z-profiles of the FEASTS data with the InnerG profile plus the outer profile for a comparison in the bottom row of Figure \ref{fig:modelled_pro}. The results for the two types of reconstructed \HI profiles are similar, so we will mainly discuss those in the top row based on the excess+int profiles. We calculate the average profiles for both edge-on sample (off-plane z direction) and face-on sample (in-plane radial direction) in the logarithm space, extending to the distance reached by 50\% of the galaxies in each sample. For galaxies with profiles extending to shorter distances than the mean, we extrapolate their profiles outward when calculating the average value. We then fit the outer mean profiles with $\NHI < 10^{19.5}$ \cmsq using linear functions in the logarithm space. The resulting slope ratio between the z- and radial mean profiles is 0.58, consistent with the axis ratio derived from the elliptical fitting.

As an alternative method without fitting, we scale up the radius of the z-profiles for edge-on galaxies until they overlap with the radial profiles for relatively face-on galaxies (see Figure \ref{fig:modelled_scaled_pro} in Appendix \ref{appendix:scaling}), yielding a scaling factor of 2.3, or an axis ratio of 0.43, roughly consistent with the axis ratio derived from the elliptical fitting.

In summary, the three methods exploited here result in similar z-to-radial axis ratios, with an average of $0.56\pm0.11$, which is far from unity.

\section{Discussion}\label{sec:discuss}

The results obtained here provide a first-order 3D view of \HI distribution in the CGM around galaxies. We find that, at the characteristic column density of $10^{18}$ \cmsq, the \HI perpendicular to the disks extends to 20\text{-}50 \kpc with a strong dependence on the \HI mass, further than extension of the conventional thick disk (typically 5\text{-}10 \kpc at $10^{18}$ \cmsq according to \citealt{marascoHALOGASPropertiesExtraplanar2019a}) and likely tracing the ISM-CGM transition regime. The associated extraplanar \HI is an obvious excess to the single Gaussian model or the interferometric observation. But the iso-column-density \HI distribution is still far from being spherical, with an average z-to-radial axis ratio of $0.56\pm0.11$.
Below, we discuss the implication of our results by quantitatively comparing with observational results in the literature, particularly those from absorption line studies, and qualitatively with theoretical predictions from hydrodynamical simulations.

\subsection{The shape of \HI extending into the CGM}\label{sec:discuss_CGM}

The  z-profiles of our sample detect \HI down to the limit of $10^{17.7}$ \cmsq and extend to $20\text{-}50$ \kpc at $10^{18}$ \cmsq above the plane, making it possible to link with the absorption-based results. While absorption measurements can detect \HI down to lower column density limit than FEASTS, they are limited by sparse sampling, typically one line of sight per galaxy for those beyond the Local Group (but see \citealt{weng2023a,weng2023} for the complexity of one-to-one association). 
Combining these two types of data---spatially resolved profiles of individual galaxies and along one line-of-sight but statistical samplings around galaxies of similar types---provides us an opportunity to more deeply characterize each of them.

The absorption line sample used here includes the COS-Halos survey \citep{tumlinson2013} and data compiled in \citet{weng2023}, which incorporate MUSE-ALMA Halos \citep{weng2023}, BASIC \citep{berg2023}, and CUBS \citep{chen2020a}. The absorbing galaxies are divided into star-forming and quenched galaxies based on a threshold of the specific star formation rate $\rm sSFR=10^{-11} \unit{\per\yr}$, and only star-forming galaxies are included for comparison. The COS-Halos survey studies gaseous halos of 44 $z=0.15\text{-}0.35$ galaxies using background QSOs and \cite{prochaskaCOSHalosSurveyMetallicities2017} re-estimate the \HI column densities for 14 Lyman limit systems using new far-ultraviolet spectra. The stellar mass of galaxies spans $\log M_*/M_\odot=9.5\mbox{-}11.5$ and the \HI column density ranges from $10^{14}$ \cmsq to $10^{19}$ \cmsq. The compiled dataset from \citet{weng2023} includes 24 galaxies at $z=0.15\text{-}0.90$, covering $\log M_*/M_\odot=8.0\mbox{-}11.0$ and \HI column density of $10^{16}\mbox{-}10^{20}$ \cmsq. The stellar mass and redshift distribution for the absorption line sample are shown in Appendix \ref{appendix:absorption}.

We recalculate the transverse distance $R_\perp$ for the absorbing galaxies based on our assumed cosmological parameters and calculate the virial radii using the stellar-to-halo mass relation from \citet{moster2010} for both our sample and the absorption line sample. Since there is no orientation information for absorbing galaxies relative to background QSOs, the result of the \HI column density for inner region (i.e. within a few tens \kpc, the radial range detected by FEASTS) of galaxies may be contaminated by \HI from the disk.  We compare the absorption line results with the reconstructed z-profile (see Section \ref{sec:axis_ratio}) and also the deprojected disk-plane radial profile of relatively face-on galaxies from \citet{wang2024a}.

The FEASTS profiles (both z- and radial ones) are consistent with the \HI distributions detected by QSO-absorption in the region where FEASTS have detections ($R_\perp \leq 50$ \kpc $(0.2R_{\rm vir})$, or $\NHI \geq 10^{17.7}$ \cmsq), particularly when measured in \kpc instead of $R_{\rm vir}$. 
Noticeably, the absorption detected \NHI values spread a wide dynamic range (2\text{-}3 dex) at a given transverse distance, and are roughly between z- and radial profiles from FEASTS.
The widely distributed $\NHI$ values at a given radius imply a possible disk-orientation dependence of $\NHI$ near and within the radius of $0.3R_{\rm vir}$ or $\NHI$ of $\sim 10^{17.3}$ \cmsq. 
The comparison indicates a high possibility for absorption line detections of the $\geq 10^{17.7}$ \cmsq \HI to sample the projected disks, and interpreting any central galaxy property dependent trend of absorption strengths in the CGM at/within a given transverse distance should firstly consider the \HI disk projection effect.

\subsubsection{More on a comparison with absorption results}\label{sec:discuss_absorption}
It is worth noticing that absorption line studies have revealed a bimodal concentration of \HI along and perpendicular to the disks at large transverse distances (i.e., close to $R_{\rm vir}$, \citealt{peroux2020a}), while no such trend is observed at intermediate transverse distances (i.e., $\sim 0.5 R_{\rm vir}$, \citealt{borthakur2024}). These results seem to conflict with our findings at the $10^{18}$ \cmsq column density level, showing enhanced \HI along the major axis rather than the minor axis.
The former seemingly tension with the bimodal distribution at $\sim R_{\rm vir}$ in absorption line studies may be attributed to different dominant mechanisms of the same baryonic cycle working at different distances from galaxies. Specifically, smaller distances may preferentially select gas with lower kinetic energy, gravitational energy, and angular momentum. The second tension with lack of azimuthal dependence at intermediate radii in absorption line studies may arise because azimuthal dependence is smoothed by the many factors affecting the \NHI distribution (e.g., \MHI, local density, \citealt{borthakur2024}) and hidden in stochastic sampling with a limited absorption line dataset (see discussion in \citealt{weng2023a}). These two seemingly tensions iterate the complex multi-physical drivers of \HI distribution in the CGM, the importance of combining different observations, and the necessity for larger sample compilations in the future.

\subsection{The vertical extension of \HI in the CGM}\label{sec:discuss_extension}
Recent absorption line studies find that the \HI disk size \RHI (inferred from the \HI mass based on the size-mass relation) is a much better indicator of Ly $\alpha$ equivalent width distribution as a function of transverse distance in the CGM, compared to the combination of halo mass, stellar mass, and SFR \citep{borthakur2024}. This is consistent with planar size-mass relation of \HI in the literature \citep{wangNewLessonsSizemass2016}, and the strong correlation between \MHI and $z_{18}$ found in this study. While these results consistently indicate that low column density \HI in the CGM is closely linked to processes building up the high column density part of the \HI disk, our finding specifically supports that some CGM gas may arrive from the perpendicular direction, possibly related to fountain flows \citep{fraternali2006}, in addition to disk-aligned hot mode accretion \citep{hafen2022}. This is in line with recent findings of simulation-based studies suggesting that gas accretion at low redshift is unlikely to be dominated by any single channel \citep{grand2019b}.

The \HI perpendicular to the disk typically extends out to 20\textsc{-}50 \kpc at the column density level of $10^{18}$ \cmsq in our study. It seems inconsistent with recent GBT observations by \citet{dasDetectionDiffuseEmission2020} and \citet{das2024}, which reported edge-on \HI extending to 50\textsc{-}65 \kpc at $\NHI \sim 10^{18}$ \cmsq and 70\textsc{-}95 \kpc at $\NHI \sim 1.2\times 10^{17}$ \cmsq. The two galaxies studied in their work are also included in our sample and in HALOGAS, allowing for a direct comparison. The details are presented in Appendix \ref{appendix:gbt_spec}. Based on that comparison, we speculate that these discrepancies could be due to the inaccurate beam shape assumed in \citet{das2024}, or significant pointing uncertainties therein.
The findings from \citet{das2024} should thus be taken with caution. In contrast, the HALOGAS and FEASTS results consistently indicate that the edge-on \HI generally extends to only 20\textsc{-}50 \kpc, even when including extreme cases such as intensive starburst galaxies (NGC 1055, \citealt{topal2024}), galaxies with prominent fountain features (NGC 891, \citealt{oosterlooColdGaseousHalo2007}), and strongly interacting systems (NGC 5775, \citealt{irwin1994}).

We note that the diffuse \HI of one FEASTS observed galaxy, NGC 4631, as discovered in \citetalias{wangFEASTSIGMCooling2023b}, indeed extends to 70 \kpc. But this galaxy has experienced multiple encounters with several companions, and is an extremely rare case of FEASTS (see \citetalias{wang2024} for a discussion). It is possible that instances similar to NGC 4631 may be more common at redshift z$>$1, when the Universe is \HI richer and dynamically more active, but such cases should be rare at low redshifts.

\subsubsection{Comparing to \HI thickness predicted in simulations}
Both simulations and observations suggest that the neutral gas along and perpendicular to galaxy disks have different origins, with gas along the disk more associated with the cooling of the hot CGM and gas perpendicular to the disk more linked to the circulation and accretion of feedback-driven fountains \citep{fraternali2017,hafen2022,vayner2023}. Therefore, it may be expected that the $\NHI$ profiles exhibit different shapes in these two directions.

\begin{figure*}
    \begin{tabular*}{\linewidth}{@{}c@{}c@{}}
        \centering
        \includegraphics[width=0.5\linewidth]{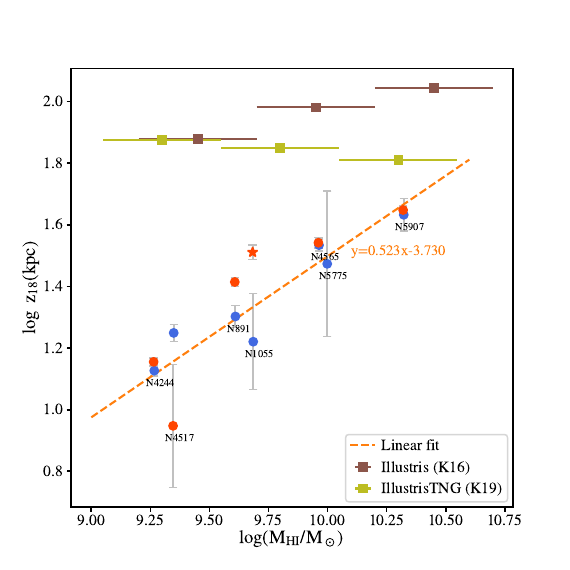} &
        \includegraphics[width=0.47\linewidth]{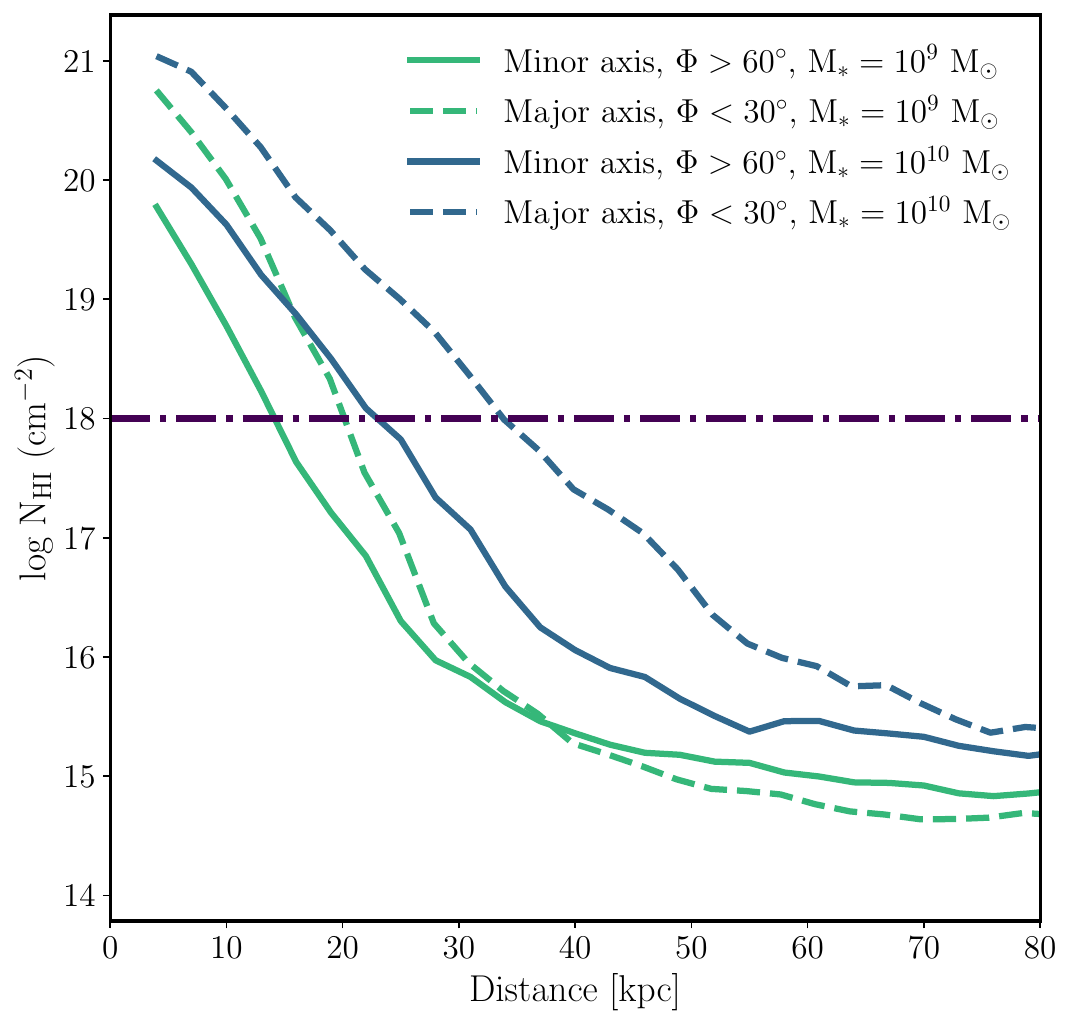}
    \end{tabular*}
    \caption{Left: the comparison of the \HI mass and outer $z_{18}$ distributions between our sample and the Milky-Way type galaxies from Illustris and IllustrisTNG simulations from \citet{kauffmann2016,kauffmann2019}. The symbols are the same with Figure \ref{fig:scale_z18}. Right: the \HI column density distribution versus impact parameters in TNG50 mocks from \citet{weng2024d}. Two colors represent two stellar mass bins (blue for $10^9$ \uMsun, and green for $10^{10}$ \uMsun). The solid and dashed lines correspond to the distribution along the minor and major axis, respectively.}\label{fig:tng50}    
\end{figure*}

However, the predicted symmetry and extensions of \HI along the minor axis on the two sides of gas-rich galaxy disks vary across simulations, depending on the specific tuning of feedback and cooling prescriptions. For example, in the Illustris simulation, AGN feedback in Milky Way (MW) mass galaxies is implemented as 50 \kpc-scale bubbles of hot gas effectively evacuating the CGM \citep{vogelsberger2014}, whereas in IllustrisTNG and all following TNG series, the feedback is set as duty-cycled, randomly oriented injection of momentum \citep{pillepich2018}. Due to these differing feedback recipes, the vertical extension of cool gas around galaxies does not vary with gas mass in the IllustrisTNG simulation, while it does in Illustris \citep{kauffmann2019}. We take \HI mass and $z_{18}$ values of Milky-Way type galaxies selected from Illustris and IllustrisTNG from \citet{kauffmann2016,kauffmann2019} and overplot them with our measurements in the diagram of $z_{18}$ versus \MHI in the left panel of Figure \ref{fig:tng50}. Our finding that $z_{18}$ correlates with \MHI seems to qualitatively support the Illustris prediction. On the other hand, \HI appears to extend too far from the disk at the column density of $10^{18}$ \cmsq in Illustris ($z_{18} \sim 110$ \kpc, indicating an almost spherical distribution), whereas the predictions from IllustrisTNG are closer to the observed values ($z_{18} \sim$ 65 \kpc, indicating much flattener distribution) \citep{kauffmann2016,kauffmann2019, peroux2020}.  Note that \HI is post-processed in both Illustris and TNG simulations, with cold gas ($T=10^4$ \unit{\kelvin}) not split by \HI and $\rm H_{2}$. We quickly check with mock data of moderately inclined galaxies from TNG50 produced in \citet{weng2024d}, and confirm that at the $10^{18}$ \cmsq column density level, \HI along the minor axis does not extend as far as along the major axis of galaxies, as shown in the right panel of Figure \ref{fig:tng50} (See Appendix \ref{appendix:tng} for details). 

On the whole, neither AGN feedback recipe perfectly reproduces the observed edge-on \HI distribution, suggesting that a mixture of the two recipes or an alternative model may be necessary. Our results could provide crucial constraints for the next generation of cosmological simulations.

\section{Summary and conclusion}\label{sec:summary} 
We analyze the \HI distribution for 7 edge-on galaxies using FEASTS data and archival interferometric data. We focus on the vertical distribution of \HI along the minor axis and build models to fit the profiles. Compared to the interferometric data and the disk \HI models, the FEASTS data exhibit clear extraplanar features in the outer region for our sample. The extraplanar \HI extends far to $20\text{-}50$ \kpc away into the CGM, much larger than the typical scale height of the thin and thick \HI disk. However the \HI distribution remains much flattener than a spherical distribution, with an average axis ratio of $0.56\pm 0.11$. The extension $z_{18}$ where the average \HI column density reaches $10^{18}$ \cmsq is found to be tightly positively correlated with the total \HI mass. 

We find the FEASTS vertical profiles for our sample are consistent with the high-column-density inner-halo QSO absorption studies \citep{prochaskaCOSHalosSurveyMetallicities2017,weng2023}, though the slopes are steeper and the \NHI values at a given transverse distance are close to the lower envelop of the distribution, compared to both the \HI distributions from absorption and the in-plane \HI radial profiles for relatively face-on galaxies from \citet{wang2024a}. We compare the edge-on \HI distribution with the results of hydrodynamical simulations and find that Illustris and TNG could only partly explain either the extension of 20\text{-}50 \kpc for the \HI at the $10^{18}$ \cmsq level or the positive correlation between $z_{18}$ and \MHI. Specifically, we check the TNG50 and find that it is consistent with our findings that \HI along the minor axis does not extend as far as the major axis at the $10^{18}$ \cmsq column density level.

The diffuse volume-filling \HI down to the $\NHI$ level of $5\times10^{17}$ \cmsq is for the first time directly mapped for these nearby edge-on galaxies. This extraplanar \HI seems to be ubiquitous around gas-rich galaxies and serves as the interface between cold ISM and hot CGM during gas accretion/ejection. With the upcoming FEASTS data for more galaxies, we will investigate in more details the origin and influence of the extraplanar \HI gas. 

\section{Acknowledgements}
We thank the anonymous referee for useful discussions. We thank Sanskriti Das for providing the GBT data and useful suggestions.

J.W. acknowledges research grants and support from the Ministry of Science and Technology of the People's Republic of China (No. 2022YFA1602902), the National Natural Science Foundation of China (No. 12073002), and the China Manned Space Project (No. CMS-CSST-2021-B02). Parts of this research were supported by the Australian Research Council Centre of Excellence for All Sky Astrophysics in 3 Dimensions (ASTRO 3D), through project number CE170100013. L.C. acknowledges support from the Australian Research Council Discovery Project funding scheme (DP210100337). L.C.H. was supported by the National Science Foundation of China (11991052, 12233001), the National Key R\&D Program of China (2022YFF0503401), and the China Manned Space Project (CMS-CSST-2021-A04, CMS-CSST-2021-A06).

This work has used the data from the Five-hundred-meter Aperture Spherical radio Telescope (FAST, \url{https://cstr.cn/resolver?identifier=31116.02.FAST}).
FAST is a Chinese national mega-science facility, operated by the National Astronomical Observatories, Chinese Academy of Sciences (NAOC)\@.

\facilities{FAST:\@500 m, VLA.}
\software{Astropy 6.1.6 \citep{astropycollaboration2013,astropycollaboration2018,astropycollaboration2022}, 
NumPy 2.1.3 \citep{vanderwalt2011}, Python 3.11.10, SciPy 1.14.1 \citep{virtanen2020}, SoFiA 2.5.0 \citep{2015MNRAS.448.1922S,westmeierSoFiAAutomatedParallel2021}.}

\appendix
\section{FAST simulation observation}\label{append:simulator}
\begin{figure}[ht!]
    \centering
    \includegraphics[width=0.48\textwidth]{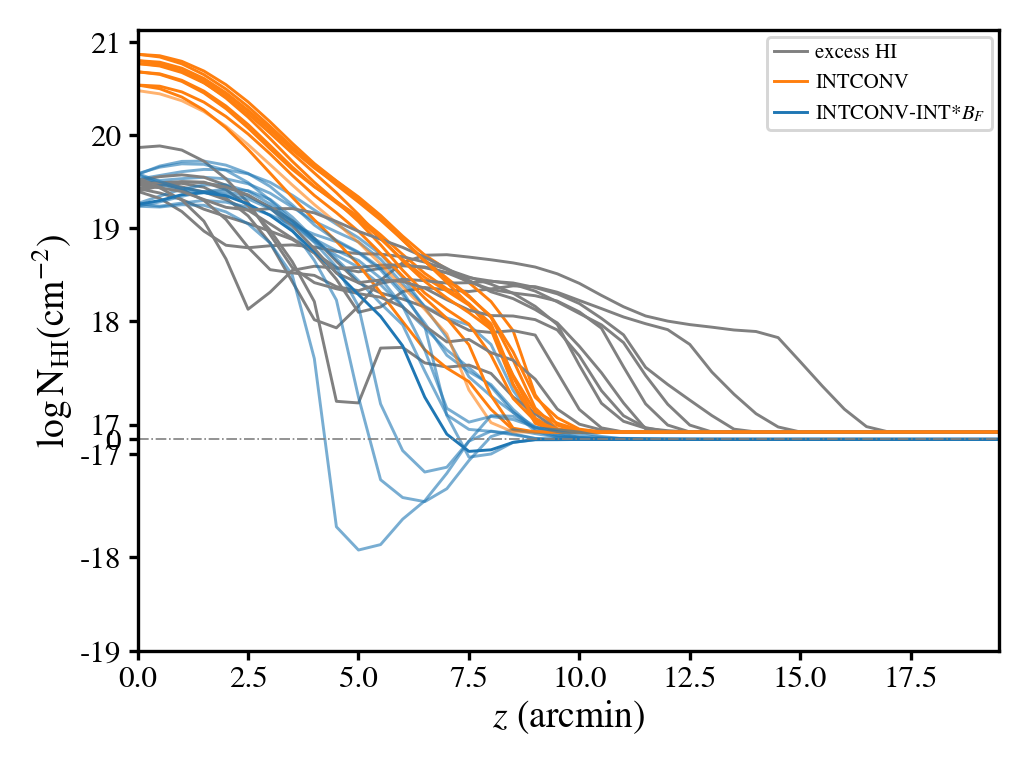}
    \caption{The difference between INTCONV (interferometric data convolved with FEASTS beam through the FEASTS observing simulator) and INT$*\rm B_F$ (interferometric data directly convolved with the average FEASTS beam) profiles compared to excess \HI and INTCONV profiles.}\label{fig:INTCONV_diff}
\end{figure}

We evaluate the necessity of using the FAST observing simulator by comparing the products with the result of convolving the average FEASTS beam directly. We use the same average FEASTS beam as \citetalias{wang2024}, which accounts for the rotation of the FAST feeds during observation. The direct convolution result of the interferometric data with the average FEASTS beam is labeled as ``INT$*\rm B_F$''. We calculate the z-profiles for both INTCONV and INT$*\rm B_F$ and quantify the difference, as shown in Figure \ref{fig:INTCONV_diff}. 

The residuals of INTCONV$-$INT$*\rm B_F$ show mostly positive values in the main lobe, sometimes decreasing to negative values outward. But in the outermost regions the INTCONV becomes larger again. Note that the convolution with the average beam already accounts for average sidelobe effects during observation. The excess in z-profiles, particularly in the outer regions, emphasizes the importance of using FAST observing simulator, which better characterizes the uneven sidelobes across the observation coverage. If direct convolution results are used, the excess of $\text{FEASTS}-\text{INT}*\rm B_F$ would likely be more significant.

\section{The effect of interferometric data depth in modeling the disk characteristic thickness}\label{append:h_disk}
\begin{figure*}[ht!]
    \includegraphics[width=0.5\textwidth]{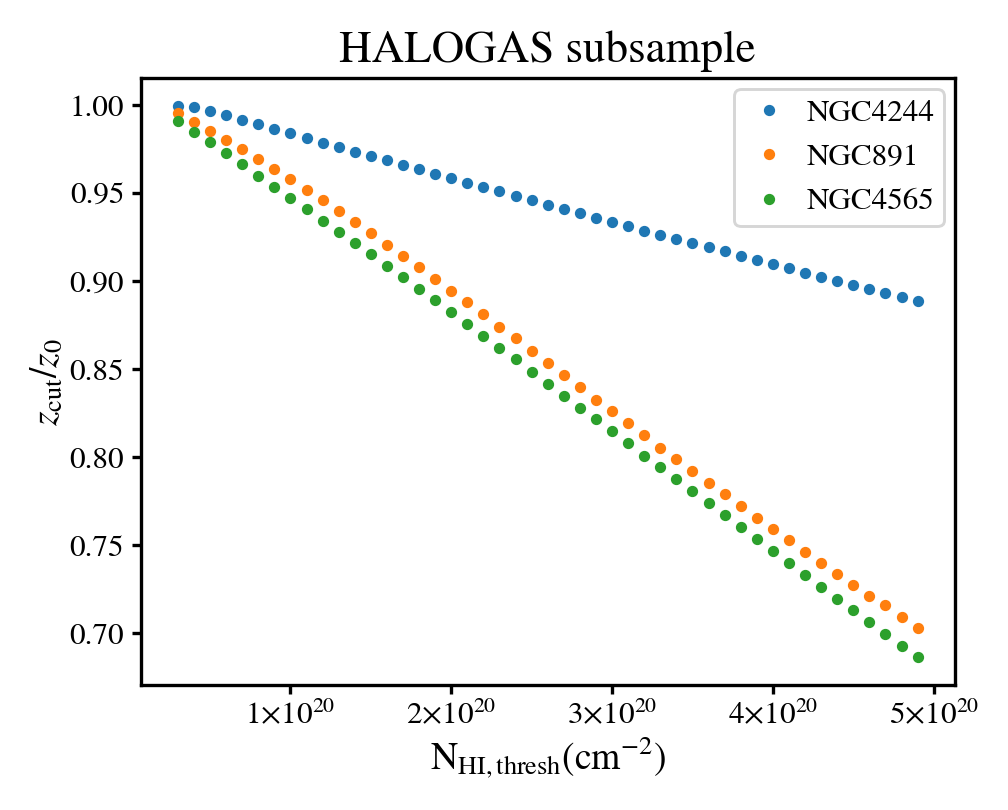}
    \includegraphics[width=0.5\textwidth]{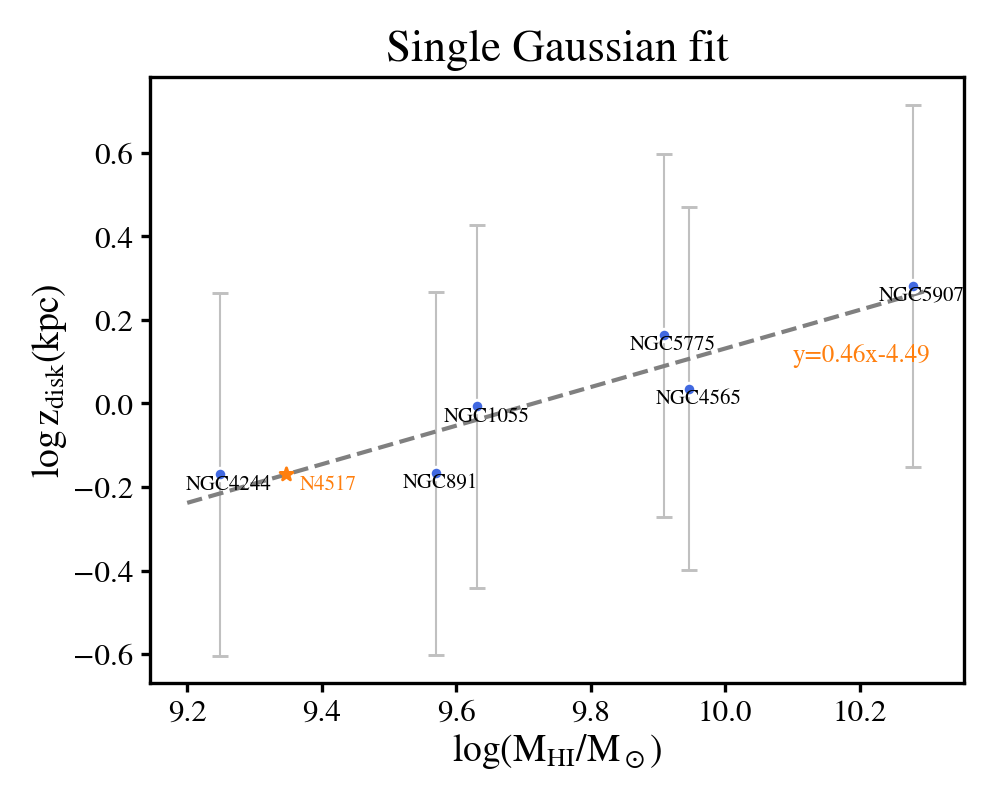}
    \caption{The relation between the cut threshold and the varying fraction of the disk characteristic thickness $z_{\rm disk}$ calculated after manually changing the threshold for the HALOGAS subsample (left) and the relation between the disk characteristic thickness $\rm z_{disk}$ and the interferometric \HI mass (right). The orange star in the right panel marks the value of $z_{disk}$ used in our modeling for NGC 4517 which has no available interferometric data.}\label{fig:h_disk_thresh}
\end{figure*}

The interferometric data for our sample come from different surveys with observation depths ranging from $3\times 10^{19}$ \cmsq to $5\times 10^{20}$ \cmsq. It is therefore necessary to assess how observation depth affects the estimation of the characteristic thickness of the disk. To do this, we manually apply a threshold to cut the interferometric data cube of the HALOGAS subsample (NGC 4244, NGC 891 and NGC 4565) and then fit the characteristic thickness of the z-profile derived from this cut cube. By varying the cut threshold,  we establish the relationship between the best-fit characteristic thickness and observation depth, as shown in the left panel of Figure~\ref{fig:h_disk_thresh}. The characteristic thickness is weakly influenced by observation depth, decreasing by $\le 30\%$ when the $\NHI$ threshold decreases by more than 1 dex. Thus, the disk characteristic thickness can be relatively uniformly estimated from different interferometric data. We find that the characteristic thickness of the \HI disk is positively correlated with the interferometric \HI mass for our sample, as shown in the right panel of Figure \ref{fig:h_disk_thresh}, consistent with the result of \citet{randriamampandry2021}. This relationship allows us to estimate the characteristic thickness for NGC 4517, which lacks available interferometric data.

We use bootstrapping method to estimate the errors in the PSF-free model profiles and the vertical extension $z_{18}$. Specifically, we generate 10000 sampling profiles by randomly selecting the parameter values of Eq. \ref{eq:triplegauss} from the distributions of model fitting results. To account for uncertainties introduced by underestimation of the characteristic thickness for the shallower interferometric data (NGC 1055, NGC 5907 and NGC 5775), we evenly sample the characteristic thickness of the disk from $z_0$ to $1.3z_0$. The errors are then derived by calculating the standard deviation of the sampled profiles.

\section{The influence of FEASTS beam in the extension of excess or outer \HI profiles}\label{appendix:beam_smoothing}
The outer and excess \HI profiles are both smoothed by FEASTS beam. We test the beam smoothing effects in the extension of extraplanar \HI gas by convolving the outer and excess profiles with FEASTS beam once again and then evaluate the change in $z_{18}$ values. The results of the $z_{18}$ values in arcmin before and after convolution are shown in Table \ref{tab:z18_conv}. The increase fractions of $z_{18}$ are mostly less than 10\%, indicating a weak influence of the FEASTS beam smoothing effects on $z_{18}$ values. The exceptions include the lower side of NGC 1055 and NGC 5775 and the upper side of NGC 4517, which are relatively close to the FEASTS beam as shown in Figure \ref{fig:all_profile}. Thus the extensions of the extraplanar \HI are more significantly affected by the FEASTS beam. 

\begin{deluxetable*}{c|cc|cc|cc|cc}
    \tablecaption{The changes in $z_{18}$ values for outer and excess \HI before and after convolution}\label{tab:z18_conv}
    \tablehead{
        ~ & 
        \multicolumn{2}{c|}{Excess \HI} & 
        \multicolumn{2}{c|}{(Excess \HI) $*\rm B_F$} & 
        \multicolumn{2}{c|}{Outer \HI} & 
        \multicolumn{2}{c}{(Outer \HI) $*\rm B_F$} \\
        Galaxy & 
        \colhead{$z_{18,\rm lower}$} & 
        $z_{18,\rm upper}$ & 
        \colhead{$z_{18,\rm lower}$} & 
        $z_{18,\rm upper}$ &
        \colhead{$z_{18,\rm lower}$} & 
        $z_{18,\rm upper}$ &
        \colhead{$z_{18,\rm lower}$} & 
        \colhead{$z_{18,\rm upper}$} \\
        ~ & (arcmin) & (arcmin) & (arcmin) & (arcmin) & (arcmin) & (arcmin) & (arcmin) & (arcmin) \\
        (1) & (2) & (3) & (4) & (5) & (6) & (7) & (8) & (9)}
    \startdata
    NGC 4244 & 10.9 & 11.6 & 11.1 (2.7\%)  & 12.4 (7.2\%) & 10.9 & 11.6 & 11.3 (3.9\%)  & 12.5 (7.5\%)  \\ 
    NGC 4517 & /    & /    & /             & /            & 7.4  & 3.7  & 7.6 (3.0\%)   & 5.1 (38.8\%)  \\
    NGC 891  & 6.3  & 9.8  & 7.5 (19.1\%)  & 10.0 (2.2\%) & 7.6  & 9.8  & 8.3 (9.6\%)   & 10.2 (4.2\%)  \\
    NGC 4565 & 9.8  & 10.1 & 10.3 (4.4\%)  & 10.4 (3.3\%) & 9.9  & 10.1 & 10.4 (4.7\%)  & 10.5 (4.3\%)  \\
    NGC 1055 & 2.6  & 8.8  & 5.5 (115.2\%) & 9.4 (6.3\%)  & 4.5  & 8.8  & 6.1 (34.8\%)  & 9.4 (6.3\%)  \\
    NGC 5907 & 8.5  & 8.9  & 8.5 (0.5\%)   & 9.4 (5.9\%)  & 8.7  & 9.0  & 8.7 (0.1\%)   & 9.9 (9.4\%)  \\
    NGC 5775 & 5.5  & /    & 6.8 (24.2\%)  & /            & 5.4  & /    & 6.5 (20.1\%)  & /        
    \enddata
    \tablecomments{
        Column~(1): galaxy name.
        Column~(2\text{-}3): the $z_{18}$ values where the \HI column density reaches $10^{18}$ \cmsq for lower and upper side of the excess \HI z-profiles.
        Column~(4\text{-}5): the $z_{18}$ values for lower and upper side of the excess \HI z-profiles after convolution of FEASTS beam again. The increasing fractions are also listed in parentheses.
        Column~(6\text{-}7): the $z_{18}$ values for lower and upper side of the outer \HI z-profiles.
        Column~(8\text{-}9): the $z_{18}$ values for lower and upper side of the outer \HI z-profiles after convolution of FEASTS beam again. The increasing fractions are also listed in parentheses.}
\end{deluxetable*}

\section{The difference velocity map}\label{appendix:mom1}
We present the moment 1 maps of FEASTS, INTCONV, and their difference for our sample with available interferometric data in Figure \ref{fig:mom1}. The difference velocity maps are generally close to zero in the disk regions, but they show clear counter-rotating features in the outer regions, suggesting a lagged rotation of the extraplanar \HI. The velocity lag is roughly 40 \kms, consistent with the findings of \citet{wang2024} and the extraplanar gas characterized with deep interferometry data \citet{marascoHALOGASPropertiesExtraplanar2019a}.    
\begin{figure*}
    \includegraphics{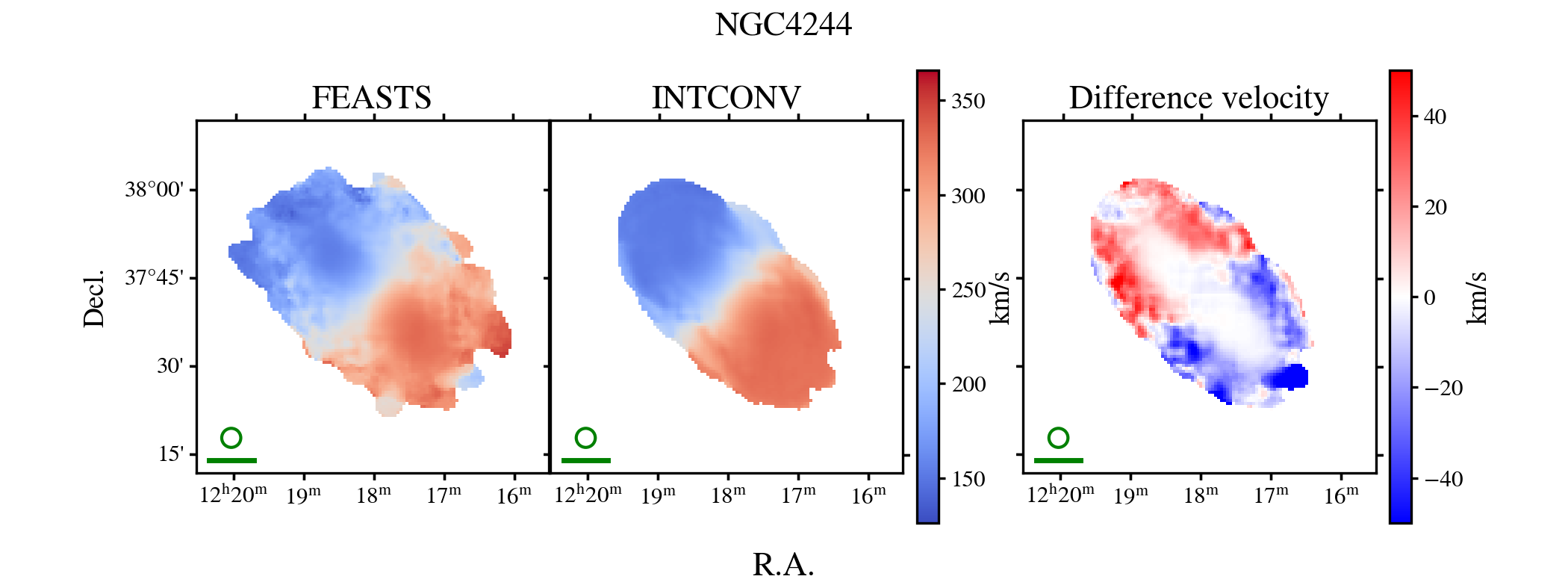}
    \includegraphics{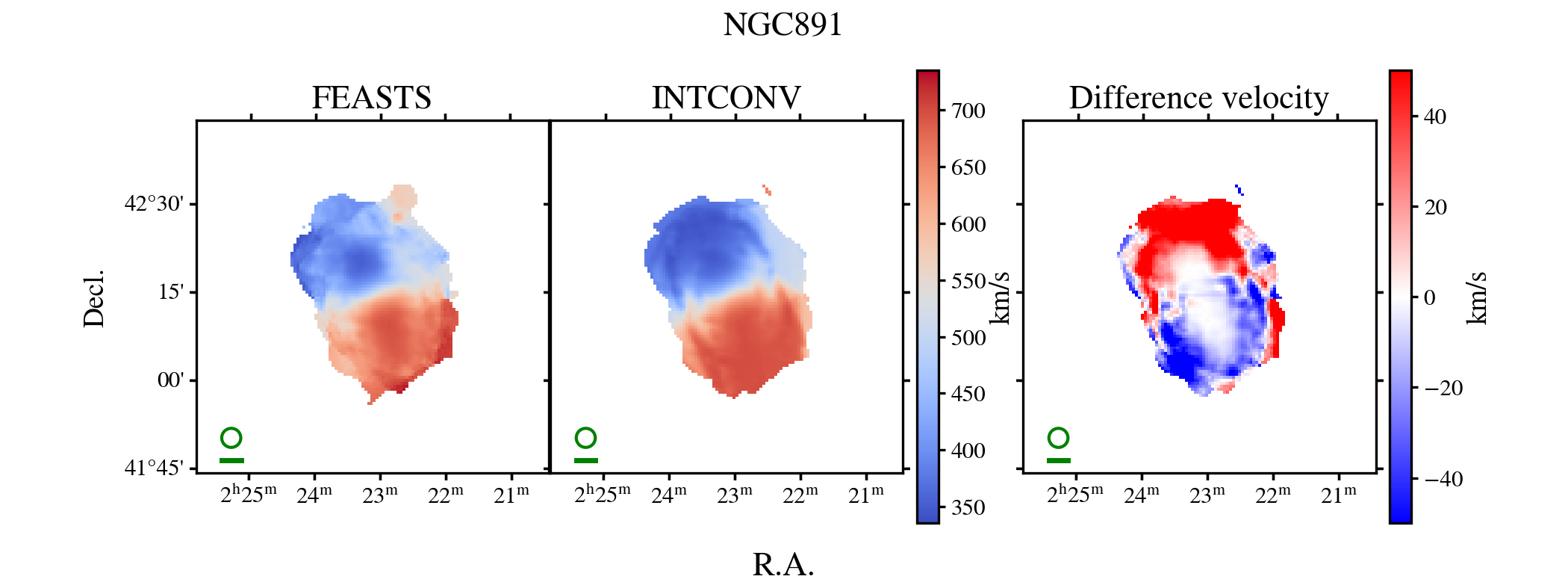}
    \includegraphics{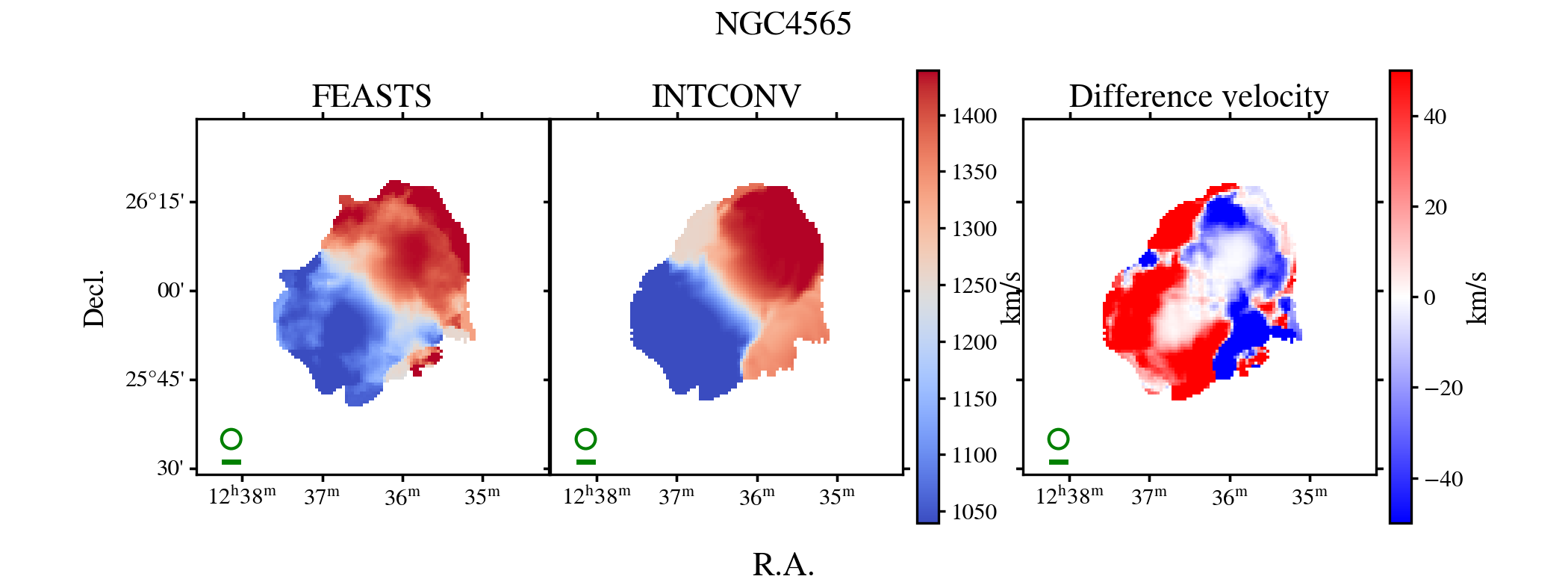}
    \caption{The moment 1 map for FEASTS (left), INTCONV (middle) and the difference of these two (right) for galaxy NGC 4244, NGC 891, and NGC4565. The green circle and horizontal bar at the lower left corner correspond to the FEASTS beam and a length of 10 \kpc, respectively.}\label{fig:mom1}
\end{figure*}

\addtocounter{figure}{-1}
\begin{figure*}
    \includegraphics{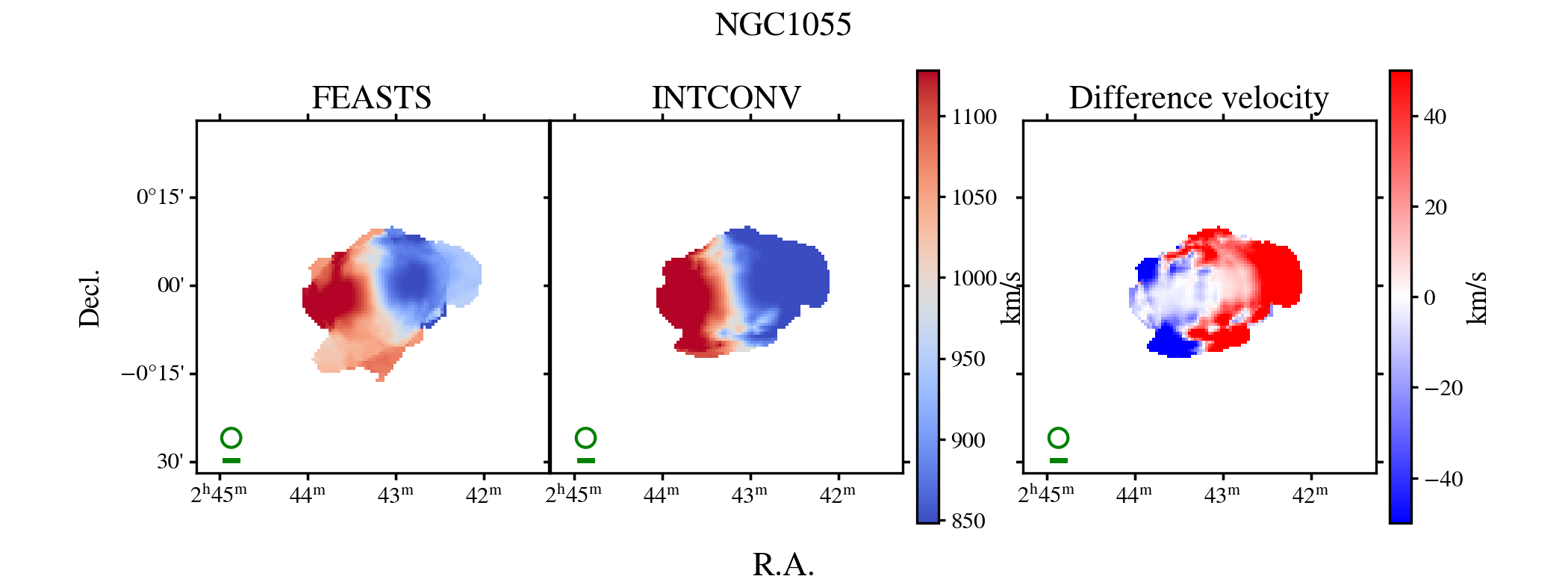}
    \includegraphics{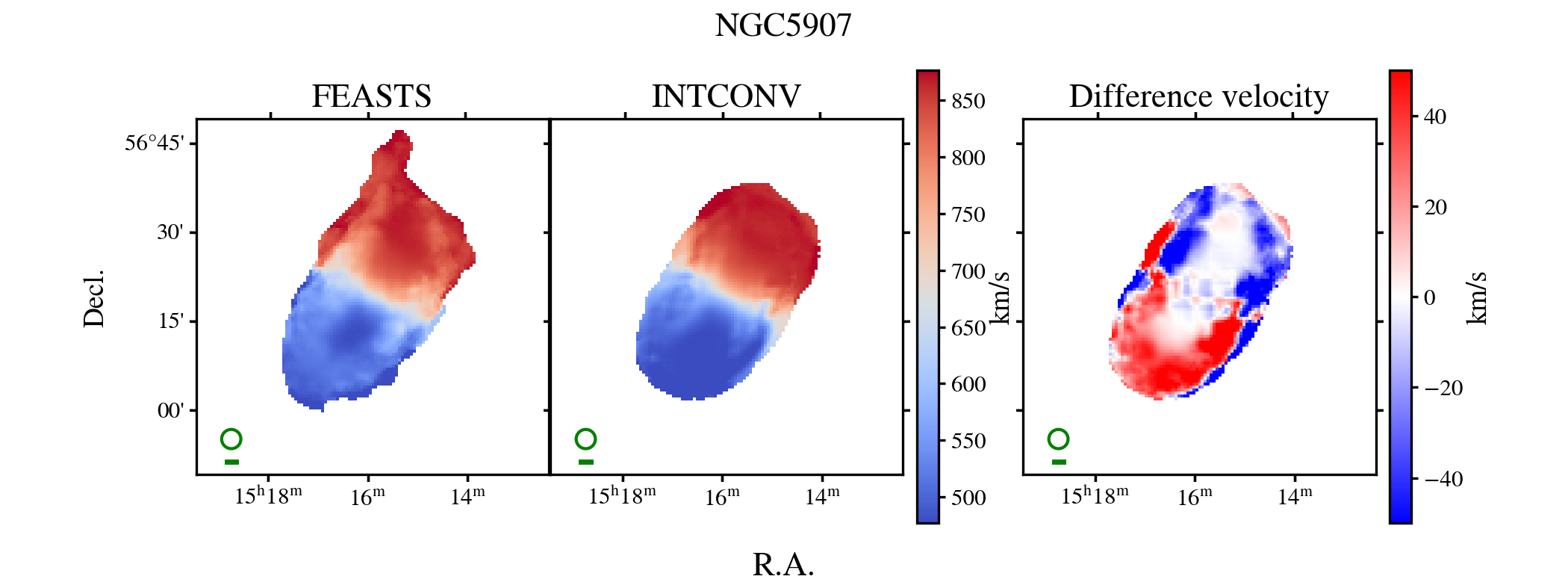}
    \includegraphics{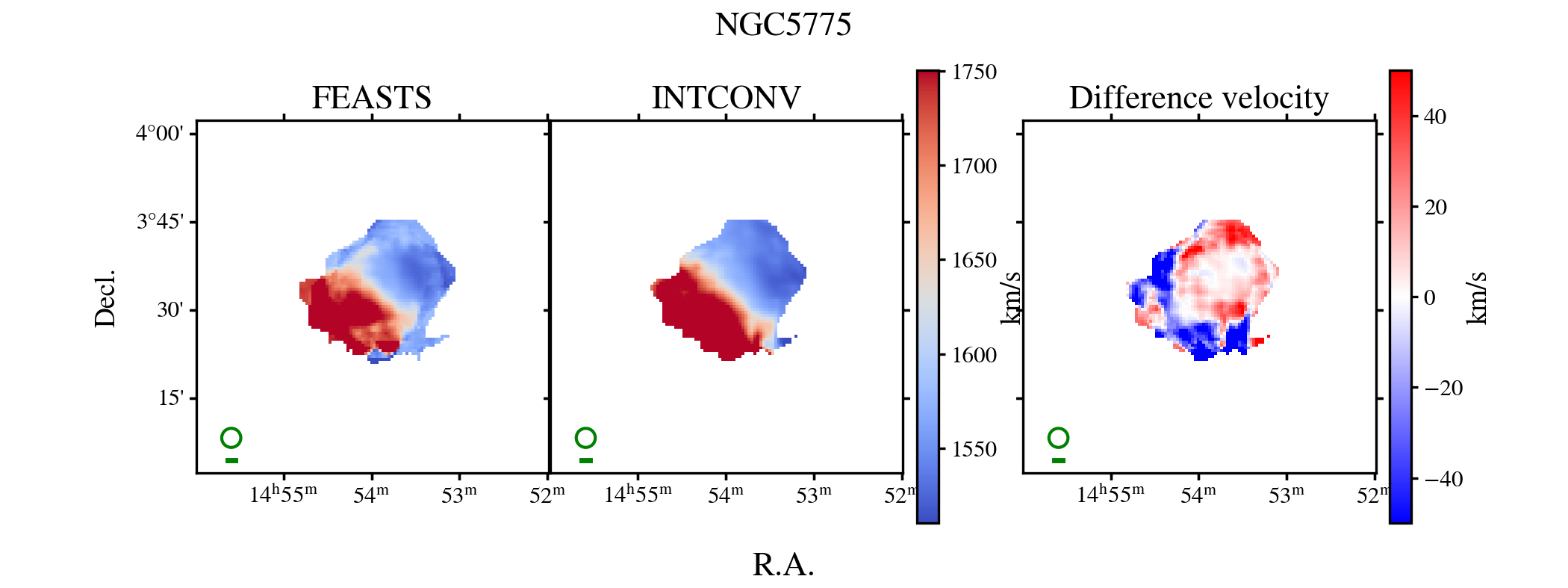}
    \caption{(Continued.) The moment 1 map for FEASTS (left), INTCONV (middle) and the difference of these two (right) for galaxy NGC 1055, NGC 5907, and NGC 5775.}
\end{figure*}

\section{The individual z-profiles and model fitting results}\label{appendix:individual}
\begin{figure*}
    \begin{tabular*}{\linewidth}{@{}c@{}c@{}}
        \includegraphics[width=0.49\linewidth]{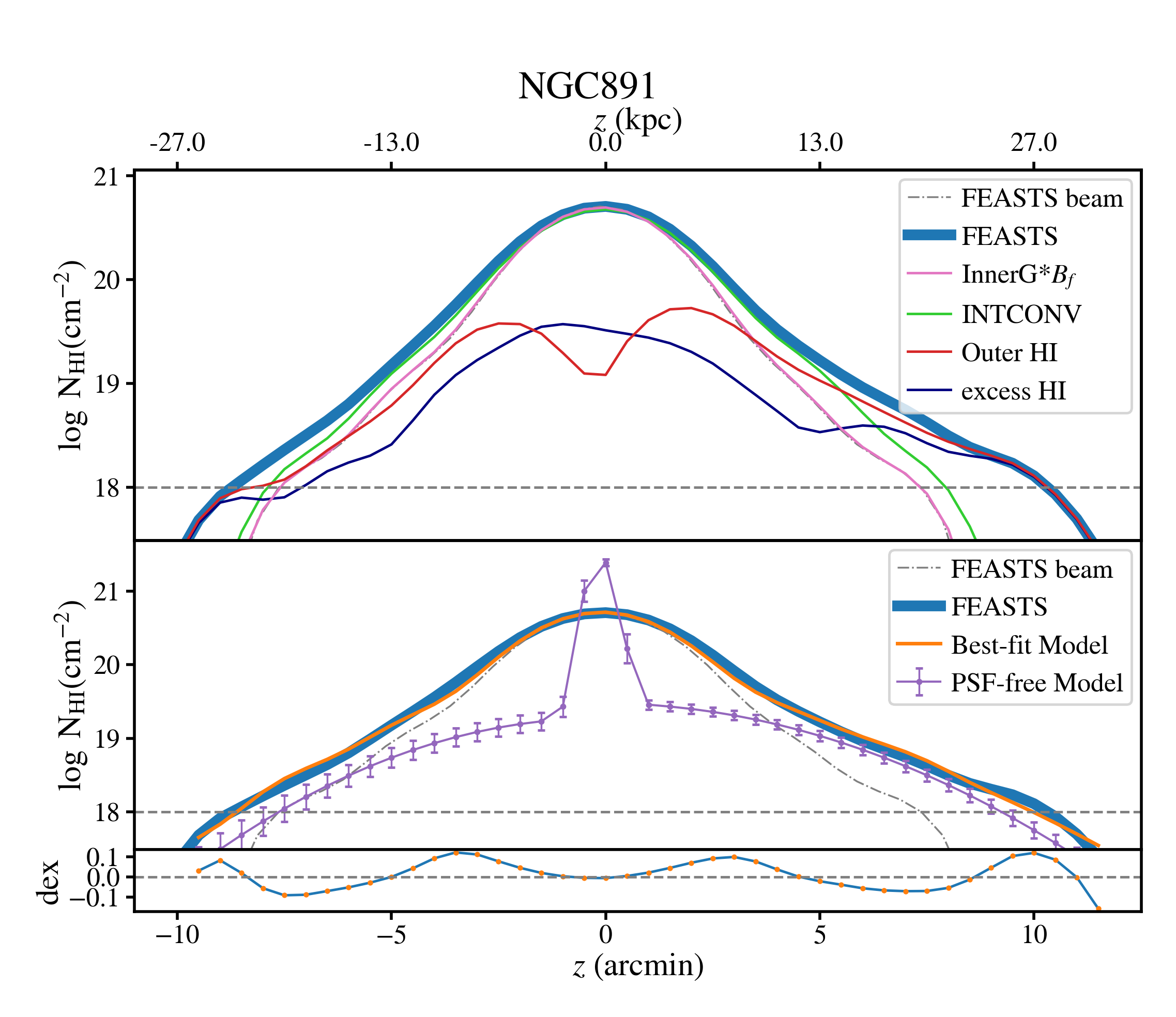} &
        \includegraphics[width=0.49\linewidth]{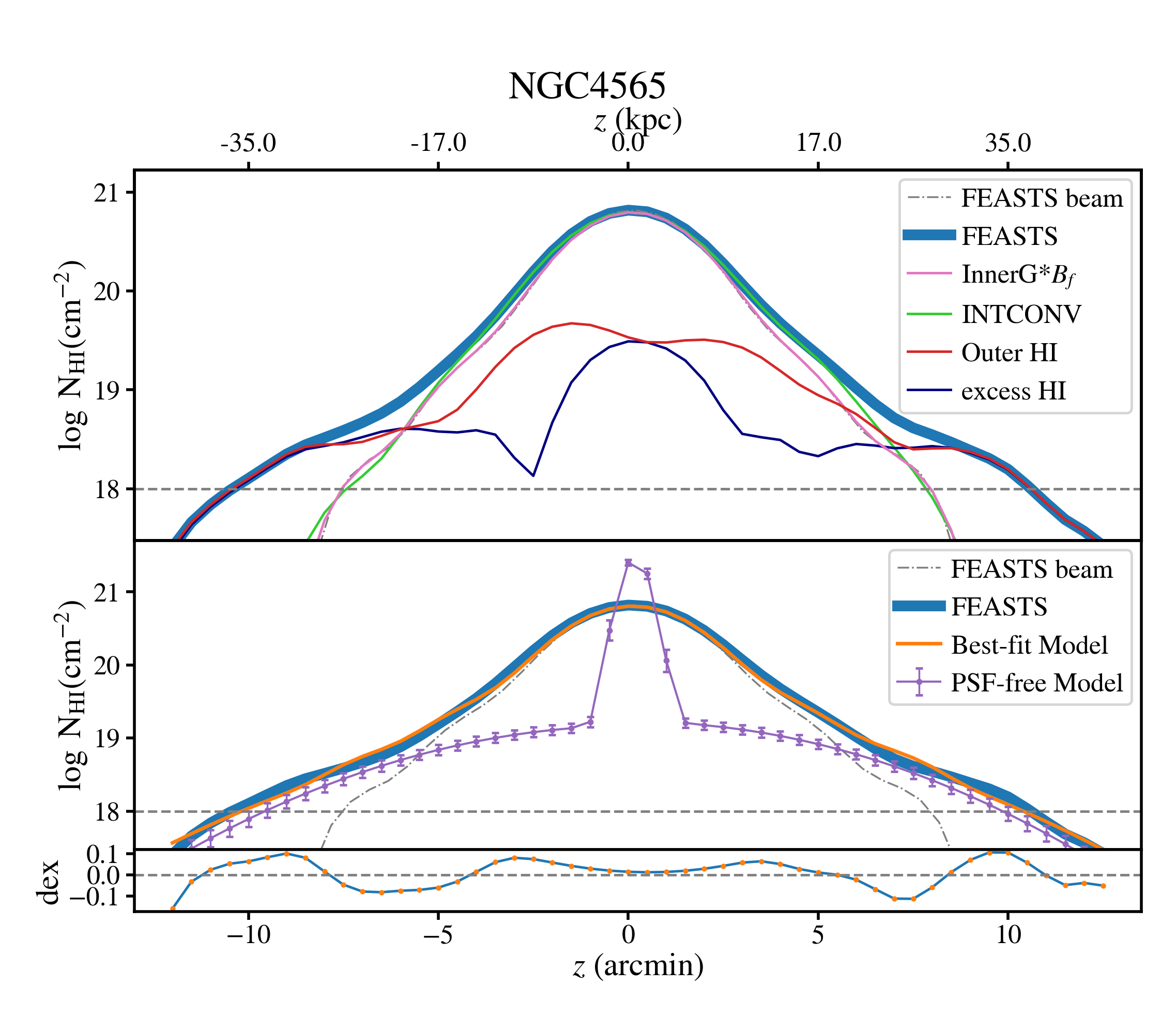} \\
        \includegraphics[width=0.49\linewidth]{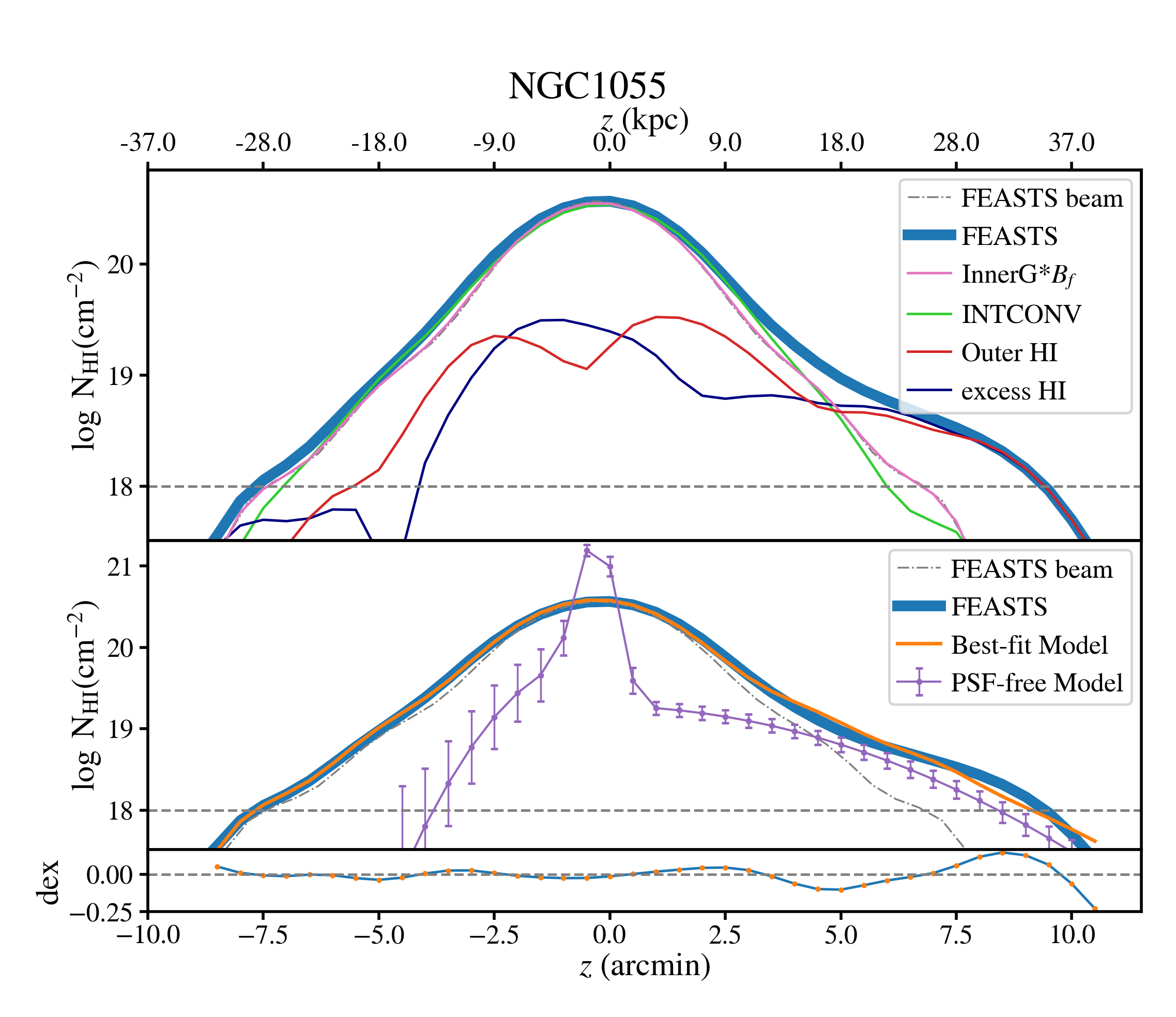} &
        \includegraphics[width=0.49\linewidth]{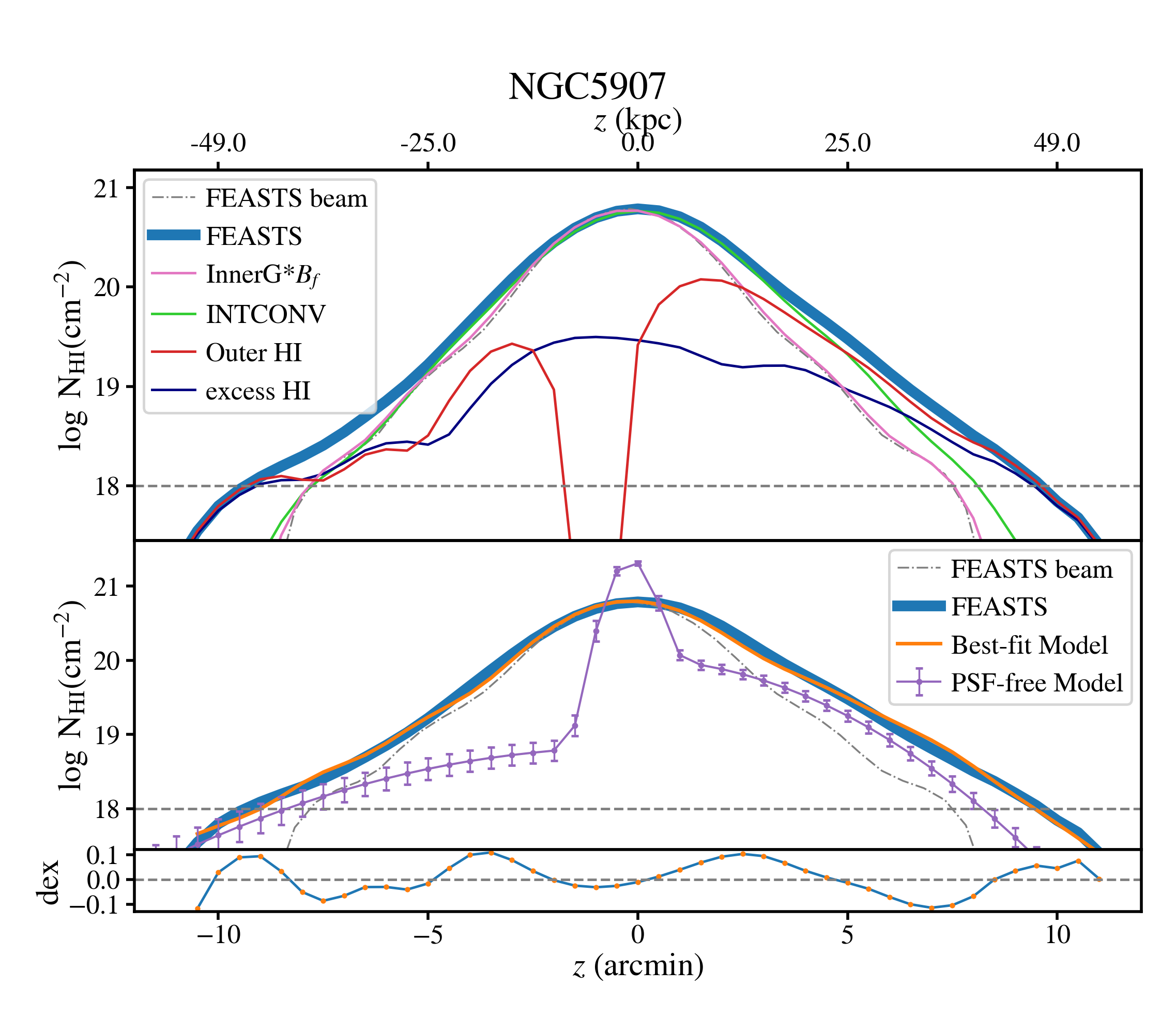} \\
        \includegraphics[width=0.49\linewidth]{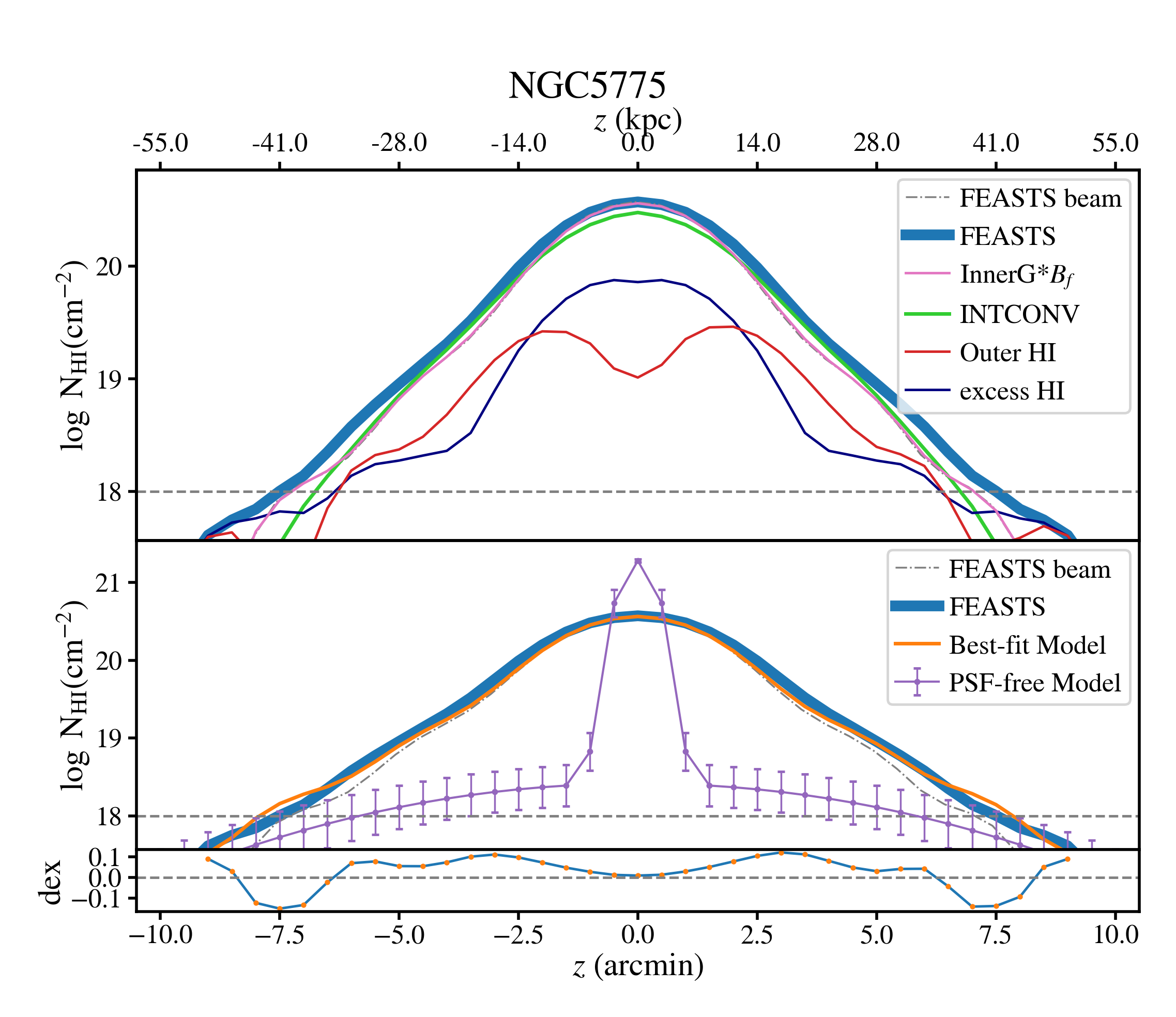} &
        \includegraphics[width=0.49\linewidth]{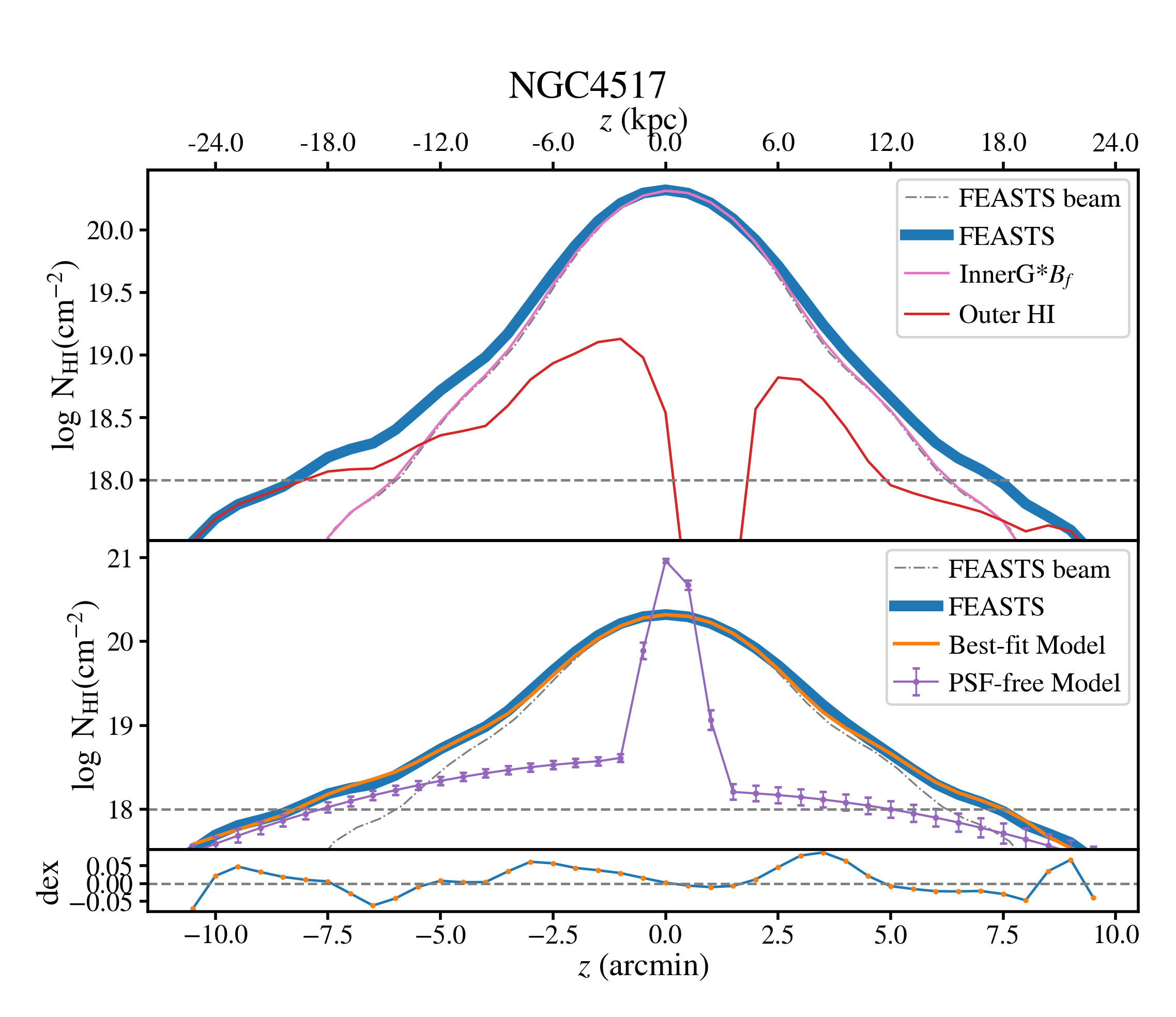}
    \end{tabular*}
    \caption{The profiles same as Figure \ref{fig:all_profile} but for other 6 galaxies. Note that for NGC 5775, only the profile on the left part (below the disk) is plotted (the right part is symmetric for visualization) as it is blended with NGC 5774 above the disk. For NGC 4517, only modeled profiles are plotted because it has no available interferometric data. }\label{fig:append_profile}
\end{figure*}

We show the z-profiles and the model fitting results for NGC 891, NGC 4565, NGC 1055, NGC 5907, NGC 5775, and NGC 4517 in Figure \ref{fig:append_profile}. For NGC 5775, only the distribution below the disk is shown because it blends with NGC 5774 on the other side. For NGC 4517, only the modeled profiles are shown because it has no available interferometric data. 

\section{The QSO-absorption sample}\label{appendix:absorption}
\begin{figure}
    \includegraphics{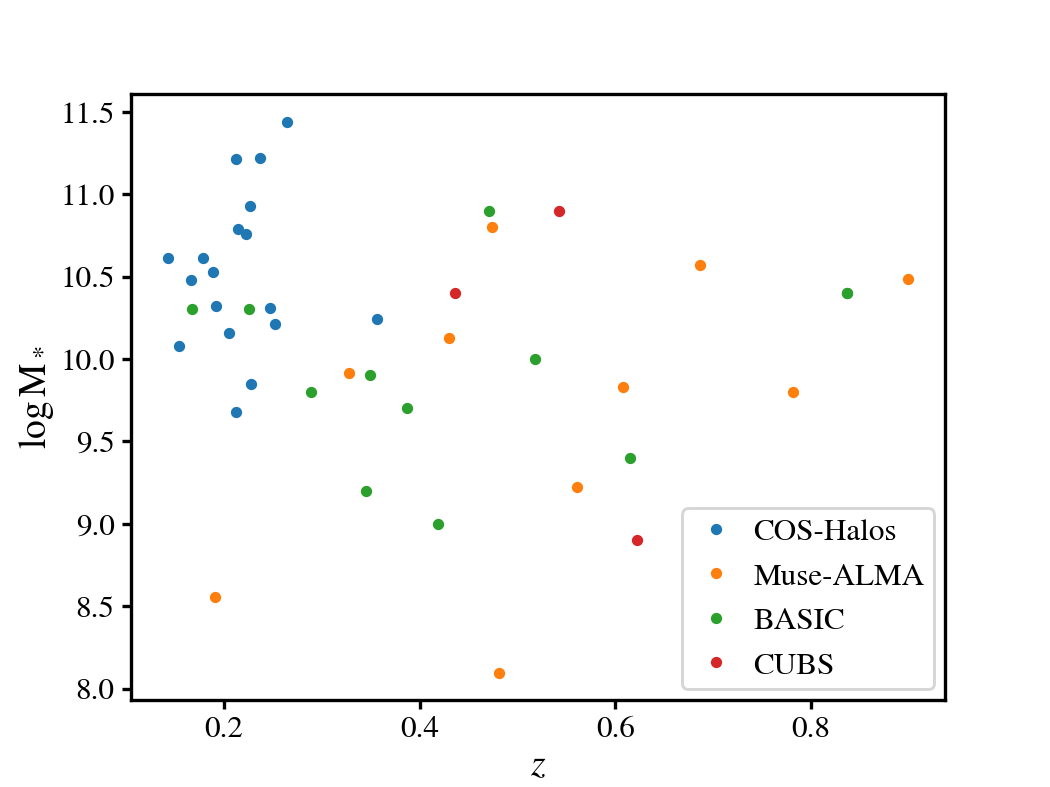}
    \caption{The stellar mass and redshift distribution of the absorption line sample used in this study. Only star-forming galaxies with $\rm sSFR>10^{-11} \unit{\per\yr}$ are included.}\label{fig:absorption_sample}
\end{figure}

The QSO-absorption sample used in this study comes from various surveys, with the redshift and stellar mass distribution shown in Figure \ref{fig:absorption_sample}. The COS-Halos sample \citep{tumlinson2013} contains the most nearby galaxies, with stellar mass similar to that of the Milky Way. The sample compiled in \citet{weng2023} is complementary to COS-Halos, covering a wider range of stellar mass and redshifts.

\section{The scaled-up FEASTS z-profiles}\label{appendix:scaling}
\begin{figure*}
    \includegraphics{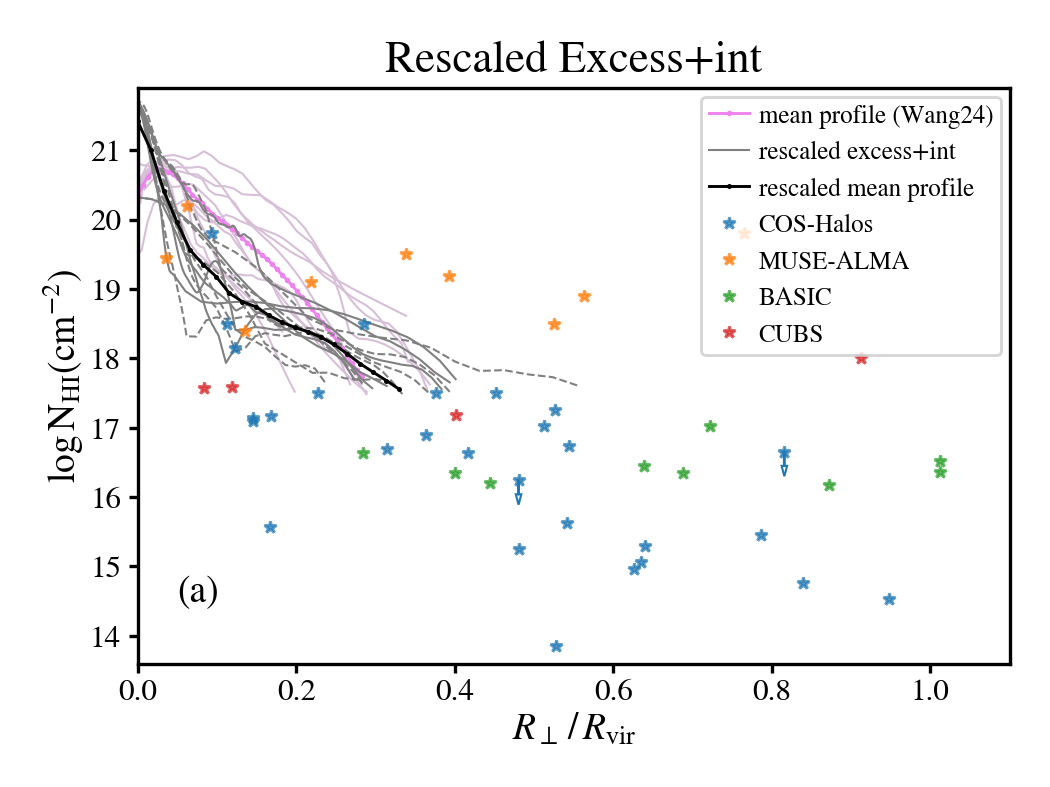}
    \includegraphics{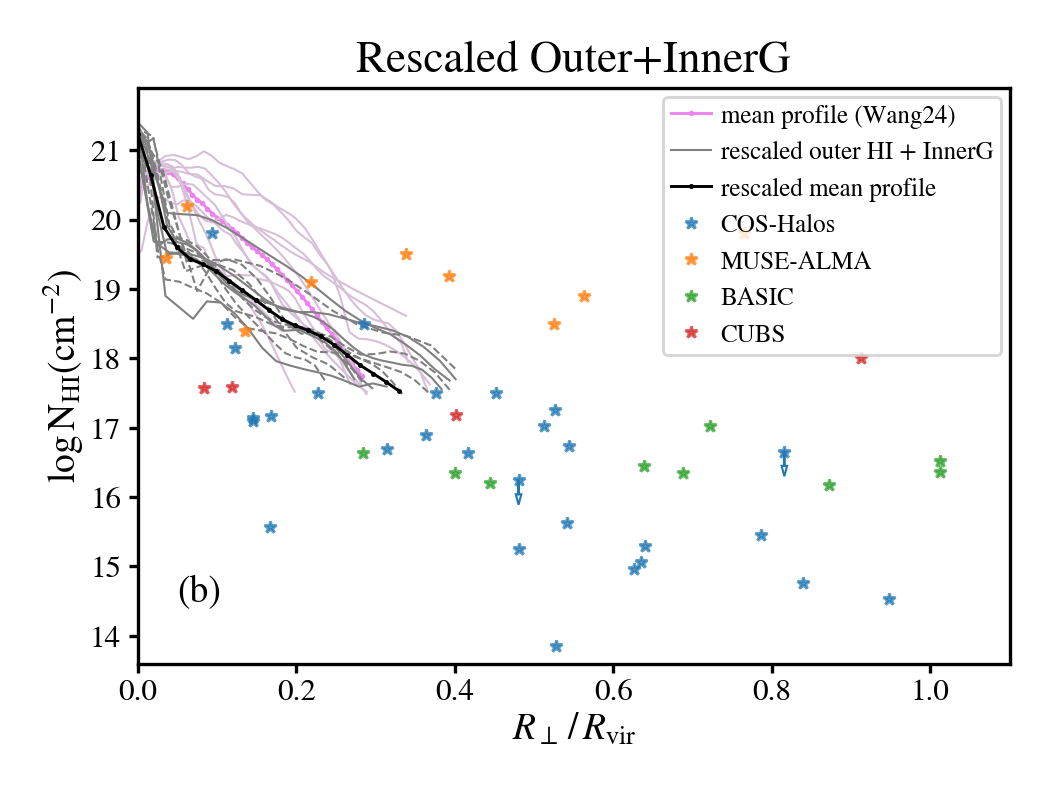}
    \caption{The scaled-up z-profiles for the FEASTS data. Profiles are same with Figure \ref{fig:modelled_pro}, except that the radius of excess+int \HI (outer \HI + InnerG) is scaled up 2.3 times on the left (right) panel.}\label{fig:modelled_scaled_pro}
\end{figure*}

Our result shows that, although the extraplanar \HI extends much further along the z-axis than the thin and thick \HI disk, it is still much flattener than a spherical distribution down to the $\NHI \sim 10^{17.7}$ \cmsq level. While the exact axis ratio for the \HI distribution requires detailed modeling and subtraction of the beam smoothing effects, we can roughly estimate the ratio by linearly scaling up the radius of the z-profiles in the logarithm space until they roughly overlap with the upper envelop of absorption measurements as well as the planar radial profiles of the relatively face-on sample from \citet{wang2024a}. The results are shown in Figure \ref{fig:modelled_scaled_pro}. The resulting scaling factor is 2.3, which indicates an axis ratio of 0.43. 

\section{Comparison to GBT-based results}\label{appendix:gbt_spec}
Recent studies ~\citet{dasDetectionDiffuseEmission2020} and~\citet{das2024} have observed the \HI gas in the CGM for the galaxies NGC 891 and NGC 4565 with GBT\@. For each galaxy, they take 7 pointings along the major axis (including the center) and 10 pointings along the minor axis, evenly spaced according to the GBT beamsize (0, 1, 2, 3 GBT beams along the major axis and 0.5, 1, 1.5, 2, 3 GBT beams along the minor axis, see Figure \ref{fig:gbt_j_N891} and Figure \ref{fig:gbt_j_N4565}, and ~\citet{das2024} for the distribution of pointings). The FWHM of the GBT beam is \ang{;9.1}, corresponding to 24 (31) \kpc for NGC 891 (NGC 4565). The GBT observations have achieved 5$\sigma$ sensitivity of $6.1 \times 10^{16}$ \cmsq per 20 \kms\ velocity channel and successfully detected \HI flux at most pointings. The furthest detections are out to 78 (93) \kpc, at a column density of $10^{17.2}$($10^{17.2}$) \cmsq for NGC 891 (NGC 4565). We compare the \HI spectra and fluxes detected at the same positions between the GBT and FEASTS observations.

The FEASTS data are convolved with the GBT PSF before comparison with GBT data. 
We convolve the FEASTS data with the circularized GBT PSF from~\citet{dasDetectionDiffuseEmission2020}.  After convolution, the FEASTS data and the GBT data are both rebinned to 20 \kms\ for comparison. 
The convolved and rebinned FEASTS data are linearly interpolated at the GBT pointings to get the \HI spectra. We apply the same analysis to the interferometric HALOGAS data. 
The rms for the convolved and rebinned FEASTS data is 0.652/0.832 \mjy\ per GBT beam for NGC 891 and NGC 4565, corresponding to 3$\sigma$ $\NHI$ detection limit of $6.9/8.7 \times 10^{16}$ \cmsq per 20 \kms, which is slightly shallower than the GBT data. Therefore we compare the GBT fluxes to both fluxes and upper limits from the FEASTS data.

\begin{figure*}
    \includegraphics{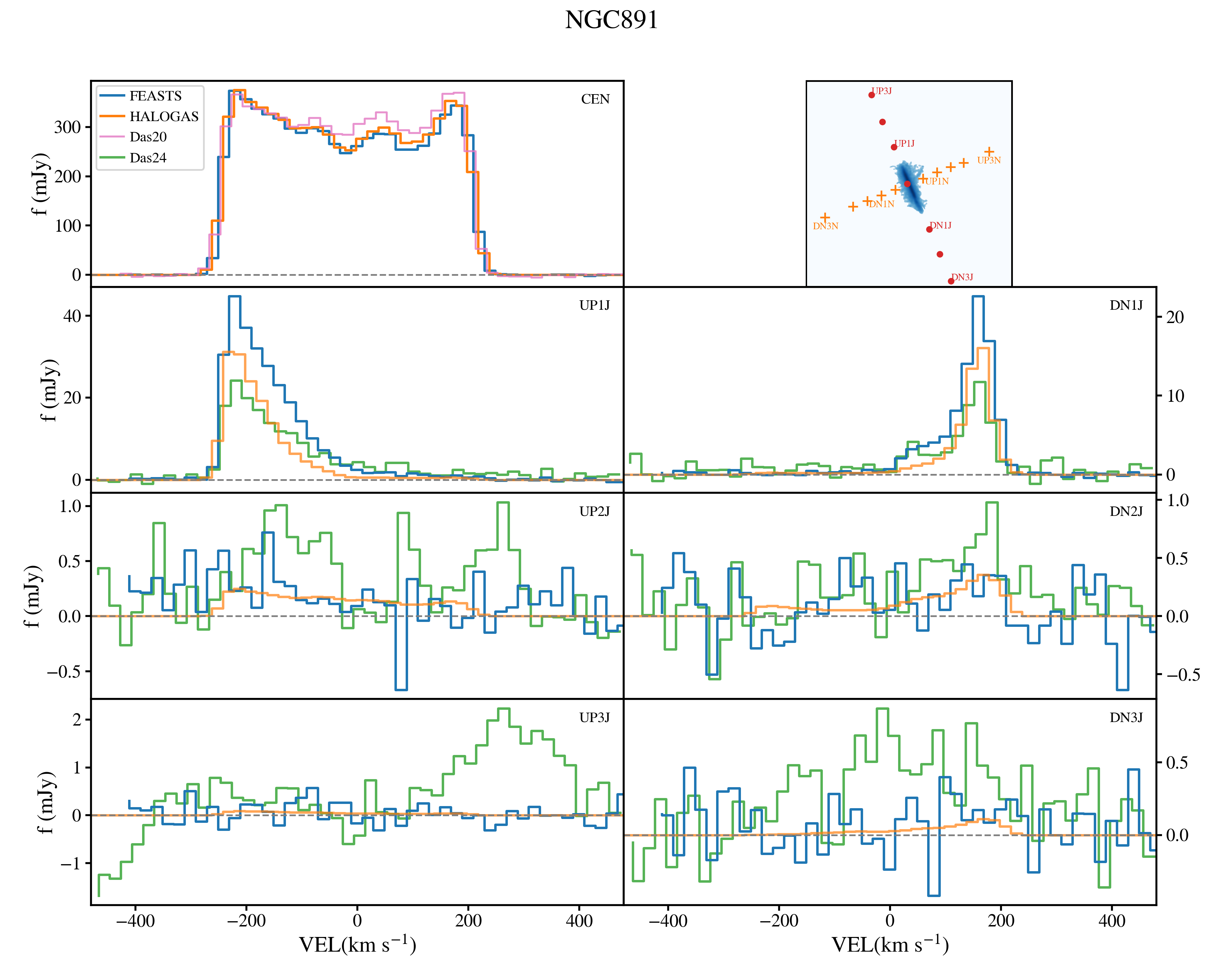}
    \caption{The distribution and spectra for 17 GBT pointings for NGC 891. The top right panel shows the positions of 17 pointings overlaid on the \HI moment 0 map of the HALOGAS data. The spectra from top to bottom are for pointings 0, 1, 2, 3 GBT beam away respectively relative to the galaxy center along the major axis. The spectra of FEASTS, HALOGAS and GBT observations are all rebinned to 20 \kms for comparison. The pointing names are listed on the top right corner, same with \citet{das2024}.}\label{fig:gbt_j_N891}
\end{figure*}

\addtocounter{figure}{-1}
\begin{figure*}
    \includegraphics{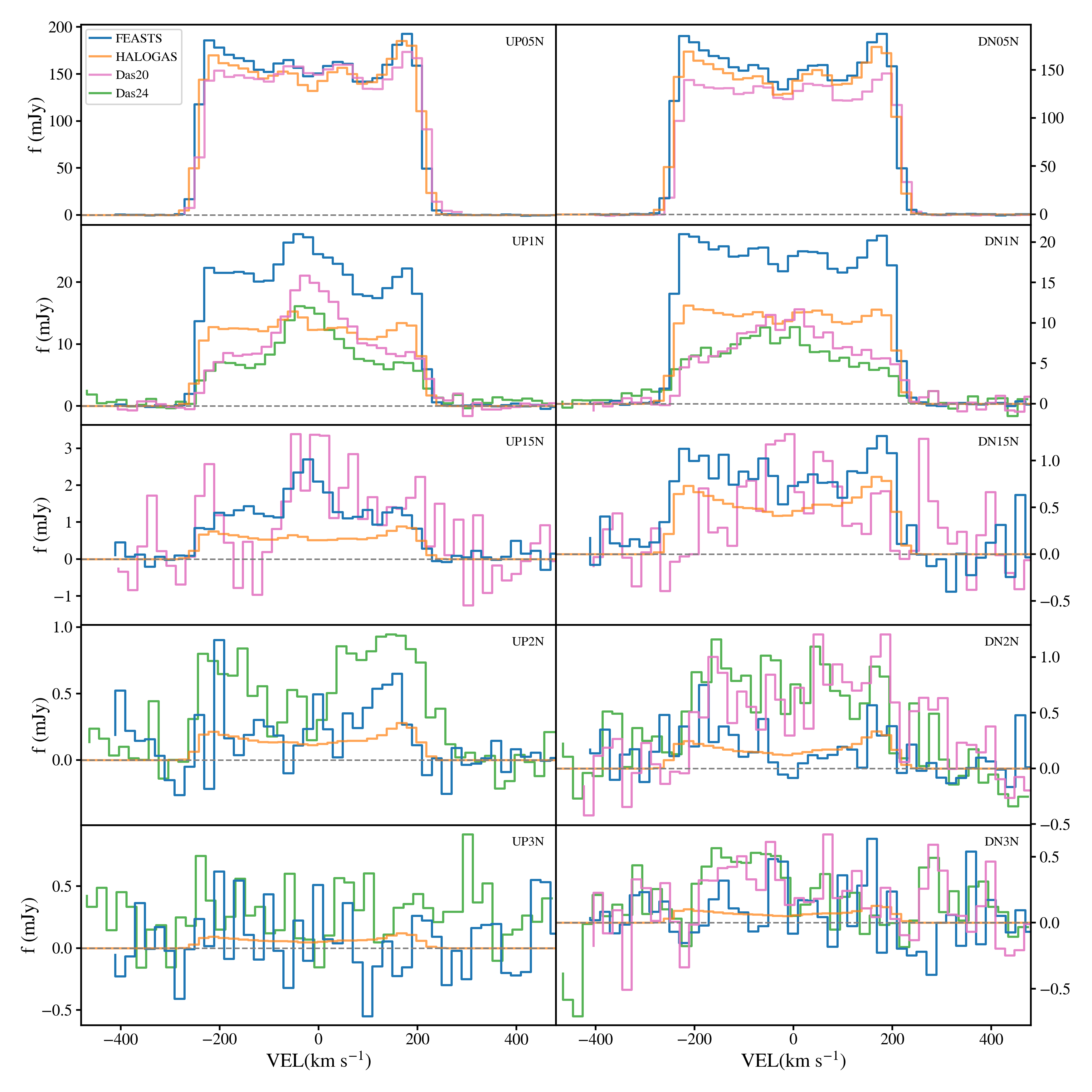}
    \caption{(Continued.) The spectra of the 10 pointings along the minor axis for NGC 891. The pointings from top to bottom are 0.5, 1, 1.5, 2, 3 GBT beam away relative to the galaxy center.}\label{fig:gbt_n_N891}
\end{figure*}

\begin{figure*}
    \includegraphics{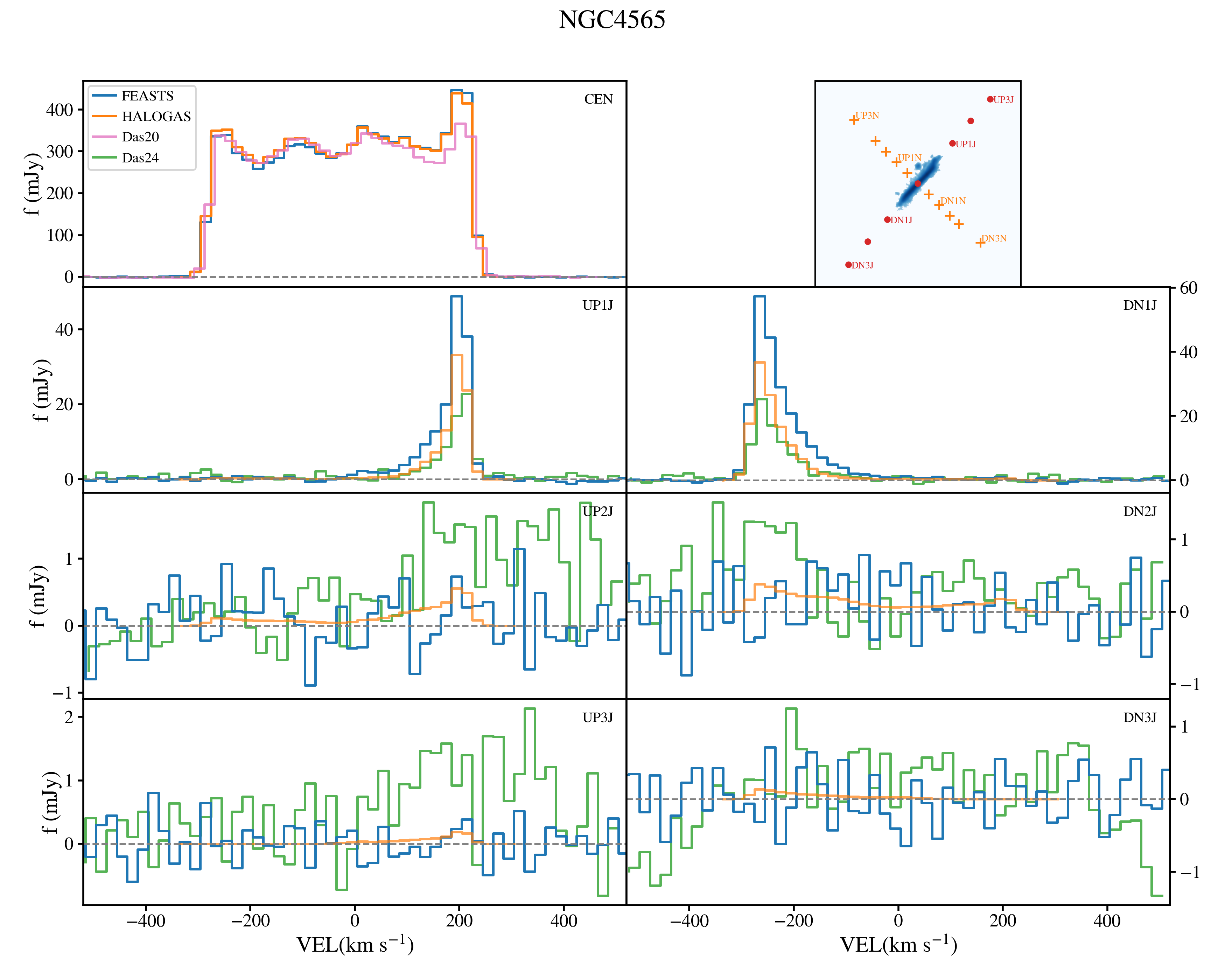}
    \caption{Similar to Figure \ref{fig:gbt_j_N891} but for NGC 4565.}\label{fig:gbt_j_N4565}
\end{figure*}

\addtocounter{figure}{-1}
\begin{figure*}
    \includegraphics{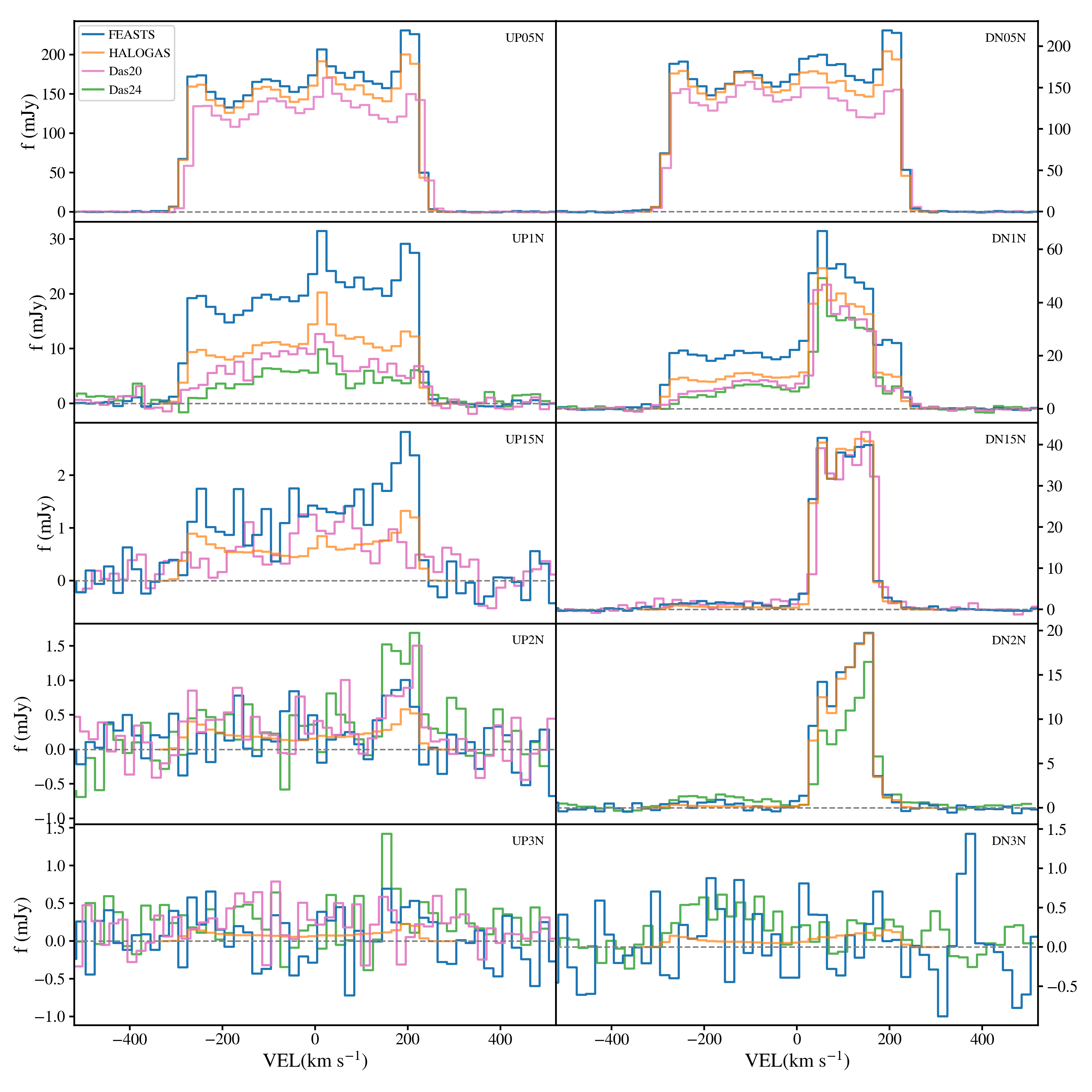}
    \caption{(Continued.) Similar to Figure \ref{fig:gbt_n_N891} but for NGC 4565.}\label{fig:gbt_n_N4565}
\end{figure*}

\begin{figure*}
    \includegraphics{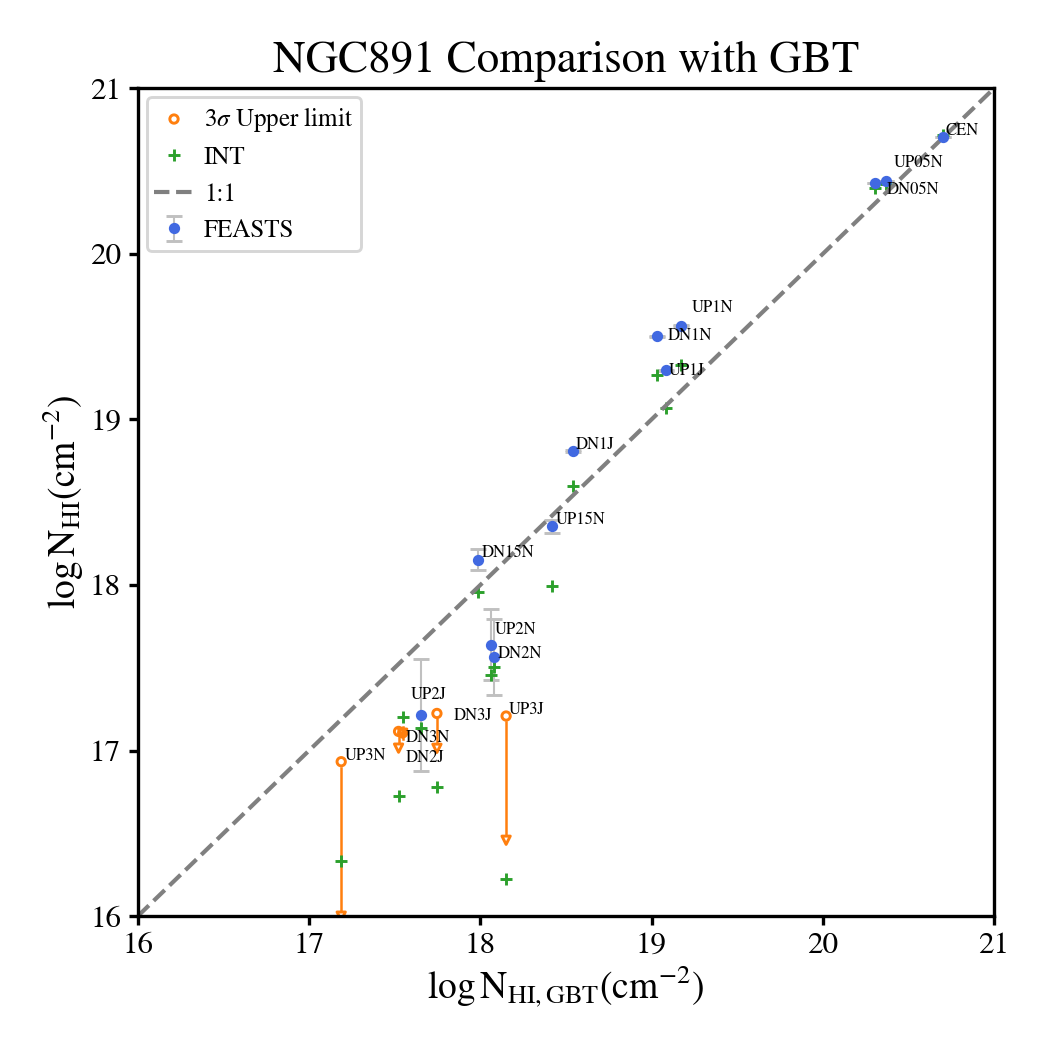}
    \includegraphics{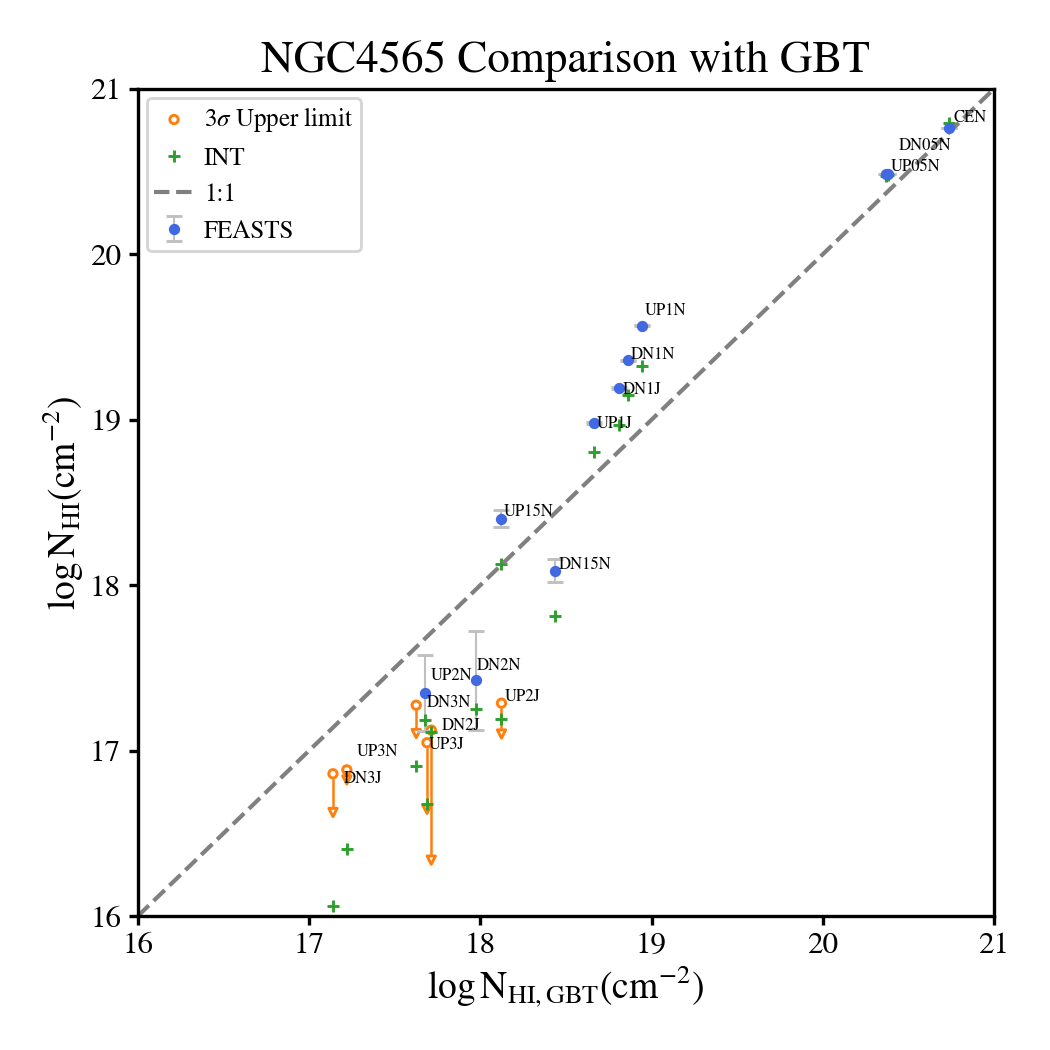}
    \caption{The flux comparison between FEASTS, HALOGAS and GBT of 17 pointings for NGC 891 (left) and NGC 4565 (right). The upper limits are plotted for the pointings with flux below $3\sigma$. }\label{fig:gbt_com}
\end{figure*}

The spectra for FEASTS, GBT and HALOGAS data are shown in Figure \ref{fig:gbt_j_N891} and Figure \ref{fig:gbt_j_N4565}. We sum up the flux in the same velocity range as~\citet{das2024} and convert it to the column density. The comparison in column densities is shown in Figure~\ref{fig:gbt_com}. For the GBT detection but FEASTS non-detection, we calculate the $3\sigma$ upper limit assuming the same velocity width as the spectra of the GBT data. 
The column densities calculated with the FEASTS data for most of the pointings 2 GBT-beam or further away from the disk are either below $3\sigma$ detection limits or lower than GBT values. We check the spectral comparison in Figure \ref{fig:gbt_j_N891} and Figure \ref{fig:gbt_j_N4565} and reach consistent conclusion. 

Because convolving FEASTS data with the GBT PSF results in the final PSF slightly larger than the actual GBT PSF, we further test by convolving the FEASTS and HALOGAS data with a Gaussian beam whose FWHM equals $\sqrt{\rm FWHM_{GBT}^2-FWHM_{raw}^2}$ to approximate the central main lobe of the GBT PSF. The real situation should be between these two types of convolutions. We find qualitatively similar results as when using the circularized real GBT PSF. 

The discrepancy between FEASTS and GBT data for these pointings 2 or 3 GBT beam away from the disk plane motivates us to further compare with the HALOGAS data. The expectation is that these single-dish observations should detect similar or slightly more fluxes near the disk, and moderately or significantly more fluxes far away from the disk than the interferometric observation of HALOGAS. The FEASTS data show similar spectra shape and higher \HI flux when compared to the HALOGAS data in all positions where fluxes are detected in the HALOGAS data. Instead, the GBT spectra are different in shape and the spectral flux densities are sometimes significantly lower than the HALOGAS in the positions DN05N, UP1J, DN1J, UP1N, DN1N for NGC 891 and UP05N, DN05N, UP1N, DN1J, DN1N for NGC 4565. The better consistency of FEASTS data in comparison with HALOGAS data supports their accuracy of \HI measurements above the detection limit. It also supports the FEASTS upper limit of $10^{17.1}$ ($10^{17.3}$) \cmsq in $\NHI$ at distances of 3 GBT beam, or 79 (93) \kpc, away from the disk of NGC 891 (NGC 4565), which is lower than the GBT measurements at the same positions. 

In summary, at the same locations, the spectral flux intensities of FEASTS and HALOGAS are in good agreement, with FEASTS data typically matching or exceeding the HALOGAS values. However the GBT results show significant inconsistencies: at the same location near the disk, GBT spectral intensities can be much lower than those from HALOGAS in many channels. On the other hand, at the same location far away from the disk, the integrated spectral fluxes of GBT are much higher than the upper limits of FEASTS in several cases.
We thus conclude that, the excess signals detected by GBT for the points 2/3 GBT beam away from the disk are unlikely to be real. They may be due to beam under-estimation, sidelobe contamination that is not captured by their assumed circularized beam, and/or pointing errors during the GBT observations. The exact explanation may need to await further experiments at the GBT.

\section{\HI distribution in mocks of TNG50}\label{appendix:tng}
We use TNG50 mock simulations \citep{weng2024d} to study the \HI distribution both along and perpendicular to the disk. We select galaxies at redshift of $z=0.5$ with an axis ratio of $b/a < 0.35$ (corresponding to inclinations $i \ge \ang{70}$). The sample is drawn from two stellar mass bins of $10^9$ and $10^{10}$ \uMsun, each with a scatter of 0.2 dex. To approximate gas distribution along the minor axis, we adopt a conical opening angle of \ang{35}. The resulting \HI distribution is shown in Figure \ref{fig:tng50}. Although sample selection and measurements differ slightly from our observational study, the TNG50 mocks show a similar trend in \HI extension at $10^{18}$ \cmsq, with \HI extending further along the major axis than the minor axis. This is in qualitative consistency with our results discussed in Section \ref{sec:discuss_absorption}. Additionally, the results indicate that \HI gas of galaxies with higher stellar mass tends to extend further. A closer comparison of prediction from different simulations with the observational results of this study will be conducted in the future.

\bibliography{EdgeonHI}{}
\bibliographystyle{aasjournal}

\end{CJK*}
\end{document}